\documentclass[nojss]{jss}\usepackage[]{graphicx}\usepackage[]{color}
\makeatletter
\def\maxwidth{ %
  \ifdim\Gin@nat@width>\linewidth
    \linewidth
  \else
    \Gin@nat@width
  \fi
}
\makeatother

\definecolor{fgcolor}{rgb}{0.345, 0.345, 0.345}
\newcommand{\hlnum}[1]{\textcolor[rgb]{0.686,0.059,0.569}{#1}}%
\newcommand{\hlstr}[1]{\textcolor[rgb]{0.192,0.494,0.8}{#1}}%
\newcommand{\hlcom}[1]{\textcolor[rgb]{0.678,0.584,0.686}{\textit{#1}}}%
\newcommand{\hlopt}[1]{\textcolor[rgb]{0,0,0}{#1}}%
\newcommand{\hlstd}[1]{\textcolor[rgb]{0.345,0.345,0.345}{#1}}%
\newcommand{\hlkwa}[1]{\textcolor[rgb]{0.161,0.373,0.58}{\textbf{#1}}}%
\newcommand{\hlkwb}[1]{\textcolor[rgb]{0.69,0.353,0.396}{#1}}%
\newcommand{\hlkwc}[1]{\textcolor[rgb]{0.333,0.667,0.333}{#1}}%
\newcommand{\hlkwd}[1]{\textcolor[rgb]{0.737,0.353,0.396}{\textbf{#1}}}%

\usepackage{framed}
\makeatletter
\newenvironment{kframe}{%
 \def\at@end@of@kframe{}%
 \ifinner\ifhmode%
  \def\at@end@of@kframe{\end{minipage}}%
  \begin{minipage}{\columnwidth}%
 \fi\fi%
 \def\FrameCommand##1{\hskip\@totalleftmargin \hskip-\fboxsep
 \colorbox{shadecolor}{##1}\hskip-\fboxsep
     \hskip-\linewidth \hskip-\@totalleftmargin \hskip\columnwidth}%
 \MakeFramed {\advance\hsize-\width
   \@totalleftmargin\z@ \linewidth\hsize
   \@setminipage}}%
 {\par\unskip\endMakeFramed%
 \at@end@of@kframe}
\makeatother

\definecolor{shadecolor}{rgb}{.97, .97, .97}
\definecolor{messagecolor}{rgb}{0, 0, 0}
\definecolor{warningcolor}{rgb}{1, 0, 1}
\definecolor{errorcolor}{rgb}{1, 0, 0}
\newenvironment{knitrout}{}{} 

\usepackage{alltt}
\usepackage[T1]{fontenc}
\usepackage[utf8]{inputenc}%
\usepackage{lmodern}
\usepackage[american]{babel}  
\usepackage{bm,amsmath,thumbpdf,amsfonts}
\usepackage{blkarray}
\newcommand{\github}{Github}
\DeclareMathOperator{\tr}{tr}
\DeclareMathOperator{\VEC}{vec}

\shortcites{bolker_strategies_2013,sleepstudy,gelman2013bayesian}

\author{Douglas Bates\\U. of Wisconsin - Madison\And
  Martin M\"achler\\ETH Zurich\And
  Benjamin M. Bolker\\McMaster University\And
  Steven C. Walker\\McMaster University
}
\Plainauthor{Douglas Bates, Martin M\"achler, Ben Bolker, Steve Walker}
\title{Fitting linear mixed-effects models using \pkg{lme4}}
\Plaintitle{Fitting Linear Mixed-Effects Models using lme4}
\Shorttitle{Linear Mixed Models with lme4}
\Abstract{%

  Maximum likelihood or restricted maximum likelihood (REML) estimates
  of the parameters in linear mixed-effects models can be determined
  using the \code{lmer} function in the \pkg{lme4} package for
  \proglang{R}. As for most model-fitting functions in \proglang{R},
  the model is described in an \code{lmer} call by a formula, in this
  case including both fixed- and random-effects
  terms. The formula and data together determine a numerical
  representation of the model from which the profiled deviance or the
  profiled REML criterion can be evaluated as a function of some of
  the model parameters.  The appropriate criterion is optimized, using
  one of the constrained optimization functions in \proglang{R}, to
  provide the parameter estimates.  We describe the structure of the
  model, the steps in evaluating the profiled deviance or REML
  criterion, and the structure of classes or types that represents such
  a model.  Sufficient detail is included to allow specialization of
  these structures by users who wish to write functions to fit
  specialized linear mixed models, such as models incorporating
  pedigrees or smoothing splines, that are not easily expressible in
  the formula language used by \code{lmer}.}

\Keywords{%
  sparse matrix methods, linear mixed models, penalized least squares,
  Cholesky decomposition} \Address{
  Douglas Bates\\
  Department of Statistics, University of Wisconsin\\
  1300 University Ave.\\
  Madison, WI 53706, U.S.A.\\
  E-mail: \email{bates@stat.wisc.edu}
  \par\bigskip
  Martin M\"achler\\
  Seminar f\"ur Statistik, HG G~16\\
  ETH Zurich\\
  8092 Zurich, Switzerland\\
  E-mail: \email{maechler@stat.math.ethz.ch}\\
  \par\bigskip
  Benjamin M. Bolker\\
  Departments of Mathematics \& Statistics and Biology \\
  McMaster University \\
  1280 Main Street W \\
  Hamilton, ON L8S 4K1, Canada \\
  E-mail: \email{bolker@mcmaster.ca}
  \par\bigskip
  Steven C. Walker\\
  Department of Mathematics \& Statistics \\
  McMaster University \\
  1280 Main Street W \\
  Hamilton, ON L8S 4K1, Canada \\
  E-mail: \email{scwalker@math.mcmaster.ca }
}

\newcommand{\betavec}{{\bm\beta}}
\newcommand{\Var}{\operatorname{Var}}

\newcommand{\bLt}{\ensuremath{\bm\Lambda_{\bm\theta}}}
\newcommand{\mc}[1]{\ensuremath{\mathcal{#1}}}
\newcommand{\trans}{\ensuremath{^\top}} 
\newcommand{\yobs}{\ensuremath{\bm y_{\mathrm{obs}}}}
\newcommand*{\eq}[1]{eqn.~\ref{#1}}

\setkeys{Gin}{width=\textwidth}
\setkeys{Gin}{height=3.5in}
\IfFileExists{upquote.sty}{\usepackage{upquote}}{}
\begin{document}

Submitted to \emph{Journal of Statistical Software}

\section{Introduction}
\label{sec:intro}

The \pkg{lme4} package for \proglang{R} provides functions to fit and
analyze linear mixed models, generalized linear mixed models and
nonlinear mixed models.  In each of these names, the term ``mixed''
or, more fully, ``mixed effects'', denotes a model that incorporates
both fixed- and random-effects terms in a linear predictor expression
from which the conditional mean of the response can be evaluated.  In
this paper we describe the formulation and representation of linear
mixed models.  The techniques used for generalized linear and
nonlinear mixed models will be described separately, in a future
paper.

The development of general software for fitting mixed models remains
an active area of research with many open problems.  Consequently, the
\pkg{lme4} package has evolved since it was first released, and
continues to improve as we learn more about mixed models. However, we
recognize the need to maintain stability and backward compatibility of
\pkg{lme4} so that it continues to be broadly useful. In order to
maintain stability while continuing to advance mixed-model
computation, we have developed several additional frameworks that draw
on the basic ideas of \code{lme4} but modify its structure or
implementation in various ways. These descendants include the
\mbox{\pkg{MixedModels}} package in \proglang{Julia}, the
\pkg{lme4pureR} package in \proglang{R}, and the \code{flexLambda}
development branch of \pkg{lme4}.  The current article is largely
restricted to describing the current stable version of the \pkg{lme4}
package (1.1-7), with Appendix~\ref{sec:modularExamples} describing
hooks into the computational machinery that are designed for extension
development. The \pkg{gamm4} \citep{gamm4} and \pkg{blme} \citep{blme}
packages currently make use of these hooks.

Another goal of this article is to contrast the approach used by
\pkg{lme4} with previous formulations of mixed models.  The
expressions for the profiled log-likelihood and profiled REML
(restricted maximum likelihood) criteria derived in
Section~\ref{sec:profdev} are similar to those presented in
\citet{bates04:_linear} and, indeed, are closely related to
``Henderson's mixed-model equations''~\citep{henderson_1982}.
Nonetheless there are subtle but important changes in the formulation
of the model and in the structure of the resulting penalized least
squares (PLS) problem to be solved (Section~\ref{sec:PLSpureR}).  We
derive the current version of the PLS problem
(Section~\ref{sec:plsMath}) and contrast this result with earlier
formulations (Section~\ref{sec:previous_lmm_form}).

This article is organized into four main sections (\ref{sec:lFormula};
\ref{sec:mkLmerDevfun}; \ref{sec:optimizeLmer}; \ref{sec:mkMerMod}),
each of which corresponds to one of the four largely separate modules
that comprise \pkg{lme4}.  Before describing the details of each
module, we describe the general form of the linear mixed model
underlying \pkg{lme4} (Section~\ref{sec:LMMs}); introduce the
\code{sleepstudy} data that will be used as an example throughout
(Section~\ref{sec:sleepstudy}); and broadly outline \pkg{lme4}'s
modular structure (Section~\ref{sec:modular}).

\subsection{Linear mixed models}
\label{sec:LMMs}

Just as a linear model is described by the distribution of a
vector-valued random response variable, $\mc{Y}$, whose observed value
is $\yobs$, a linear mixed model is described by the distribution of
two vector-valued random variables: $\mc{Y}$, the response, and
$\mc{B}$, the vector of random effects.  In a linear model the
distribution of $\mc Y$ is multivariate normal,
\begin{equation}
  \label{eq:linearmodel}
  \mc Y\sim\mc{N}(\bm X\bm\beta+\bm o,\sigma^2\bm W^{-1}),
\end{equation}
where $n$ is the dimension of the response vector, $\bm W$ is a
diagonal matrix of known prior weights, $\bm\beta$ is a
$p$-dimensional coefficient vector, $\bm X$ is an $n\times p$ model
matrix, and $\bm o$ is a vector of known prior offset terms. The
parameters of the model are the coefficients $\bm\beta$ and the scale
parameter $\sigma$.

In a linear mixed model it is the \emph{conditional} distribution of
$\mc Y$ given $\mc B=\bm b$ that has such a form,
\begin{equation}
  \label{eq:LMMcondY}
  ( \mc Y|\mc B=\bm b)\sim\mc{N}(\bm X\bm\beta+\bm Z\bm b+\bm o,\sigma^2\bm W^{-1}), 
\end{equation}
where $\bm Z$ is the $n\times q$ model matrix for the $q$-dimensional
vector-valued random effects variable, $\mc B$, whose value we are
fixing at $\bm b$.  The unconditional distribution of $\mc B$ is also
multivariate normal with mean zero and a parameterized $q\times q$
variance-covariance matrix, $\bm\Sigma$,
\begin{equation}
  \label{eq:LMMuncondB}
  \mc B\sim\mc N(\bm0,\bm\Sigma) .
\end{equation}
As a variance-covariance matrix, $\bm\Sigma$ must be positive
semidefinite.  It is convenient to express the model in terms of a
\emph{relative covariance factor}, $\bLt$, which is a
$q\times q$ matrix, depending on the \emph{variance-component
  parameter}, $\bm\theta$, and generating the symmetric $q\times q$
variance-covariance matrix, $\bm\Sigma$, according to
\begin{equation}
  \label{eq:relcovfac}
  \bm\Sigma_{\bm\theta}=\sigma^2\bLt\bLt\trans ,
\end{equation}
where $\sigma$ is the same scale factor as in the conditional
distribution (\ref{eq:LMMcondY}).

Although equations \ref{eq:LMMcondY}, \ref{eq:LMMuncondB}, and
\ref{eq:relcovfac} fully describe the class of linear mixed models
that \pkg{lme4} can fit, this terse description
hides many important details.  Before moving on to these details, we
make a few observations:
\begin{itemize}
\item This formulation of linear mixed models allows for a relatively
  compact expression for the profiled log-likelihood of $\bm\theta$
  (Section~\ref{sec:profdev}, Equation~\ref{eq:profiledDeviance}).
\item The matrices associated with random effects, $\bm Z$ and $\bLt$,
  typically have a sparse structure with a sparsity pattern that
  encodes various model assumptions.  Sections~\ref{sec:LMMmatrix} and
  \ref{sec:CSCmats} provide details on these structures, and how to
  represent them efficiently.
\item The interface provided by \pkg{lme4}'s \code{lmer} function 
  is slightly less general than the model described by equations
  \ref{eq:LMMcondY}, \ref{eq:LMMuncondB}, and \ref{eq:relcovfac}.  To
  take advantage of the entire range of possibilities, one may use the
  modular functions (Sections~\ref{sec:modular} and
  Appendix~\ref{sec:modularExamples}) or explore the
  experimental \code{flexLambda} branch of \pkg{lme4} on \github.
\end{itemize}

\subsection{Example}
\label{sec:sleepstudy}

Throughout our discussion of \pkg{lme4}, we will work with a data set
on the average reaction time per day for subjects in a sleep
deprivation study \citep{sleepstudy}. On day 0 the subjects had their
normal amount of sleep.  Starting that night they were restricted to 3
hours of sleep per night.  The response variable, \code{Reaction},
represents average reaction times in milliseconds (ms) on a series of tests given each
\code{Day} to each \code{Subject} (Figure~\ref{fig:sleepPlot}),
\begin{knitrout}
\definecolor{shadecolor}{rgb}{0.969, 0.969, 0.969}\color{fgcolor}\begin{kframe}
\begin{alltt}
\hlstd{> }\hlkwd{str}\hlstd{(sleepstudy)}
\end{alltt}
\begin{verbatim}
'data.frame':	180 obs. of  3 variables:
 $ Reaction: num  250 259 251 321 357 ...
 $ Days    : num  0 1 2 3 4 5 6 7 8 9 ...
 $ Subject : Factor w/ 18 levels "308","309","310",..: 1 1 1 1 1 1 ..
\end{verbatim}
\end{kframe}
\end{knitrout}

\begin{knitrout}
\definecolor{shadecolor}{rgb}{0.969, 0.969, 0.969}\color{fgcolor}\begin{figure}[tb]

{\centering \includegraphics[width=\maxwidth]{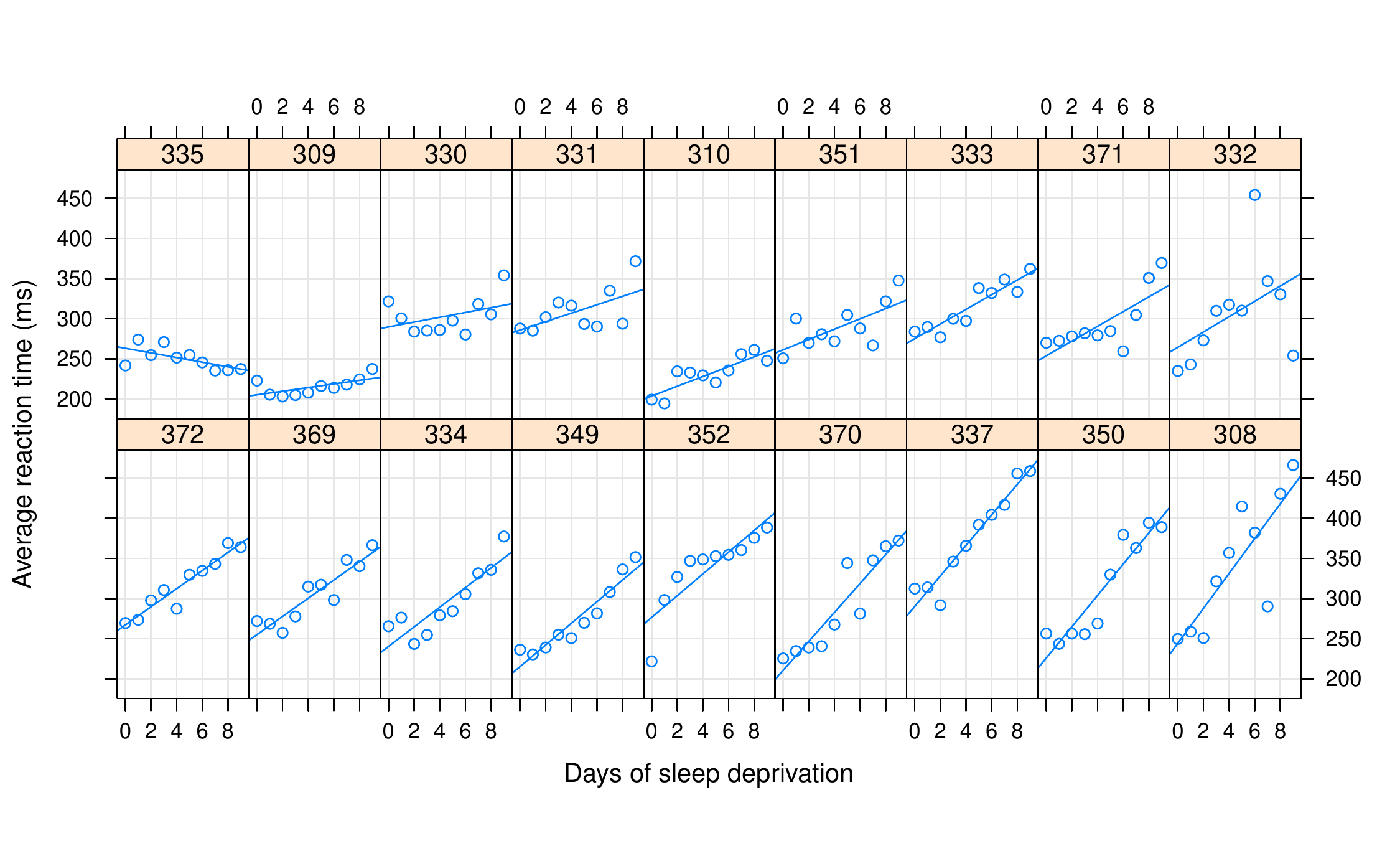} 

}

\caption[Reaction time versus days by subject]{Average reaction time versus days of sleep deprivation by subject.  Subjects ordered by increasing slope of subject-specific linear regressions.\label{fig:sleepPlot}}
\end{figure}

\end{knitrout}

Each subject's reaction time increases approximately linearly
with the number of sleep-deprived days.
However, subjects also appear to vary in the slopes and intercepts
of these relationships, which suggests a model with random slopes and intercepts.
As we shall see, such a model may be fitted by minimizing the
REML criterion (Equation~\ref{eq:REMLdeviance}) using
\begin{knitrout}
\definecolor{shadecolor}{rgb}{0.969, 0.969, 0.969}\color{fgcolor}\begin{kframe}
\begin{alltt}
\hlstd{> }\hlstd{fm1} \hlkwb{<-} \hlkwd{lmer}\hlstd{(Reaction} \hlopt{~} \hlstd{Days} \hlopt{+} \hlstd{(Days} \hlopt{|} \hlstd{Subject), sleepstudy)}
\end{alltt}
\end{kframe}
\end{knitrout}
The estimates of the standard deviations of the random effects for the
intercept and the slope are
24.74 ms 
and
5.92 ms/day. 
The fixed-effects coefficients, $\betavec$, are
251.4 ms and
10.47 ms/day
for the intercept and slope.  In this model, one interpretation of
these fixed effects is that they are the estimated population mean values of the
random intercept and slope (Section~\ref{sec:intuitiveFormulas}).

\subsection{High level modular structure}
\label{sec:modular}

The \code{lmer} function is composed of four largely independent
modules.  In the first module, a mixed-model formula is parsed and
converted into the inputs required to specify a linear mixed model
(Section~\ref{sec:lFormula}).  The second module uses these inputs to
construct an \proglang{R} function which takes the covariance
parameters, $\bm\theta$, as arguments and returns negative twice the log
profiled likelihood or the REML criterion
(Section~\ref{sec:mkLmerDevfun}).  The third module optimizes this
objective function to produce maximum likelihood (ML) or REML
estimates of $\bm\theta$ (Section~\ref{sec:optimizeLmer}).  Finally,
the fourth module provides utilities for interpreting the optimized
model (Section~\ref{sec:mkMerMod}).

\begin{table}[tb]
  \centering
  \begin{tabular}{lllp{2.1in}}
    \hline
    Module & & \proglang{R} function & Description \\
    \hline
    Formula module & (Section~\ref{sec:lFormula}) & \code{lFormula} &
    Accepts a mixed-model formula, data, and other user inputs, and returns
    a list of objects required to fit a linear mixed model. \\
    Objective function module & (Section~\ref{sec:mkLmerDevfun}) &
    \code{mkLmerDevfun} &
    Accepts the results of \code{lFormula} and returns a function to
    calculate the deviance (or restricted deviance) as a function of the
    covariance parameters, $\bm\theta$.\\
    Optimization module & (Section~\ref{sec:optimizeLmer}) &
    \code{optimizeLmer} &
    Accepts a deviance function returned by \code{mkLmerDevfun} and
    returns the results of the optimization of that deviance function. \\
    Output module & (Section~\ref{sec:mkMerMod}) & \code{mkMerMod} &
    Accepts an optimized deviance function and packages the results
    into a useful object.  \\
    \hline
  \end{tabular}
  \caption{The high-level modular structure of \code{lmer}.}
  \label{tab:modular}
\end{table}

To illustrate this modularity, we recreate the \code{fm1} object by a
series of four modular steps,
\begin{knitrout}
\definecolor{shadecolor}{rgb}{0.969, 0.969, 0.969}\color{fgcolor}\begin{kframe}
\begin{alltt}
\hlstd{> }                                        \hlcom{# formula module}
\hlstd{> }\hlstd{parsedFormula} \hlkwb{<-} \hlkwd{lFormula}\hlstd{(}\hlkwc{formula} \hlstd{= Reaction} \hlopt{~} \hlstd{Days} \hlopt{+} \hlstd{(Days}\hlopt{|}\hlstd{Subject),}
\hlstd{+ }                             \hlkwc{data} \hlstd{= sleepstudy)}
\hlstd{> }
\hlstd{> }                                        \hlcom{# objective function module}
\hlstd{> }\hlstd{devianceFunction} \hlkwb{<-} \hlkwd{do.call}\hlstd{(mkLmerDevfun, parsedFormula)}
\hlstd{> }
\hlstd{> }                                        \hlcom{# optimization module}
\hlstd{> }\hlstd{optimizerOutput} \hlkwb{<-} \hlkwd{optimizeLmer}\hlstd{(devianceFunction)}
\hlstd{> }
\hlstd{> }                                        \hlcom{# output module}
\hlstd{> }\hlkwd{mkMerMod}\hlstd{(}   \hlkwc{rho} \hlstd{=} \hlkwd{environment}\hlstd{(devianceFunction),}
\hlstd{+ }            \hlkwc{opt} \hlstd{= optimizerOutput,}
\hlstd{+ }         \hlkwc{reTrms} \hlstd{= parsedFormula}\hlopt{$}\hlstd{reTrms,}
\hlstd{+ }             \hlkwc{fr} \hlstd{= parsedFormula}\hlopt{$}\hlstd{fr)}
\end{alltt}
\end{kframe}
\end{knitrout}

\section{Formula module}
\label{sec:lFormula}

\subsection{Mixed-model formulas}
\label{sec:formulas}

Like most model-fitting functions in \proglang{R},
\code{lmer} takes as its first two arguments a \emph{formula}
specifying the model and the \emph{data} with which to evaluate the
formula. This second argument, \code{data}, is optional but
recommended and is usually the name of an \proglang{R} data frame.
In the \proglang{R} \code{lm} function for fitting linear models,
formulas take the form \verb+resp ~ expr+,
where \code{resp} determines the response variable and \code{expr}
is an expression that specifies the columns of the model matrix.
Formulas for the \code{lmer} function contain special random-effects
terms,
\begin{knitrout}
\definecolor{shadecolor}{rgb}{0.969, 0.969, 0.969}\color{fgcolor}\begin{kframe}
\begin{alltt}
\hlstd{> }\hlstd{resp} \hlopt{~} \hlstd{FEexpr} \hlopt{+} \hlstd{(REexpr1}\hlopt{|}\hlstd{factor1)} \hlopt{+} \hlstd{(REexpr2}\hlopt{|}\hlstd{factor2)} \hlopt{+} \hlstd{...}
\end{alltt}
\end{kframe}
\end{knitrout}
where \code{FEexpr} is an expression determining the columns of the
fixed-effects model matrix, $\bm X$, and the random-effects terms,
\code{(REexpr1 | factor1)} and \code{(REexpr2 | factor2)}, determine
both the random-effects model matrix, $\bm Z$ (Section~\ref{sec:mkZ}),
and the structure of the relative covariance factor,
$\bLt$ (Section~\ref{sec:mkLambdat}).  In principle, a
mixed-model formula may contain arbitrarily many random-effects terms,
but in practice the number of such terms is typically low.

\subsection{Understanding mixed-model formulas}
\label{sec:intuitiveFormulas}

Before describing the details of how \pkg{lme4} parses mixed-model
formulas (Section~\ref{sec:LMMmatrix}), we provide an informal
explanation and then some examples. Our discussion assumes familiarity
with the standard \proglang{R} modelling paradigm
\citep{Chambers:1993}.

Each random-effects term is of the form \code{(expr|factor)}.
The expression \code{expr} is evaluated as a linear model formula,
producing a model matrix following the same rules used in standard
\proglang{R} modelling functions (e.g., \code{lm} or \code{glm}). The
expression \code{factor} is evaluated as an \proglang{R} factor.  One
way to think about the vertical bar operator is as a special kind of
interaction between the model matrix and the grouping factor. This
interaction ensures that the columns of the model matrix have
different effects for each level of the grouping factor.  What makes
this a special kind of interaction is that these effects are modelled
as unobserved random variables, rather than unknown fixed parameters.
Much has been written about important practical and philosophical
differences between these two types of interactions
(e.g., \citet{henderson_1982}; \citet{gelman2005analysis}).  For
example, the random-effects implementation of such interactions can be
used to obtain shrinkage estimates of regression coefficients
(e.g., \citet{1977EfronAndMorris}), or account
for lack of independence in the residuals due to block structure or
repeated measurements (e.g., \citet{laird_ware_1982}).

Table~\ref{tab:formulas} provides several examples of the
right-hand-sides of mixed-model formulas. The first example,
\code{(1|g)}, 
is the simplest possible mixed-model formula, where each level of the
grouping factor, \code{g}, has its own random intercept. The mean and
standard deviation of these intercepts are parameters to be estimated.
Our description of this model incorporates any non-zero mean of the
random effects as fixed-effects parameters.  If one wishes to specify
that a random intercept has \emph{a priori} known means, one may use
the \code{offset} function as in the second  model in
Table~\ref{tab:formulas}.  This model contains no fixed effects, or
more accurately the fixed-effects model matrix, $\bm X$, has zero
columns and $\bm\beta$ has length zero.

\begin{table}[tb]
  \centering
  \begin{tabular}{lll}
    \hline
    Formula 			& Alternative & Meaning \\
    \hline
    \code{(1|g)}		& \code{1+(1|g)} 
    & Random intercept with fixed mean \\
    \code{0+offset(o)+(1|g)}    & \code{-1+offset(o)+(1|g)}
    & Random intercept with \emph{a priori} means \\
    \code{(1|g1/g2)} 		& \code{(1|g1)+(1|g1:g2)}   
    & Intercept varying among \code{g1} and \code{g2} within \code{g1} \\
    \code{(1|g1)+(1|g2)} 	& \code{1+(1|g1)+(1|g2)} &
    Intercept varying among \code{g1} and \code{g2} \\
    \code{x+(x|g)} 		& \code{1+x+(1+x|g)} &
    Correlated random intercept and slope \\
    \code{x+(x||g)} 		& \code{1+x+(1|g)+(0+x|g)}
    & Uncorrelated random intercept and slope \\
    \hline
  \end{tabular}
  \caption{Examples of the right-hand sides of mixed-effects model
    formulas. The names of grouping factors are denoted \code{g},
    \code{g1}, and \code{g2}, and covariates and \emph{a priori} known
    offsets as \code{x} and \code{o}.}
  \label{tab:formulas}
\end{table}

We may also construct models with multiple grouping factors. For
example, if the observations are grouped by \code{g2}, which is nested
within \code{g1}, then the third formula in Table \ref{tab:formulas}
can be used to model variation in the intercept.  A common objective
in mixed modeling is to account for such nested (or hierarchical)
structure.  However, one of the most useful aspects of \pkg{lme4} is
that it can be used to fit random-effects associated with non-nested
grouping factors.  For example, suppose the data are grouped by fully
crossing two factors, \code{g1} and \code{g2}, then the fourth formula
in Table \ref{tab:formulas} may be used.  Such models are common in
item response theory, where \code{subject} and \code{item} factors are
fully crossed \citep{doran2007estimating}. In addition to varying
intercepts, we may also have varying slopes (e.g., the
\code{sleepstudy} data, Section~\ref{sec:sleepstudy}).  The fifth
example in Table~\ref{tab:formulas} gives a model where both the
intercept and slope vary among the levels of the grouping factor.

\subsubsection{Specifying uncorrelated random effects}
\label{sec:uncor}

By default, \pkg{lme4} assumes that all coefficients associated with
the same random-effects term are correlated. To specify an
uncorrelated slope and intercept (for example), one may either use
double-bar notation, \code{(x||g)}, or equivalently use multiple
random effects terms, \code{x+(1|g)+(0+x|g)}, as in the final example
of Table~\ref{tab:formulas}.  For example, from the printout of model
\code{fm1} of the \code{sleepstudy} data
(Section~\ref{sec:sleepstudy}), we see that the estimated correlation
between the slope for \code{Days} and the intercept is fairly low
(0.066) 
(See Section~\ref{sec:summary} below for more on how to extract the
random effects covariance matrix.)  We may use double-bar notation to
fit a model that excludes a correlation parameter:
\begin{knitrout}
\definecolor{shadecolor}{rgb}{0.969, 0.969, 0.969}\color{fgcolor}\begin{kframe}
\begin{alltt}
\hlstd{> }\hlstd{fm2} \hlkwb{<-} \hlkwd{lmer}\hlstd{(Reaction} \hlopt{~} \hlstd{Days} \hlopt{+} \hlstd{(Days} \hlopt{||} \hlstd{Subject), sleepstudy)}
\end{alltt}
\end{kframe}
\end{knitrout}

Although mixed models where the random slopes and intercepts are
assumed independent are commonly used to reduce the complexity of
random-slopes models, they do have one subtle drawback.  Models in
which the slopes and intercepts are allowed to have a non-zero
correlation (e.g., \code{fm1}) are invariant to additive shifts of the
continuous predictor (\code{Days} in this case).  This invariance
breaks down when the correlation is constrained to zero; any shift in
the predictor will necessarily lead to a change in the estimated
correlation, and in the likelihood and predictions of the model.  For
example, we can eliminate the correlation in \code{fm1} simply by
shifting \code{Days} by an amount equal to the ratio of the estimated
among-subject standard deviations multiplied by the estimated
correlation (i.e., $\sigma_{\text{\small slope}}/\sigma_{\text{\small
    intercept}} \cdot \rho_{\text{\small slope:intercept}}$).  The use
of such models should ideally be restricted to cases where the
predictor is measured on a ratio scale (i.e., the zero point on the
scale is meaningful, not just a location defined by convenience or
convention).


\subsection{Algebraic and computational account of mixed-model formulas}
\label{sec:LMMmatrix}

The fixed-effects terms of a mixed-model formula are parsed to produce
the fixed-effects model matrix, $\bm X$, in the same way that the
\proglang{R} \code{lm} function generates model matrices.  However, a
mixed-model formula incorporates $k\ge1$ random-effects terms of the
form \code{(r|f)} as well. 
These $k$ terms are used to produce the random effects model matrix,
$\bm Z$ (Equation~\ref{eq:LMMcondY}; Section~\ref{sec:mkZ}), and the
structure of the relative covariance factor, $\bLt$
(Equation~\ref{eq:relcovfac}; Section~\ref{sec:mkLambdat}), which are
matrices that typically have a sparse structure.  We now describe how
one might construct these matrices from the random-effects terms,
considering first a single term, \code{(r|f)}, 
and then generalizing to multiple terms. Tables~\ref{tab:dim} and
\ref{tab:algebraic} summarize the matrices and vectors that determine
the structure of $\bm Z$ and
$\bLt$.

\begin{table}[tb]
  \centering
  \begin{tabular}{lll}
    \hline
    Symbol & Size \\
    \hline
    $n$ & Length of the response vector, $\mc{Y}$ \\
    $p$ & Number of columns of fixed effects model matrix, $\bm X$ \\
    $q = \sum_i^k q_i$ & Number of columns of random effects model matrix, $\bm Z$ \\
    $p_i$ &  Number of columns of the raw model matrix, $\bm X_i$ \\
    $\ell_i$ &  Number of levels of the grouping factor indices, $\bm i_i$ \\
    $q_i = p_i\ell_i$  & Number of columns of the term-wise model matrix, $\bm Z_i$ \\
    $k$ &  Number of random effects terms \\
    $m_i = \binom{p_i+1}{2}$ & Number of covariance parameters for term
    $i$ \\
    $m = \sum_i^k m_i$ & Total number of covariance parameters \\
    \hline
  \end{tabular}
  \caption{Dimensions of linear mixed models.  The
    subscript $i = 1, \dots, k$ denotes a specific random effects term.}
  \label{tab:dim}
\end{table}

\begin{table}[tb]
  \centering
  \begin{tabular}{lll}
    \hline
    Symbol  & Size & Description \\
    \hline
    $\bm X_i$ & $n\times p_i$ &  Raw random effects model matrix \\
    $\bm J_i$  & $n\times \ell_i$ & Indicator matrix of grouping factor indices\\
    $\bm X_{ij}$ & $p_i\times 1$ & Column vector containing $j$th row of $\bm X_i$ \\
    $\bm J_{ij}$ & $\ell_i\times 1$ & Column vector containing $j$th row of $\bm J_i$ \\
    $\bm i_i$  & $n$ & Vector of grouping factor indices \\
    $\bm Z_i$ & $n\times q_i$ & Term-wise random effects model matrix \\
    $\bm\theta$ & $m$ & Covariance parameters \\
    $\bm T_i$ & $p_i\times p_i$ & Lower triangular template matrix \\
    $\bm\Lambda_i$ & $q_i\times q_i$ & Term-wise relative covariance factor \\
    \hline
  \end{tabular}
  \caption{Symbols used to describe the structure of the random
    effects model matrix and the relative covariance factor. The
    subscript $i = 1, \dots, k$ denotes a specific random effects term.}
  \label{tab:algebraic}
\end{table}

The expression, \code{r}, is a linear model formula that evaluates to
an \proglang{R} model matrix, $\bm X_i$, of size $n\times p_i$, called
the \emph{raw random effects model matrix} for term $i$. A term is
said to be a \emph{scalar} random-effects term when $p_i=1$, otherwise
it is \emph{vector-valued}.  For a \emph{simple, scalar}
random-effects term of the form \code{(1|f)}, $\bm X_i$ is the    
$n\times 1$ matrix of ones, which implies a random intercept model.

The expression \code{f} evaluates to an \proglang{R} factor, called
the \emph{grouping factor}, for the term. For the $i$th term, we
represent this factor mathematically with a vector $\bm i_i$ of
\emph{factor indices}, which is an $n$-vector of values from
$1,\dots,\ell_i$.\footnote{In practice,
  fixed-effects model matrices and random-effects terms are evaluated
  with respect to a \emph{model frame}, ensuring that any expressions
  for grouping factors have been coerced to factors and any unused
  levels of these factors have been dropped.  That is, $\ell_i$, the
  number of levels in the grouping factor for the $i$th random-effects
  term, is well-defined.}
Let $\bm J_i$ be the $n\times \ell_i$ matrix of
indicator columns for $\bm i_i$.  Using the \pkg{Matrix} package 
\citep{Matrix_pkg} in
\proglang{R}, we may construct the transpose of $\bm J_i$ from a
factor vector, \code{f}, by coercing \code{f} to a \code{sparseMatrix}
object.  For example,

\begin{knitrout}
\definecolor{shadecolor}{rgb}{0.969, 0.969, 0.969}\color{fgcolor}\begin{kframe}
\begin{alltt}
\hlstd{> }\hlstd{(f} \hlkwb{<-} \hlkwd{gl}\hlstd{(}\hlnum{3}\hlstd{,}\hlnum{2}\hlstd{))}
\end{alltt}
\begin{verbatim}
[1] 1 1 2 2 3 3
Levels: 1 2 3
\end{verbatim}
\begin{alltt}
\hlstd{> }\hlstd{(Ji} \hlkwb{<-} \hlkwd{t}\hlstd{(}\hlkwd{as}\hlstd{(f,} \hlkwc{Class}\hlstd{=}\hlstr{"sparseMatrix"}\hlstd{)))}
\end{alltt}
\begin{verbatim}
6 x 3 sparse Matrix of class "dgCMatrix"
     1 2 3
[1,] 1 . .
[2,] 1 . .
[3,] . 1 .
[4,] . 1 .
[5,] . . 1
[6,] . . 1
\end{verbatim}
\end{kframe}
\end{knitrout}
When $k>1$ we order the random-effects terms so that
$\ell_1\ge\ell_2\ge\dots\ge\ell_k$; in general, this ordering reduces
``fill-in'' (i.e., the proportion of elements that are zero in the
lower triangle of $\bLt\trans\bm Z\trans\bm W\bm Z\bLt+\bm I$ but not in
the lower triangle of its left Cholesky factor, $\bm L_{\bm\theta}$,
described below in Equation~\ref{eq:blockCholeskyDecomp}). This reduction
in fill-in provides more efficient matrix operations within the
penalized least squares algorithm (Section~\ref{sec:plsMath}).

\subsubsection{Constructing the random effects model matrix}
\label{sec:mkZ}

The $i$th random effects term contributes $q_i=\ell_ip_i$ columns to
the model matrix $\bm Z$.  We group these columns into a matrix, $\bm
Z_i$, which we refer to as the \emph{term-wise model matrix} for the
$i$th term.  Thus $q$, the number of columns in $\bm Z$ and the
dimension of the random variable, $\mc{B}$, is
\begin{equation}
  \label{eq:qcalc}
  q=\sum_{i=1}^k q_i = \sum_{i=1}^k \ell_i\,p_i .
\end{equation}
Creating the matrix $\bm Z_i$ from $\bm X_i$ and $\bm J_i$ is a
straightforward concept that is, nonetheless, somewhat awkward to
describe.  Consider $\bm Z_i$ as being further decomposed into
$\ell_i$ blocks of $p_i$ columns.  The rows in the first block are the
rows of $\bm X_i$ multiplied by the 0/1 values in the first column of
$\bm J_i$.  Similarly for the subsequent blocks.  With these
definitions we may define the term-wise random-effects model matrix,
$\bm Z_i$, for the $i$th term as a transposed Khatri-Rao product,
\begin{equation}
  \label{eq:Zi}
  \bm Z_i = (\bm J_i\trans * \bm X_i\trans)\trans =
  \begin{bmatrix}
    \bm J_{i1}\trans \otimes \bm X_{i1}\trans \\
    \bm J_{i2}\trans \otimes \bm X_{i2}\trans \\
    \vdots \\
    \bm J_{in}\trans \otimes \bm X_{in}\trans \\
  \end{bmatrix}
\end{equation}
where $*$ and $\otimes$ are the Khatri-Rao\footnote{Note that the
  original definition of the Khatri-Rao product is more general than
  the definition used in \pkg{Matrix} package, which is the definition
  we use here.} \citep{khatri1968solutions} and Kronecker products,
and $\bm J_{ij}\trans$ and $\bm X_{ij}\trans$ are row vectors of the
$j$th rows of $\bm J_i$ and $\bm X_i$.  These rows correspond to the
$j$th sample in the response vector, $\mc Y$, and thus $j$ runs from
$1 \dots n$.  The \pkg{Matrix} package for \proglang{R} contains a
\code{KhatriRao} function, which can be used to form $\bm Z_i$.  For
example, if we begin with a raw model matrix,
\begin{knitrout}
\definecolor{shadecolor}{rgb}{0.969, 0.969, 0.969}\color{fgcolor}\begin{kframe}
\begin{alltt}
\hlstd{> }\hlstd{(Xi} \hlkwb{<-} \hlkwd{cbind}\hlstd{(}\hlnum{1}\hlstd{,}\hlkwd{rep.int}\hlstd{(}\hlkwd{c}\hlstd{(}\hlopt{-}\hlnum{1}\hlstd{,}\hlnum{1}\hlstd{),}\hlnum{3L}\hlstd{)))}
\end{alltt}
\begin{verbatim}
     [,1] [,2]
[1,]    1   -1
[2,]    1    1
[3,]    1   -1
[4,]    1    1
[5,]    1   -1
[6,]    1    1
\end{verbatim}
\end{kframe}
\end{knitrout}
then the term-wise random effects
model matrix is,
\begin{knitrout}
\definecolor{shadecolor}{rgb}{0.969, 0.969, 0.969}\color{fgcolor}\begin{kframe}
\begin{alltt}
\hlstd{> }\hlstd{(Zi} \hlkwb{<-} \hlkwd{t}\hlstd{(}\hlkwd{KhatriRao}\hlstd{(}\hlkwd{t}\hlstd{(Ji),}\hlkwd{t}\hlstd{(Xi))))}
\end{alltt}
\begin{verbatim}
6 x 6 sparse Matrix of class "dgCMatrix"
                   
[1,] 1 -1 .  . .  .
[2,] 1  1 .  . .  .
[3,] .  . 1 -1 .  .
[4,] .  . 1  1 .  .
[5,] .  . .  . 1 -1
[6,] .  . .  . 1  1
\end{verbatim}
\end{kframe}
\end{knitrout}

In particular, for a simple, scalar term, $\bm Z_i$ is exactly $\bm
J_i$, the matrix of indicator columns.  For other scalar terms, $\bm
Z_i$ is formed by element-wise multiplication of the single column of
$\bm X_i$ by each of the columns of $\bm J_i$.

Because each $\bm Z_i$ is generated from indicator columns, its
cross-product, $\bm Z_i\trans\bm Z_i$ is block-diagonal consisting of
$\ell_i$ diagonal blocks each of size $p_i$.\footnote{To see this,
  note that by the properties of Kronecker products we may write the
  cross-product matrix $Z_i\trans Z_i$ as $\sum_{j=1}^n \bm J_{ij} \bm
  J_{ij}\trans \otimes \bm X_{ij} \bm X_{ij}\trans$. Because $\bm
  J_{ij}$ is a unit vector along a coordinate axis, the cross product
  $\bm J_{ij} \bm J_{ij}\trans$ is a $p_i\times p_i$ matrix of all
  zeros except for a single $1$ along the diagonal. Therefore, the
  cross products, $\bm X_{ij} \bm X_{ij}\trans$, will be added to one
  of the $\ell_i$ blocks of size $p_i\times p_i$ along the diagonal of
  $Z_i\trans Z_i$.} Note that this means that when $k=1$ (i.e., there
is only one random-effects term, and $\bm Z_i = \bm Z$), $\bm
Z\trans\bm Z$ will be block diagonal.  These block-diagonal properties
allow for more efficient sparse matrix computations
(Section~\ref{sec:CSCmats}).

The full random effects model matrix, $\bm Z$, is constructed from
$k\ge 1$ blocks,
\begin{equation}
  \label{eq:Z}
  \bm Z =
  \begin{bmatrix}
    \bm Z_1 &
    \bm Z_2 &
    \hdots &
    \bm Z_k \\
  \end{bmatrix}
\end{equation}
By transposing Equation~\ref{eq:Z} and substituting in Equation~\ref{eq:Zi}, we
may represent the structure of the transposed random effects model
matrix as follows,
\begin{equation}
  \label{eq:Zt}
  \bm Z\trans =
  \begin{blockarray}{ccccc}
    \text{sample 1} & \text{sample 2} & \hdots & \text{sample n} & \\
    \begin{block}{[cccc]c}
    \bm J_{11} \otimes \bm X_{11} &
    \bm J_{12} \otimes \bm X_{12} &
    \hdots &
    \bm J_{1n} \otimes \bm X_{1n} &
    \text{term 1} \\
    \bm J_{21} \otimes \bm X_{21} &
    \bm J_{22} \otimes \bm X_{22} &
    \hdots &
    \bm J_{2n} \otimes \bm X_{2n} &
    \text{term 2} \\
    \vdots & \vdots & \ddots & \vdots & \vdots \\
    \end{block}
  \end{blockarray}
\end{equation}
Note that the proportion of elements of $Z\trans$ that are structural
zeros is
\begin{equation}
  \label{eq:ZtSparsity}
  \frac{\sum_{i=1}^k p_i(\ell_i - 1)}{\sum_{i=1}^k p_i} \qquad .
\end{equation}
Therefore, the sparsity of $\bm Z\trans$ increases with the number of
grouping factor levels. As the number of levels is often large in
practice, it is essential for speed and efficiency to take account of
this sparsity, for example by using sparse matrix methods, when
fitting mixed models (Section~\ref{sec:CSCmats}).

\subsubsection{Constructing the relative covariance factor}
\label{sec:mkLambdat}

The $q\times q$ covariance factor, $\bLt$, is a block diagonal matrix
whose $i$th diagonal block, $\bm\Lambda_i$, is of size
$q_i,i=1,\dots,k$.  We refer to $\bm\Lambda_i$ as the \emph{term-wise
  relative covariance factor}.  Furthermore, $\bm\Lambda_i$ is a
homogeneous block diagonal matrix with each of the $\ell_i$
lower-triangular blocks on the diagonal being a copy of a $p_i\times
p_i$ lower-triangular \emph{template matrix}, $\bm T_i$.  The
covariance parameter vector, $\bm\theta$, of length $m_i
=\binom{p_i+1}{2}$, consists of the elements in the lower triangle of
$\bm T_i,i=1,\dots,k$.  To provide a unique representation we require
that the diagonal elements of the $\bm T_i,i=1,\dots,k$ be
non-negative.

The template, $\bm T_i$, can be constructed from the number $p_i$
alone.  In \proglang{R} code we denote $p_i$ as \code{nc}.  For
example, if we set \code{nc <- 3}, we could create the
template for term $i$ as,

\begin{knitrout}
\definecolor{shadecolor}{rgb}{0.969, 0.969, 0.969}\color{fgcolor}\begin{kframe}
\begin{alltt}
\hlstd{> }\hlstd{(rowIndices} \hlkwb{<-} \hlkwd{rep}\hlstd{(}\hlnum{1}\hlopt{:}\hlstd{nc,} \hlnum{1}\hlopt{:}\hlstd{nc))}
\end{alltt}
\begin{verbatim}
[1] 1 2 2 3 3 3
\end{verbatim}
\begin{alltt}
\hlstd{> }\hlstd{(colIndices} \hlkwb{<-} \hlkwd{sequence}\hlstd{(}\hlnum{1}\hlopt{:}\hlstd{nc))}
\end{alltt}
\begin{verbatim}
[1] 1 1 2 1 2 3
\end{verbatim}
\begin{alltt}
\hlstd{> }\hlstd{(template} \hlkwb{<-} \hlkwd{sparseMatrix}\hlstd{(rowIndices, colIndices,}
\hlstd{+ }                          \hlkwc{x} \hlstd{=} \hlnum{1}\hlopt{*}\hlstd{(rowIndices}\hlopt{==}\hlstd{colIndices)))}
\end{alltt}
\begin{verbatim}
3 x 3 sparse Matrix of class "dgCMatrix"
          
[1,] 1 . .
[2,] 0 1 .
[3,] 0 0 1
\end{verbatim}
\end{kframe}
\end{knitrout}
Note that the \code{rowIndices} and \code{colIndices} fill the entire
lower triangle, which contains the initial values of the covariance
parameter vector, $\bm\theta$,
\begin{knitrout}
\definecolor{shadecolor}{rgb}{0.969, 0.969, 0.969}\color{fgcolor}\begin{kframe}
\begin{alltt}
\hlstd{> }\hlstd{(theta} \hlkwb{<-} \hlstd{template}\hlopt{@}\hlkwc{x}\hlstd{)}
\end{alltt}
\begin{verbatim}
[1] 1 0 0 1 0 1
\end{verbatim}
\end{kframe}
\end{knitrout}

(because the \code{@x} slot of the sparse matrix \code{template} is a
numeric vector containing the non-zero elements). This template
contains three types of elements: structural zeros (denoted
by~\code{.}), off-diagonal covariance parameters (initialized at
\code{0}), and diagonal variance parameters (initialized at \code{1}).
The next step in the construction of the relative covariance factor is
to repeat the template once for each level of the grouping factor to
construct a sparse block diagonal matrix.  For example, if we set the
number of levels, $\ell_i$, to two, \code{nl <- 2}, we
could create the transposed term-wise relative covariance factor,
$\bm\Lambda_i\trans$, using the \code{.bdiag} function in the
\pkg{Matrix} package,

\begin{knitrout}
\definecolor{shadecolor}{rgb}{0.969, 0.969, 0.969}\color{fgcolor}\begin{kframe}
\begin{alltt}
\hlstd{> }\hlstd{(Lambdati} \hlkwb{<-} \hlkwd{.bdiag}\hlstd{(}\hlkwd{rep}\hlstd{(}\hlkwd{list}\hlstd{(}\hlkwd{t}\hlstd{(template)), nl)))}
\end{alltt}
\begin{verbatim}
6 x 6 sparse Matrix of class "dgTMatrix"
                
[1,] 1 0 0 . . .
[2,] . 1 0 . . .
[3,] . . 1 . . .
[4,] . . . 1 0 0
[5,] . . . . 1 0
[6,] . . . . . 1
\end{verbatim}
\end{kframe}
\end{knitrout}
For a model with a single random effects term, $\bm\Lambda_i\trans$
would be the initial transposed relative covariance factor itself.

The transposed relative covariance factor, $\bLt\trans$, that arises
from parsing the formula and data is set at the initial value of the
covariance parameters, $\bm\theta$.  However, during model fitting, it
needs to be updated to a new $\bm\theta$ value at each iteration (see
Section~\ref{sec:plsI}).  This is achieved by constructing a vector of
indices, \code{Lind}, that identifies which elements of \code{theta}
should be placed in which elements of \code{Lambdat},
\begin{knitrout}
\definecolor{shadecolor}{rgb}{0.969, 0.969, 0.969}\color{fgcolor}\begin{kframe}
\begin{alltt}
\hlstd{> }\hlstd{LindTemplate} \hlkwb{<-} \hlstd{rowIndices} \hlopt{+} \hlstd{nc}\hlopt{*}\hlstd{(colIndices} \hlopt{-} \hlnum{1}\hlstd{)} \hlopt{-} \hlkwd{choose}\hlstd{(colIndices,} \hlnum{2}\hlstd{)}
\hlstd{> }\hlstd{(Lind} \hlkwb{<-} \hlkwd{rep}\hlstd{(LindTemplate, nl))}
\end{alltt}
\begin{verbatim}
 [1] 1 2 4 3 5 6 1 2 4 3 5 6
\end{verbatim}
\end{kframe}
\end{knitrout}
For example, if we randomly generate a new value for \code{theta},
\begin{knitrout}
\definecolor{shadecolor}{rgb}{0.969, 0.969, 0.969}\color{fgcolor}\begin{kframe}
\begin{alltt}
\hlstd{> }\hlstd{thetanew} \hlkwb{<-} \hlkwd{round}\hlstd{(}\hlkwd{runif}\hlstd{(}\hlkwd{length}\hlstd{(theta)),} \hlnum{1}\hlstd{)}
\end{alltt}
\end{kframe}
\end{knitrout}
we may update \code{Lambdat} as follows,
\begin{knitrout}
\definecolor{shadecolor}{rgb}{0.969, 0.969, 0.969}\color{fgcolor}\begin{kframe}
\begin{alltt}
\hlstd{> }\hlstd{Lambdati}\hlopt{@}\hlkwc{x} \hlkwb{<-} \hlstd{thetanew[Lind]}
\end{alltt}
\end{kframe}
\end{knitrout}
Section~\ref{sec:plsI} describes the process of
updating the relative covariance factor in more detail.

\section{Objective function module}
\label{sec:mkLmerDevfun}

\subsection{Model reformulation for improved computational stability}
\label{sec:stability}

In our initial formulation of the linear mixed model
(Equations~\ref{eq:LMMcondY}, \ref{eq:LMMuncondB}, and \ref{eq:relcovfac}),
the covariance parameter vector, $\bm\theta$, appears only in the
marginal distribution of the random effects (Equation~\ref{eq:LMMuncondB}).
However, from the perspective of computational stability and
efficiency, it is advantageous to reformulate the model such that
$\bm\theta$ appears only in the conditional distribution for the
response vector given the random effects.  Such a reformulation allows
us to work with singular covariance matrices, which regularly arise in
practice (e.g., during intermediate steps of the nonlinear optimizer,
Section~\ref{sec:optimizeLmer}).

The reformulation is made by defining a \emph{spherical}%
\footnote{$\mathcal{N}(\bm\mu,\sigma^2\bm I)$ distributions are called
  ``spherical'' because contours of the probability density are
  spheres.}  \emph{random effects} variable, $\mc U$, with
distribution
\begin{equation}
  \label{eq:distU}
  \mc U\sim\mathcal{N}(\bm 0,\sigma^2\bm I_q).
\end{equation}
If we set,
\begin{equation}
  \label{eq:BequalsLamU}
  \mc B=\bLt\mc U,
\end{equation}
then $\mc B$ will have the desired $\mathcal{N}(\bm
0,\bm\Sigma_{\bm\theta})$ distribution (Equation~\ref{eq:LMMuncondB}).
Although it may seem more natural to define $\mc U$ in terms of $\mc
B$ we must write the relationship as in \eq{eq:BequalsLamU} to allow
for singular $\bLt$. The conditional distribution
(Equation~\ref{eq:LMMcondY}) of the response vector given the random
effects may now be reformulated as,
\begin{equation}
  \label{eq:distYgivenU}
  (\mc Y|\mc U=\bm u)\sim\mc{N}(\bm\mu_{\mc Y|\mc U=\bm u},\sigma^2\bm W^{-1})
\end{equation}
where
\begin{equation}
  \label{eq:meanYgivenU}
  \bm\mu_{\mc Y|\mc U=\bm u} = \bm X \bm\beta + \bm Z\bLt\bm u + \bm o
\end{equation} 
is a vector of linear predictors, which can be interpreted as a
conditional mean (or mode).  Similarly, we also define $\bm\mu_{\mc
  U|\mc Y = \yobs}$ as the conditional mean (or mode) of the spherical
random effects given the observed value of the response vector. Note
also that we use the $\bm u$ symbol throughout to represent a specific
value of the random variable, $\mc U$.

\subsection{Penalized least squares}
\label{sec:plsMath}

Our computational methods for maximum likelihood fitting of the linear
mixed model involve repeated applications of the method of penalized
least-squares (PLS). In particular, the penalized least squares
problem is to minimize the penalized weighted residual sum-of-squares,
\begin{equation}
  \label{eq:pwrss}
  r^2(\bm\theta, \bm\beta, \bm u)
  = \rho^2(\bm\theta, \bm\beta, \bm u)
  + \left\| \bm u \right\|^2,
\end{equation}
over $\begin{bmatrix} \bm u \\
  \bm\beta \\ \end{bmatrix}$, where,
\begin{equation}
  \label{eq:wrss}
  \rho^2(\bm\theta, \bm\beta, \bm u) = \left\|
  \bm W^{1/2}\left[\yobs - \bm\mu_{\mc Y|\mc U=\bm u} \right]\right\|^2,
\end{equation} 
is the weighted residual sum-of-squares. This notation makes explicit
the fact that $r^2$ and $\rho^2$ both depend on $\bm\theta$,
$\betavec$, and $\bm u$.  The reason for the word `penalized' is that
the term, $\left\| \bm u \right\|^2$, in Equation~\ref{eq:pwrss} penalizes models with
larger magnitude values of $\bm u$.

In the so-called ``pseudo-data'' approach we write the penalized
weighted residual sum-of-squares as the squared length of a block
matrix equation,
\begin{equation}
  \label{eq:pwrssPseudoData}
  r^2(\bm\theta,\bm\beta,\bm u) =
  \left\|
  \begin{bmatrix}
  \bm W^{1/2}(\yobs - \bm o) \\
  \bm 0
  \end{bmatrix} -
  \begin{bmatrix}
  \bm W^{1/2}\bm Z\bm\bLt & \bm W^{1/2}\bm X \\
  \bm I_q & \bm 0 \\
  \end{bmatrix}
  \begin{bmatrix}
  \bm u \\
  \bm\beta \\
  \end{bmatrix}
  \right\|^2.
\end{equation}
This pseudo-data approach shows that the PLS problem may also be
thought of as a standard least squares problem for an extended
response vector, which implies that the minimizing value,
$\begin{bmatrix}
  \bm\mu_{\mc U|\mc Y = \yobs} \\
  \widehat{\bm\beta}_{\bm\theta} \\ \end{bmatrix}$, satisfies the
normal equations,
\begin{equation}
  \label{eq:normalEquations}
  \begin{bmatrix}
    \bLt\trans \bm Z\trans \bm W (\yobs - \bm o) \\
    \bm X\trans \bm W (\yobs - \bm o) \\
  \end{bmatrix} =
  \begin{bmatrix}
    \bLt\trans \bm Z\trans \bm W \bm Z \bLt + \bm I &
    \bLt\trans \bm Z\trans \bm W \bm X \\
    \bm X\trans \bm W \bm Z \bLt &
    \bm X\trans \bm W \bm X \\
  \end{bmatrix}
  \begin{bmatrix}
    \mu_{\mc U | \mc Y = \yobs} \\ \widehat{\betavec}_{\bm\theta}
  \end{bmatrix} ,
\end{equation} 
where $\mu_{\mc U | \mc Y = \yobs}$ is the conditional mean of $\mc U$
given that $\mc Y = \yobs$.  Note that this conditional mean depends
on $\bm\theta$, although we do not make this dependency explicit in
order to reduce notational clutter.

The cross-product matrix in Equation~\ref{eq:normalEquations} can be
Cholesky decomposed,
\begin{equation}
  \label{eq:blockCholeskyDecomp}
  \begin{bmatrix}
    \bLt\trans \bm Z\trans \bm W \bm Z \bLt + \bm I &
    \bLt\trans \bm Z\trans \bm W \bm X \\
    \bm X\trans \bm W \bm Z \bLt &
    \bm X\trans \bm W \bm X \\
  \end{bmatrix} =
  \begin{bmatrix}
    \bm L_{\bm\theta} & \bm 0 \\
    \bm R\trans_{ZX} & \bm R\trans_{X} \\
  \end{bmatrix}
  \begin{bmatrix}
    \bm L\trans_{\bm\theta} & \bm R_{ZX} \\
    \bm 0 & \bm R_{X}
  \end{bmatrix} .
\end{equation}
We may use this decomposition to rewrite the penalized weighted
residual sum-of-squares as,
\begin{equation}
  \label{eq:pwrssIdentity}
  \begin{aligned}
  r^2(\bm\theta,\bm\beta,\bm u) = r^2(\bm\theta) &
  + \left\|\bm L\trans_{\bm\theta} (\bm u - \bm\mu_{\mc U|\mc Y = \yobs})
  +    \bm R_{ZX} (\bm\beta - \widehat{\bm\beta}_{\bm\theta})\right\|^2
  + \left\|\bm R_{X} (\bm\beta - \widehat{\bm\beta}_{\bm\theta})\right\|^2
  \end{aligned} ,
\end{equation} 
where we have simplified notation by writing
$r^2(\bm\theta,\widehat{\bm\beta}_{\bm\theta},\bm\mu_{\mc U|\mc Y =
  \yobs})$ as $r^2(\bm\theta)$. This is an important expression in the
theory underlying \pkg{lme4}. It relates the penalized weighted
residual sum-of-squares, $r^2(\bm\theta,\bm\beta,\bm u)$, with its
minimum value, $r^2(\bm\theta)$.  This relationship is useful in the
next section where we integrate over the random effects, which is
required for maximum likelihood estimation.

\subsection{Probability densities}
\label{sec:densities}

The residual sums-of-squares discussed in the previous section can be
used to express various probability densities, which are required for
maximum likelihood estimation of the linear mixed model\footnote{These
  expressions only technically hold at the observed value, $\yobs$, of
  the response vector, $\mc Y$.},
\begin{equation}
  \label{eq:densityYgivenU}
  f_{\mc Y | \mc U}(\yobs|\bm u) =
  \frac{|\bm W|^{1/2}}{(2\pi\sigma^2)^{n/2}}
  \exp\left[\frac{-\rho^2(\bm\theta,\bm\beta,\bm u)}{2\sigma^2}\right] ,
\end{equation}
\begin{equation}
  \label{eq:densityU}
  f_{\mc U}(\bm u) =
  \frac{1}{(2\pi\sigma^2)^{q/2}}
  \exp\left[\frac{-\left\| \bm u \right\|^2}{2\sigma^2}\right] ,
\end{equation}
\begin{equation}
  \label{eq:densityYU}
  f_{\mc Y , \mc U}(\yobs,\bm u) =
  \frac{|\bm W|^{1/2}}{(2\pi\sigma^2)^{(n+q)/2}}
  \exp\left[\frac{-r^2(\bm\theta,\bm\beta,\bm u)}{2\sigma^2}\right] ,
\end{equation}
\begin{equation}
  \label{eq:densityUgivenY}
  f_{\mc U | \mc Y}(\bm u | \yobs) =
  \frac{f_{\mc Y , \mc U}(\yobs,\bm u)}{f_{\mc Y}(\yobs)} ,
\end{equation}
where,
\begin{equation}
  \label{eq:densityY}
  f_{\mc Y}(\yobs) =
  \int f_{\mc Y , \mc U}(\yobs,\bm u)d\bm u ,
\end{equation}
The log-likelihood to be maximized can therefore be expressed as,
\begin{equation}
  \label{eq:likelihood}
  \mc L(\bm\theta,\bm\beta,\sigma^2 | \yobs) = \log f_{\mc Y}(\yobs)
\end{equation} 

The integral in Equation~\ref{eq:densityY} may be more explicitly written
as,
\begin{equation}
  \label{eq:densityYexpand}
  \begin{aligned}
  f_{\mc Y}(\yobs) =
  \frac{|\bm W|^{1/2}}{(2\pi\sigma^2)^{(n+q)/2}} &
  \exp\left[\frac{
  -r^2(\bm\theta)
  - \left\|\bm R_{X} (\bm\beta - \widehat{\bm\beta}_{\bm\theta})\right\|^2
  }{2\sigma^2}\right] \\ &
    \int \exp\left[\frac{
  - \left\|\bm L\trans_{\bm\theta} (\bm u - \bm\mu_{\mc U|\mc Y = \yobs})
  +    \bm R_{ZX} (\bm\beta - \widehat{\bm\beta}_{\bm\theta})\right\|^2
  }{2\sigma^2}\right]
   d\bm u ,
  \end{aligned}
\end{equation} 
which can be evaluated with the change of variables,
\begin{equation}
  \label{eq:changeOfVariables}
  \bm v = \bm L\trans_{\bm\theta} (\bm u - \mu_{\mc U|\mc Y = \yobs})
  +    \bm R_{ZX} (\bm\beta - \widehat{\bm\beta}_{\bm\theta}) ,
\end{equation} 
The Jacobian determinant of the transformation from $\bm u$ to $\bm
v$ is $|\bm L_{\bm\theta} |$. Therefore we are able to write the integral
as,
\begin{equation}
  \label{eq:densityYintermediate}
  \begin{aligned}
  f_{\mc Y}(\yobs) =
  \frac{|\bm W|^{1/2}}{(2\pi\sigma^2)^{(n+q)/2}} &
  \exp\left[\frac{
  -r^2(\bm\theta)
  - \left\|\bm R_{X} (\bm\beta - \widehat{\bm\beta}_{\bm\theta})\right\|^2
  }{2\sigma^2}\right] \\ &
    \int \exp\left[\frac{
  - \left\|\bm v\right\|^2
  }{2\sigma^2}\right]
   |\bm L_{\bm\theta}|^{-1} d\bm v ,
  \end{aligned}
\end{equation}
which by the properties of exponential integrands becomes,
\begin{equation}
  \label{eq:densityYfinal}
   \exp\mc L(\bm\theta,\bm\beta,\sigma^2 | \yobs) = f_{\mc Y}(\yobs) =
  \frac{|\bm W|^{1/2}|\bm L_{\bm\theta}|^{-1}}{(2\pi\sigma^2)^{n/2}}
  \exp\left[\frac{
  -r^2(\bm\theta)
  - \left\|\bm R_{X} (\bm\beta - \widehat{\bm\beta}_{\bm\theta})\right\|^2
  }{2\sigma^2}\right] .
\end{equation}

\subsection{Evaluating and profiling the deviance and REML criterion}
\label{sec:profdev}

We are now in a position to understand why the formulation in
equations \ref{eq:LMMcondY} and \ref{eq:LMMuncondB} is particularly
useful. We are able to explicitly profile $\betavec$ and $\sigma$ out
of the log-likelihood (Equation~\ref{eq:likelihood}), to find a compact
expression for the profiled deviance (negative twice the profiled
log-likelihood) and the profiled REML criterion as a function of the
relative covariance parameters, $\bm\theta$, only.  Furthermore these
criteria can be evaluated quickly and accurately.

To estimate the parameters, $\bm\theta$, $\bm\beta$, and $\sigma^2$, we
minimize negative twice the log-likelihood, which can be written as,
\begin{equation}
  \label{eq:fullDeviance}
  -2\mc L(\bm\theta,\bm\beta,\sigma^2 | \yobs) =    
  \log\frac{|\bm L_{\bm\theta}|^2}{|\bm W|}
  + n\log(2\pi\sigma^2)
  + \frac{r^2(\bm\theta)}{\sigma^2}
  + \frac{ \left\|\bm R_{X} (\bm\beta -
    \widehat{\bm\beta}_{\bm\theta})\right\|^2}{\sigma^2} .
\end{equation}
It is very easy to profile out $\bm\beta$, because it enters into
the ML criterion only through the final term, which is zero if
$\bm\beta = \widehat{\bm\beta}_{\bm\theta}$, where
$\widehat{\bm\beta}_{\bm\theta}$ is found by solving the penalized
least-squares problem in Equation~\ref{eq:pwrssPseudoData}. Therefore we
can write a partially profiled ML criterion as,
\begin{equation}
  \label{eq:partiallyProfiledDeviance}
  -2\mc L(\bm\theta,\sigma^2 | \yobs)  =    
  \log\frac{|\bm L_{\bm\theta}|^2}{|\bm W|}
  + n\log(2\pi\sigma^2)
  + \frac{r^2(\bm\theta)}{\sigma^2} .
\end{equation}
This criterion is only partially profiled because it still depends on
$\sigma^2$. Differentiating this criterion with respect to $\sigma^2$
and setting the result equal to zero yields,
\begin{equation}
  \label{eq:residualVarianceEstimatingEquation}
  0 = \frac{n}{\widehat{\sigma}_{\bm\theta}^2} -
  \frac{r^2(\bm\theta)}{\widehat{\sigma}_{\bm\theta}^4},
\end{equation}
which leads to a maximum profiled likelihood estimate,
\begin{equation}
  \label{eq:residualVarianceEstimate}
  \widehat{\sigma}_{\bm\theta}^2 = \frac{r^2(\bm\theta)}{n}.
\end{equation}
This estimate can be substituted into the partially profiled criterion to yield the
fully profiled ML criterion,
\begin{equation}
  \label{eq:profiledDeviance}
  -2\mc L(\bm\theta | \yobs)  =           
  \log\frac{|\bm L_{\bm\theta}|^2}{|\bm W|}
  + n\left[1 + \log\left(
  \frac{2\pi r^2(\bm\theta)}{n}
  \right)\right] .
\end{equation}
This expression for the profiled deviance depends only on
$\bm\theta$. Although $q$, the number of columns in $\bm Z$ and the
size of $\bm\Sigma_{\bm\theta}$, can be very large indeed, the
dimension of $\bm\theta$ is small, frequently less than 10. The
\pkg{lme4} package uses generic nonlinear optimizers
(Section~\ref{sec:optimizeLmer}) to optimize this expression over
$\bm\theta$ to find its maximum likelihood estimate.

\subsubsection{The REML criterion}
\label{sec:reml}

The REML criterion can be obtained by
integrating the marginal density for $\mc Y$ with respect to the fixed
effects \citep{laird_ware_1982},
\begin{equation}
  \label{eq:REMLintegral}
  \int f_{\mc Y}(\yobs)d\bm\beta =
  \frac{|\bm W|^{1/2}|\bm L_{\bm\theta}|^{-1}}{(2\pi\sigma^2)^{n/2}}
  \exp\left[\frac{-r^2(\bm\theta)}{2\sigma^2}\right]
  \int \exp\left[\frac{
  - \left\|\bm R_{X} (\bm\beta - \widehat{\bm\beta}_{\bm\theta})\right\|^2
  }{2\sigma^2}\right]d\bm\beta ,
\end{equation}
which can be evaluated with the change of variables,
\begin{equation}
  \label{eq:REMLchangeOfVariables}
  \bm v = \bm R_{X} (\bm\beta - \widehat{\bm\beta}_{\bm\theta}) .
\end{equation}
The Jacobian determinant of the transformation from $\bm\beta$ to
$\bm v$ is $|\bm R_X |$. Therefore we are able to write the
integral as,
\begin{equation}
  \label{eq:REMLintegralIntermediate}
  \int f_{\mc Y}(\yobs)d\bm\beta =
  \frac{|\bm W|^{1/2}|\bm L_{\bm\theta}|^{-1}}{(2\pi\sigma^2)^{n/2}}
  \exp\left[\frac{-r^2(\bm\theta)}{2\sigma^2}\right]
  \int \exp\left[\frac{
  - \left\|\bm v\right\|^2
  }{2\sigma^2}\right]|\bm R_X|^{-1} d\bm v ,
\end{equation}
which simplifies to,
\begin{equation}
  \label{eq:REMLintermediate}
  \int f_{\mc Y}(\yobs)d\bm\beta =
  \frac{|\bm W|^{1/2}|\bm L_{\bm\theta}|^{-1}|\bm R_X|^{-1}}{(2\pi\sigma^2)^{(n-p)/2}}
  \exp\left[\frac{-r^2(\bm\theta)}{2\sigma^2}\right] .
\end{equation}
Minus twice the log of this integral is the (unprofiled) REML criterion,
\begin{equation}
  \label{eq:REMLdeviance}
  -2\mc L_R(\bm\theta,\sigma^2 | \yobs) =
  \log\frac{|\bm L_{\bm\theta}|^2|\bm R_X|^2}{|\bm W|}
  + (n-p)\log(2\pi\sigma^2)
  + \frac{r^2(\bm\theta)}{\sigma^2} .
\end{equation}
Note that because $\bm\beta$ gets integrated out, the REML criterion
cannot be used to find a point estimate of $\bm\beta$. However, we
follow others in using the maximum likelihood estimate,
$\widehat{\bm\beta}_{\widehat{\bm\theta}}$, at the optimum value of
$\bm\theta = \widehat{\bm\theta}$. The REML estimate of $\sigma^2$ is,
\begin{equation}
  \label{eq:REMLresidualVarianceEstimate}
  \widehat{\sigma}_{\bm\theta}^2 = \frac{r^2(\bm\theta)}{n-p} ,
\end{equation}
which leads to the profiled REML criterion,
\begin{equation}
  \label{eq:REMLprofiled}
  -2\mc L_R(\bm\theta | \yobs) =
  \log\frac{|\bm L_{\bm\theta}|^2|\bm R_X|^2}{|\bm W|}
  + (n-p)\left[1 + \log\left(\frac{2\pi
        r^2(\bm\theta)}{n-p}\right)\right] .
\end{equation}

\subsection{Changes relative to previous formulations}
\label{sec:previous_lmm_form}

We compare the PLS problem as formulated in Section~\ref{sec:plsMath}
with earlier versions and describe why we use this version.  What have
become known as ``Henderson's mixed-model equations'' are given as
Equation 6 of \citet{henderson_1982} and would be expressed as,
\begin{equation}
  \label{eq:henderson}
  \begin{bmatrix}
    \bm X\trans\bm X/\sigma^2& \bm X\trans\bm Z/\sigma^2\\
    \bm Z\trans\bm X/\sigma^2& \bm Z\trans\bm Z/\sigma^2 + \bm\Sigma^{-1}
  \end{bmatrix}
  \begin{bmatrix}
    \widehat{\bm\beta}_\theta \\ \bm\mu_{\mc B|\mc Y=\yobs}
  \end{bmatrix}=
  \begin{bmatrix}\bm X\trans\yobs/\sigma^2\\\bm Z\trans\yobs/\sigma^2
  \end{bmatrix} ,
\end{equation}
in our notation (ignoring weights and offsets, without loss of
generality).  The matrix written as $\bm R$ in \citet{henderson_1982}
is $\sigma^2\bm I_n$ in our formulation of the model.

\citet{bates04:_linear} modified the PLS equations to
\begin{equation}
  \label{eq:batesDebRoy}
  \begin{bmatrix}
    \bm Z\trans\bm Z+\bm\Omega&\bm Z\trans\bm X\\
    \bm X\trans\bm Z&\bm X\trans\bm X
  \end{bmatrix}
  \begin{bmatrix}
     \bm\mu_{\mc B|\mc Y=\yobs}\\\widehat{\bm\beta}_\theta
  \end{bmatrix}=
  \begin{bmatrix}\bm X\trans\yobs\\\bm Z\trans\yobs\end{bmatrix} .
\end{equation}
where
$\bm\Omega_\theta=\left(\bLt\trans\bLt\right)^{-1}=\sigma^2\bm\Sigma^{-1}$
is the \emph{relative precision matrix} for a given value of
$\bm\theta$.  They also showed that the profiled log-likelihood can be
expressed (on the deviance scale) as
\begin{equation}
  \label{eq:batesDebRoyprofdev}
  -2\mc L(\bm\theta)=
  \log\left(\frac{|\bm Z\trans\bm Z+\bm\Omega|}{|\bm\Omega|}\right) +
  n\left[1+\log\left(\frac{2\pi r^2_\theta}{n}\right)\right] .
\end{equation}

The primary difference between Equation~\ref{eq:henderson} and
Equation~\ref{eq:batesDebRoy} is the order of the blocks in the system matrix.
The PLS problem can be solved using a Cholesky factor of the system
matrix with the blocks in either order.  The advantage of using the
arrangement in Equation~\ref{eq:batesDebRoy} is to allow for evaluation of
the profiled log-likelihood.  To evaluate $|\bm Z\trans\bm
Z+\bm\Omega|$ from the Cholesky factor that block must be in the
upper-left corner, not the lower right.  Also, $\bm Z$ is sparse whereas $\bm
X$ is usually dense.  It is straightforward to exploit the sparsity of
$\bm Z\trans\bm Z$ in the Cholesky factorization when the block containing
this matrix is the first block to be factored.  If $\bm X\trans\bm
X$ is the first block to be factored it is much more difficult to
preserve sparsity.

The main change from the formulation in \citet{bates04:_linear} to the
current formulation is the use of a relative covariance factor,
$\bLt$, instead of a relative precision matrix, $\bm\Omega_\theta$,
and solving for the mean of $\mc U|\mc Y = \yobs$ instead of the mean
of $\mc B|\mc Y = \yobs$.  This change improves stability, because the
solution to the PLS problem in Section~\ref{sec:plsMath} is
well-defined when $\bLt$ is singular whereas the formulation in
Equation~\ref{eq:batesDebRoy} cannot be used in these cases because
$\bm\Omega_\theta$ does not exist.

It is important to allow for $\bLt$ to be singular because situations
where the parameter estimates, $\widehat{\bm\theta}$, produce a
singular $\bm\Lambda_{\widehat{\theta}}$ do occur in practice.  And
even if the parameter estimates do not correspond to a singular
$\bm\Lambda_\theta$, it may be desirable to evaluate the estimation
criterion at such values during the course of the numerical
optimization of the criterion.

\citet{bates04:_linear} also provided expressions for the gradient of
the profiled log-likelihood expressed as
Equation~\ref{eq:batesDebRoyprofdev}.  These expressions can be translated
into the current formulation.  From Equation~\ref{eq:profiledDeviance} we can see
that (again ignoring weights),
\begin{equation}
  \label{eq:grad}
  \begin{aligned}
    \nabla\left(-2\mc L(\theta)\right)
    &=\nabla\log(|\bm L_\theta|^2)+\nabla \left(n\log(r^2(\bm\theta))\right)\\
    &=\nabla\log(|\bLt\trans\bm Z\trans\bm Z\bLt+\bm I|)+
    n\left(\nabla r^2(\bm\theta)\right)/r^2(\bm\theta)\\
    &=\nabla\log(|\bLt\trans\bm Z\trans\bm Z\bLt+\bm I|)+
    \left(\nabla r^2(\bm\theta)\right)/(\widehat{\sigma^2})
  \end{aligned} .
\end{equation}
The first gradient is easy to express but difficult to evaluate for
the general model.  The individual elements of this gradient are
\begin{equation}
  \label{eq:graddet}
  \begin{aligned}
    \frac{\partial\log(|\bLt\trans\bm Z\trans\bm Z\bLt+\bm
      I|)}{\partial \theta_i} &=
    \tr\left[\frac{\partial\left(\bLt\trans\bm Z\trans\bm
          Z\bLt\right)}{\partial\theta_i}
      \left(\bLt\trans\bm Z\trans\bm Z\bLt+\bm I\right)^{-1}\right]\\
    &=\tr\left[
      \left(\bm L_\theta\bm L_\theta\trans\right)^{-1}
      \left(
        \bLt\trans\bm Z\trans\bm Z\frac{\partial\bLt}{\partial\theta_i} +
        \frac{\partial\bLt\trans}{\partial\theta_i}\bm Z\trans\bm Z\bLt
      \right)
    \right] .
  \end{aligned}
\end{equation}
The second gradient term can be expressed as a linear function of the
residual, with individual elements of the form
\begin{equation}
  \label{eq:gradr2}
  \frac{\partial r^2(\bm\theta)}{\partial\theta_i}=
  -2\bm u\trans\frac{\partial\bLt\trans}{\partial\theta_i}\bm
  Z\trans\left(\bm y-\bm Z\bLt\bm u-\bm X\widehat{\bm\beta}_{\bm\theta
   }\right),
\end{equation}
using the results of \citet{golub_pereyra_1973}. Although we do not
use these results in \pkg{lme4}, they are used for certain model types
in the \pkg{MixedModels} package for \proglang{Julia} and do provide
improved performance.

\subsection{Penalized least squares algorithm}
\label{sec:PLSpureR}

For efficiency, in \pkg{lme4} itself, PLS is implemented in compiled
\proglang{C++} code using the \pkg{Eigen} templated \proglang{C++}
package for numerical linear algebra.  Here however, in order to
improve readability we describe a version in pure \proglang{R}.
Section~\ref{sec:CSCmats} provides greater detail on the techniques
and concepts for computational efficiency, which is important in cases
where the nonlinear optimizer (Section~\ref{sec:optimizeLmer})
requires many iterations. 

The PLS algorithm takes a vector of covariance parameters,
$\bm\theta$, as inputs and returns the profiled deviance (Equation
\ref{eq:profiledDeviance}) or the REML criterion
(Equation \ref{eq:REMLprofiled}). This PLS algorithm consists of four main
steps:
\begin{enumerate}
\item Update the relative covariance factor (Section~\ref{sec:plsI})
\item Solve the normal equations (Section~\ref{sec:plsII})
\item Update the linear predictor and residuals (Section~\ref{sec:plsIII})
\item Compute and return the profiled deviance (Section~\ref{sec:plsIV})
\end{enumerate}
PLS also requires the objects described in Table~\ref{tab:inputsPLS},
which define the structure of the model.  These objects do not get
updated during the PLS iterations, and so it is useful to store
various matrix products involving them (Table~\ref{tab:constantPLS}).
Table~\ref{tab:updatePLS} lists the objects that do get updated over
the PLS iterations.  The symbols in this table correspond to a version
of \pkg{lme4} that is implemented entirely in \proglang{R} (i.e., no
compiled code as in \pkg{lme4} itself).  This implementation is called
\pkg{lme4pureR} and is currently available on \github\
(\url{https://github.com/lme4/lme4pureR/}).

\begin{table}[tb]
  \centering
  \begin{minipage}{\textwidth}
  \begin{tabular}{p{2.2in}llp{2in}}
    \hline
    Name/description & Pseudocode & Math & Type \\
    \hline
    Mapping from covariance parameters to relative covariance factor &
    \code{mapping} &  & function \\
    Response vector & \code{y} & $\yobs$ (Section~\ref{sec:LMMs}) & double vector \\
    Fixed effects model matrix & \code{X} & $\bm X$ (Equation~\ref{eq:LMMcondY}) & double dense\footnote{In previous versions of \pkg{lme4} a sparse $X$ matrix, useful for models with categorical fixed-effect predictors with many levels, could be specified; this feature is not currently available.}
    matrix \\
    Transposed random effects model matrix & \code{Zt} & $\bm Z\trans$
    (Equation~\ref{eq:LMMcondY}) &
    double sparse matrix \\
    Square-root weights matrix & \code{sqrtW} & $\bm W^{1/2}$
    (Equation~\ref{eq:LMMcondY}) & double diagonal matrix \\
    Offset & \code{offset} & $\bm o$ (Equation~\ref{eq:LMMcondY}) & double vector \\
    \hline
  \end{tabular}
  \renewcommand{\footnoterule}{%
    \kern -3pt
    \hrule width \textwidth height 0pt
    \kern 2pt
  }
  \end{minipage}
  \caption{Inputs into an LMM}
  \label{tab:inputsPLS}
\end{table}

\begin{table}[tb]
  \centering
  \begin{tabular}{ll}
    \hline
    Pseudocode & Math \\
    \hline
    \code{ZtW} & $\bm Z\trans\bm W^{1/2}$ \\
    \code{ZtWy} & $\bm Z\trans\bm W \yobs$ \\
    \code{ZtWX} & $\bm Z\trans\bm W \bm X$ \\
    \code{XtWX} & $\bm X\trans\bm W \bm X$ \\
    \code{XtWy} & $\bm X\trans\bm W \yobs$ \\
    \hline
  \end{tabular}
  \caption{Constant symbols involved in penalized least squares.}
  \label{tab:constantPLS}
\end{table}

\begin{table}[tb]
  \centering
  \begin{tabular}{p{2.2in}llp{2in}}
    \hline
    Name/description & Pseudocode & Math & Type \\
    \hline
    Relative covariance factor & \code{lambda} & $\bLt$
    (Equation~\ref{eq:relcovfac})
    & sparse double lower-triangular matrix  \\
    Random-effects Cholesky factor & \code{L} & $\bm L_{\bm\theta}$
    (Equation~\ref{eq:blockCholeskyDecomp})
    & double sparse triangular matrix \\
    Intermediate vector in the solution of the normal equations &
    \code{cu} & $\bm c_u$ (Equation~\ref{eq:cu}) & double vector \\
    Block in the full Cholesky factor & \code{RZX} & $\bm R_{ZX}$
    (Equation~\ref{eq:blockCholeskyDecomp})  & double dense matrix \\
    Cross-product of the fixed-effects Cholesky factor & \code{RXtRX}
    & $\bm R_X\trans\bm R_X$ (Equation~\ref{eq:RX}) & double dense matrix \\
    Fixed effects coefficients & \code{beta} & $\bm\beta$
    (Equation~\ref{eq:LMMcondY}) & double vector \\
    Spherical conditional modes & \code{u} & $\bm u$ (Section~\ref{sec:stability}) & double vector \\
    Non-spherical conditional modes & \code{b}
    & $\bm b$ (Equation~\ref{eq:LMMcondY}) & double vector \\
    Linear predictor & \code{mu} & $\bm\mu_{\mc Y|\mc U=\bm u}$
    (Equation~\ref{eq:meanYgivenU}) & double vector \\
    Weighted residuals & \code{wtres} & $\bm W^{1/2} (\yobs - \mu)$ &
    double vector \\
    Penalized weighted residual sum-of-squares & \code{pwrss} &
    $r^2(\bm\theta)$ (Equation~\ref{eq:pwrssIdentity}) &
    double \\
    Twice the log determinant random-effects Cholesky factor  &
    \code{logDet}
    & $\log|\bm L_{\bm\theta}|^2$ & double \\
    \hline
  \end{tabular}
  \caption{Quantities updated during an evaluation of the linear mixed model
    objective function.}
  \label{tab:updatePLS}
\end{table}

\subsubsection{PLS step I: update relative covariance factor}
\label{sec:plsI}

The first step of PLS is to update the relative covariance factor,
$\bLt$, from the current value of the covariance parameter vector,
$\bm\theta$.  The updated $\bLt$ is then used to update the random
effects Cholesky factor, $\bm L_{\bm\theta}$
(Equation~\ref{eq:blockCholeskyDecomp}).  The mapping from the covariance
parameters to the relative covariance factor can take many forms, but
in general involves a function that takes $\bm\theta$ into the values
of the non-zero elements of $\bLt$.  

If $\bLt$ is stored as an object of class \code{"dgCMatrix"} from the
\pkg{Matrix} package for \proglang{R}, then we may update $\bLt$ from
$\bm\theta$ by,
\begin{knitrout}
\definecolor{shadecolor}{rgb}{0.969, 0.969, 0.969}\color{fgcolor}\begin{kframe}
\begin{alltt}
\hlstd{> }\hlstd{Lambdat}\hlopt{@}\hlkwc{x}\hlstd{[]} \hlkwb{<-} \hlkwd{mapping}\hlstd{(theta)}
\end{alltt}
\end{kframe}
\end{knitrout}
where \code{mapping} is an \proglang{R} function that both accepts and
returns a numeric vector. The non-zero elements of sparse matrix
classes in \pkg{Matrix} are stored in a slot called \code{x}.

In the current version of \pkg{lme4} (v. 1.1-7) the mapping from
$\bm\theta$ to $\bLt$ is represented as an \proglang{R} integer
vector, \code{Lind}, of indices, so that
\begin{knitrout}
\definecolor{shadecolor}{rgb}{0.969, 0.969, 0.969}\color{fgcolor}\begin{kframe}
\begin{alltt}
\hlstd{> }\hlstd{mapping} \hlkwb{<-} \hlkwa{function}\hlstd{(}\hlkwc{theta}\hlstd{) theta[Lind]}
\end{alltt}
\end{kframe}
\end{knitrout}
The index vector \code{Lind} is computed during the model setup and
stored in the function's environment.  Continuing the example from
Section~\ref{sec:mkLambdat}, consider a new value for \code{theta},
\begin{knitrout}
\definecolor{shadecolor}{rgb}{0.969, 0.969, 0.969}\color{fgcolor}\begin{kframe}
\begin{alltt}
\hlstd{> }\hlstd{thetanew} \hlkwb{<-} \hlkwd{c}\hlstd{(}\hlnum{1}\hlstd{,} \hlopt{-}\hlnum{0.1}\hlstd{,} \hlnum{2}\hlstd{,} \hlnum{0.1}\hlstd{,} \hlopt{-}\hlnum{0.2}\hlstd{,} \hlnum{3}\hlstd{)}
\end{alltt}
\end{kframe}
\end{knitrout}
To put these values in the appropriate elements in \code{Lambdati},
we use \code{mapping},
\begin{knitrout}
\definecolor{shadecolor}{rgb}{0.969, 0.969, 0.969}\color{fgcolor}\begin{kframe}
\begin{alltt}
\hlstd{> }\hlstd{Lambdati}\hlopt{@}\hlkwc{x}\hlstd{[]} \hlkwb{<-} \hlkwd{mapping}\hlstd{(thetanew)}
\hlstd{> }\hlstd{Lambdati}
\end{alltt}
\begin{verbatim}
6 x 6 sparse Matrix of class "dgTMatrix"
                            
[1,] 1 -0.1  2.0 .  .    .  
[2,] .  0.1 -0.2 .  .    .  
[3,] .  .    3.0 .  .    .  
[4,] .  .    .   1 -0.1  2.0
[5,] .  .    .   .  0.1 -0.2
[6,] .  .    .   .  .    3.0
\end{verbatim}
\end{kframe}
\end{knitrout}
This \code{Lind} approach can be useful for extending the capabilities
of \pkg{lme4} by using the modular approach to fitting mixed models.
For example, Appendix~\ref{sec:homoVar} shows how to use \code{Lind}
to fit a model where a random slope and intercept are uncorrelated and
have the same variance.

The mapping from the covariance parameters to the relative covariance
factor is treated differently in other implementations of the
\code{lme4} approach to linear mixed models.  At the other extreme, the
\code{flexLambda} branch of \pkg{lme4} and the \pkg{lme4pureR} package
allows the capabilities for a completely general mapping.  This added
flexibility has the advantage of allowing a much wider variety of
models (e.g., compound symmetry, auto-regression).  However, the
disadvantage of this approach is that it becomes possible to fit a
much wider variety of ill-posed models.  Finally, if one would like
such added flexibility with the current stable version of \pkg{lme4},
it is always possible to use the modular approach to wrap the
\code{Lind}-based deviance function in a general mapping function
taking a parameter to be optimized, say $\phi$, into $\bm\theta$.
However, this approach is likely to be inefficient in many cases.

The \code{update} method from the \pkg{Matrix} package efficiently
updates the random effects Cholesky factor, $\bm L_{\bm\theta}$, from a
new value of $\bm\theta$ and the updated $\bLt$.
\begin{knitrout}
\definecolor{shadecolor}{rgb}{0.969, 0.969, 0.969}\color{fgcolor}\begin{kframe}
\begin{alltt}
\hlstd{> }\hlstd{L} \hlkwb{<-} \hlkwd{update}\hlstd{(L, Lambdat} \hlopt{%*%} \hlstd{ZtW,} \hlkwc{mult} \hlstd{=} \hlnum{1}\hlstd{)}
\end{alltt}
\end{kframe}
\end{knitrout}
The \code{mult = 1} argument corresponds to the addition of the
identity matrix to the upper-left block on the left-hand-side of
Equation~\ref{eq:blockCholeskyDecomp}.

\subsubsection{PLS step II:  solve normal equations}
\label{sec:plsII}

With the new covariance parameters installed in $\bLt$, the next step
is to solve the normal equations (Equation~\ref{eq:normalEquations}) for the
current estimate of the fixed effects coefficients,
$\widehat{\betavec}_{\bm\theta}$, and the conditional mode, $\mu_{\mc
  U|\mc Y = \yobs}$. We solve these equations using a sparse Cholesky
factorization (Equation~\ref{eq:blockCholeskyDecomp}).  In a complex model
fit to a large data set, the dominant calculation in the evaluation of
the profiled deviance (Equation~\ref{eq:profiledDeviance}) or REML criterion
(Equation~\ref{eq:REMLprofiled}) is this sparse Cholesky factorization
(Equation~\ref{eq:blockCholeskyDecomp}).  The factorization is performed in two
phases; a symbolic phase and a numeric phase.  The symbolic phase, in
which the fill-reducing permutation $\bm P$ is determined along with
the positions of the non-zeros in $\bm L_{\bm\theta}$, does not depend
on the value of $\bm\theta$.  It only depends on the positions of the
nonzeros in $\bm Z\bLt$.  The numeric phase uses $\bm\theta$ to
determine the numeric values of the non-zeros in $\bm
L_{\bm\theta}$. Using this factorization, the solution proceeds by the
following steps,
\begin{align}
    \bm L_{\bm\theta}\bm c_u&=\bm P\bLt\trans\bm Z\trans\bm W\bm y\label{eq:cu}\\
    \bm L_{\bm\theta}\bm R_{ZX}&=\bm P\bLt\trans\bm Z\trans\bm W\bm X\label{eq:RZX}\\
    \bm R_X\trans\bm R_X&=\bm X\trans\bm W\bm X-\bm R_{ZX}\trans\bm R_{ZX}\label{eq:RX}\\
    \left(\bm R_X\trans\bm R_X\right)\widehat{\bm\beta}_{\bm\theta}&=
      \bm X\trans\bm W\bm y-\bm R_{ZX}\bm c_u\label{eq:betahat}\\
    \bm L\trans\bm_{\bm\theta} \bm P\bm u&=\bm c_u-\bm R_{ZX}\widehat{\bm\beta}\label{eq:tildeu}
\end{align}
which can be solved using the \pkg{Matrix} package with,
\begin{knitrout}
\definecolor{shadecolor}{rgb}{0.969, 0.969, 0.969}\color{fgcolor}\begin{kframe}
\begin{alltt}
\hlstd{> }\hlstd{cu[]} \hlkwb{<-} \hlkwd{as.vector}\hlstd{(}\hlkwd{solve}\hlstd{(L,} \hlkwd{solve}\hlstd{(L, Lambdat} \hlopt{%*%} \hlstd{ZtWy,}
\hlstd{+ }                                 \hlkwc{system}\hlstd{=}\hlstr{"P"}\hlstd{),} \hlkwc{system}\hlstd{=}\hlstr{"L"}\hlstd{))}
\hlstd{> }\hlstd{RZX[]} \hlkwb{<-} \hlkwd{as.vector}\hlstd{(}\hlkwd{solve}\hlstd{(L,} \hlkwd{solve}\hlstd{(L, Lambdat} \hlopt{%*%} \hlstd{ZtWX,}
\hlstd{+ }                                  \hlkwc{system}\hlstd{=}\hlstr{"P"}\hlstd{),} \hlkwc{system}\hlstd{=}\hlstr{"L"}\hlstd{))}
\hlstd{> }\hlstd{RXtRX} \hlkwb{<-} \hlkwd{as}\hlstd{(XtWX} \hlopt{-} \hlkwd{crossprod}\hlstd{(RZX),} \hlstr{"dpoMatrix"}\hlstd{)}
\hlstd{> }\hlstd{beta[]} \hlkwb{<-} \hlkwd{as.vector}\hlstd{(}\hlkwd{solve}\hlstd{(RXtRX, XtWy} \hlopt{-} \hlkwd{crossprod}\hlstd{(RZX, cu)))}
\hlstd{> }\hlstd{u[]} \hlkwb{<-} \hlkwd{as.vector}\hlstd{(}\hlkwd{solve}\hlstd{(L,} \hlkwd{solve}\hlstd{(L, cu} \hlopt{-} \hlstd{RZX} \hlopt{%*%} \hlstd{beta,}
\hlstd{+ }                                \hlkwc{system} \hlstd{=} \hlstr{"Lt"}\hlstd{),} \hlkwc{system}\hlstd{=}\hlstr{"Pt"}\hlstd{))}
\hlstd{> }\hlstd{b[]} \hlkwb{<-} \hlkwd{as.vector}\hlstd{(}\hlkwd{crossprod}\hlstd{(Lambdat,u))}
\end{alltt}
\end{kframe}
\end{knitrout}
Notice the nested calls to \code{solve}.  The inner calls of the first
two assignments determine and apply the permutation matrix
(\code{system="P"}), whereas the outer calls actually solve the
equation (\code{system="L"}). In the assignment to \code{u[]}, the
nesting is reversed in order to return to the original permutation.

\subsubsection{PLS step III: update linear predictor and residuals}
\label{sec:plsIII}

The next step is to compute the linear predictor, $\mu_{\mc Y| \mc U}$
(Equation \ref{eq:meanYgivenU}), and the weighted residuals with new values for
$\widehat{\betavec}_\theta$ and $\mu_{\mc B | \mc Y = \yobs}$.  In
\pkg{lme4pureR} these quantities can be computed as,
\begin{knitrout}
\definecolor{shadecolor}{rgb}{0.969, 0.969, 0.969}\color{fgcolor}\begin{kframe}
\begin{alltt}
\hlstd{> }\hlstd{mu[]} \hlkwb{<-} \hlkwd{as.vector}\hlstd{(}\hlkwd{crossprod}\hlstd{(Zt,b)} \hlopt{+} \hlstd{X} \hlopt{%*%} \hlstd{beta} \hlopt{+} \hlstd{offset)}
\hlstd{> }\hlstd{wtres} \hlkwb{<-} \hlstd{sqrtW}\hlopt{*}\hlstd{(y}\hlopt{-}\hlstd{mu)}
\end{alltt}
\end{kframe}
\end{knitrout}
where \code{b} represents the current estimate of $\mu_{\mc B | \mc Y
  = \yobs}$.

\subsubsection{PLS step IV: compute profiled deviance}
\label{sec:plsIV}

Finally, the updated linear predictor and weighted residuals can be
used to compute the profiled deviance (or REML criterion), which in
\pkg{lme4pureR} proceeds as,
\begin{knitrout}
\definecolor{shadecolor}{rgb}{0.969, 0.969, 0.969}\color{fgcolor}\begin{kframe}
\begin{alltt}
\hlstd{> }\hlstd{pwrss} \hlkwb{<-} \hlkwd{sum}\hlstd{(wtres}\hlopt{^}\hlnum{2}\hlstd{)} \hlopt{+} \hlkwd{sum}\hlstd{(u}\hlopt{^}\hlnum{2}\hlstd{)}
\hlstd{> }\hlstd{logDet} \hlkwb{<-} \hlnum{2}\hlopt{*}\hlkwd{determinant}\hlstd{(L,} \hlkwc{logarithm} \hlstd{=} \hlnum{TRUE}\hlstd{)}\hlopt{$}\hlstd{modulus}
\hlstd{> }\hlkwa{if} \hlstd{(REML) logDet} \hlkwb{<-} \hlstd{logDet} \hlopt{+} \hlkwd{determinant}\hlstd{(RXtRX,}
\hlstd{+ }                                         \hlkwc{logarithm} \hlstd{=} \hlnum{TRUE}\hlstd{)}\hlopt{$}\hlstd{modulus}
\hlstd{> }\hlkwd{attributes}\hlstd{(logDet)} \hlkwb{<-} \hlkwa{NULL}
\hlstd{> }\hlstd{profDev} \hlkwb{<-} \hlstd{logDet} \hlopt{+} \hlstd{degFree}\hlopt{*}\hlstd{(}\hlnum{1} \hlopt{+} \hlkwd{log}\hlstd{(}\hlnum{2}\hlopt{*}\hlstd{pi}\hlopt{*}\hlstd{pwrss)} \hlopt{-} \hlkwd{log}\hlstd{(degFree))}
\end{alltt}
\end{kframe}
\end{knitrout}
The profiled deviance consists of three components: (1)
log-determinant(s) of Cholesky factorization (\code{logDet}), (2) the
degrees of freedom (\code{degFree}), and the penalized weighted
residual sum-of-squares (\code{pwrss}).

\subsection{Sparse matrix methods}
\label{sec:CSCmats}

In fitting linear mixed models, an instance of the penalized least
squares (PLS) problem (\ref{eq:normalEquations}) must be solved at each
evaluation of the objective function during the optimization
(Section~\ref{sec:optimizeLmer}) with respect to $\bm\theta$.  Because
this operation must be performed many times it is worthwhile
considering how to provide effective evaluation methods for objects
and calculations involving the sparse matrices associated with random
effects terms (Sections~\ref{sec:LMMmatrix}).

The CHOLMOD library of C functions
\citep{Chen:2008:ACS:1391989.1391995}, on which the sparse matrix
capabilities of the \pkg{Matrix} package for \proglang{R} and the
sparse Cholesky factorization in \proglang{Julia} are based, allows
for separation of the symbolic and numeric phases.  Thus we perform
the symbolic phase as part of establishing the structure representing
the model (Section~\ref{sec:lFormula}).  Furthermore, because CHOLMOD
functions allow for updating $\bm L_{\bm\theta}$ directly from the
matrix $\bLt\trans\bm Z\trans$ without actually forming $\bLt\trans\bm
Z\trans\bm Z\bLt+\bm I$ we generate and store $\bm Z\trans$ instead of
$\bm Z$ (note that we have ignored the weights matrix, $\bm W$, for
simplicity).  We can update $\bLt\trans\bm Z\trans$ directly from
$\bm\theta$ without forming $\bLt$ and multiplying two sparse
matrices.  Although such a direct approach is used in the
\pkg{MixedModels} package for \proglang{Julia}, in \pkg{lme4} we first
update $\bLt\trans$ then form the sparse product $\bLt\trans\bm
Z\trans$.  A third alternative, which we employ in \pkg{lme4pureR}, is
to compute and save the cross-products of the model matrices, $\bm X$
and $\bm Z$, and the response, $\bm y$, before starting the
iterations. To allow for case weights, we save the products $\bm
X\trans\bm W\bm X$, $\bm X\trans\bm W\bm y$, $\bm Z\trans\bm W\bm X$,
$\bm Z\trans\bm W\bm y$ and $\bm Z\trans\bm W\bm Z$ (see
Table~\ref{tab:constantPLS}).

We wish to use structures and algorithms that allow us to take a new
value of $\bm\theta$ and evaluate the $\bm L_\theta$
(eqn.~\ref{eq:blockCholeskyDecomp}) efficiently.  The key to doing so
is the special structure of $\bLt\trans\bm Z\trans\bm W^{1/2}$.  To
understand why this matrix, and not its transpose, is of interest we
describe the sparse matrix structures used in \proglang{Julia} and in
the \pkg{Matrix} package for \proglang{R}.

Dense matrices are stored in \proglang{R} and in \proglang{Julia} as a
one-dimensional array of contiguous storage locations addressed in
\emph{column-major order}.  This means that elements in the same
column are in adjacent storage locations whereas elements in the same
row can be widely separated in memory.  For this reason, algorithms
that work column-wise are preferred to those that work row-wise.

Although a sparse matrix can be stored in a \emph{triplet} format,
where the row position, column position and element value the nonzeros
are recorded as triplets, the preferred storage forms for actual
computation with sparse matrices are compressed sparse column (CSC) or
compressed sparse row (CSR)~\citep[Ch.~2]{davis06:csparse_book}. The
CHOLMOD (and, more generally, the SuiteSparse package of C libraries)
uses the CSC storage format.  In this format the non-zero elements in
a column are in adjacent storage locations and access to all the
elements in a column is much easier than access to those in a row
(similar to dense matrices stored in column-major order).

The matrices $\bm Z$ and $\bm Z\bLt$ have the property that the number
of nonzeros in each row, $\sum_{i=1}^k p_i$, is constant.  For CSC
matrices we want consistency in the columns rather than the rows,
which is why we work with $\bm Z\trans$ and $\bLt\trans\bm Z\trans$
rather than their transposes.

An arbitrary $m\times n$ sparse matrix in CSC format is expressed as
two vectors of indices, the row indices and the column pointers, and a
numeric vector of the non-zero values.  The elements of the row
indices and the nonzeros are aligned and are ordered first by
increasing column number then by increasing row number within column.
The column pointers are a vector of size $n+1$ giving the location of
the start of each column in the vectors of row indices and nonzeros.

Because the number of nonzeros in each column of $\bm Z\trans$, and in
each column of matrices derived from $\bm Z\trans$ (such as $\bLt\trans\bm Z\trans\bm W^{1/2}$) is constant, the vector of nonzeros in the
CSC format can be viewed as a dense matrix, say $\bm N$, of size
$\left(\sum_{i=1}^n p_i\right)\times n$.  We do not need to store the
column pointers because the columns of $\bm Z\trans$ correspond to
columns of $\bm N$.  All we need is $\bm N$, the dense matrix of
nonzeros, and the row indices, which are derived from the grouping
factor index vectors $\bm i_i,i=1,\dots,k$ and can also be arranged as
a dense matrix of size $\left(\sum_{i=1}^n p_i\right)\times n$.
Matrices like $\bm Z\trans$, with the property that there are the same
number of nonzeros in each column, are sometimes called \emph{regular
  sparse column-oriented} (RSC) matrices.
\section{Nonlinear optimization module}
\label{sec:optimizeLmer}

The objective function module produces a function which implements the
penalized least squares algorithm for a particular mixed model. The
next step is to optimize this function over the covariance parameters,
$\bm\theta$, which is a nonlinear optimization problem.  The
\pkg{lme4} package separates the efficient computational linear
algebra required to compute profiled likelihoods and deviances for a
given value of $\bm\theta$ from the nonlinear optimization algorithms,
which use general-purpose nonlinear optimizers.

One benefit of this separation is that it allows for experimentation
with different nonlinear optimizers.  Throughout the development of
\pkg{lme4}, the default optimizers and control parameters have changed
in response to feedback from users about both efficiency and
convergence properties.  \pkg{lme4} incorporates two built-in
optimization choices, an implementation of the Nelder-Mead simplex
algorithm adapted from Steven Johnson's \code{NLopt} C library
\citep{NLopt} and a wrapper for Powell's BOBYQA
algorithm, implemented in the \pkg{minqa} package \cite{minqa_pkg}
as a wrapper around
Powell's original FORTRAN code \citep{Powell_bobyqa}.  More generally,
\pkg{lme4} allows any user-specified optimizer that (1) can work with
an objective function (i.e., does not require a gradient function to
be specified), (2) allows box constraints on the parameters, and (3)
conforms to some simple rules about argument names and structure of
the output. An internal wrapper allows the use of the \pkg{optimx}
package \citep{optimx_pkg}, although the only optimizers provided via \pkg{optimx} that
satisfy the constraints above are the \code{nlminb} and
\code{L-BFGS-B} algorithms that are themselves wrappers around the
versions provided in base R. 
Several other algorithms from Steven Johnson's
\code{NLopt} package are also available via the \pkg{nloptr} wrapper
package (e.g., alternate implementations of Nelder-Mead and BOBYQA,
and Powell's COBYLA algorithm).

This flexibility assists with diagnosing convergence problems --- it
is easy to switch among several algorithms to determine whether the
problem lies in a failure of the nonlinear optimization stage, as
opposed to a case of model misspecification or unidentifiability or a
problem with the underlying PLS algorithm.  To date we
have only observed PLS failures, which arise if
$\bm X\trans\bm W\bm X-\bm R_{ZX}\trans\bm R_{ZX}$ becomes singular
during an evaluation of the objective function, with badly scaled
problems (i.e., problems with continuous predictors that take a
very large or very small numerical range of values).

The requirement for optimizers that can handle box constraints stems
from our decision to parameterize the variance-covariance matrix in a
constrained space, in order to allow for singular fits. In contrast to
the approach taken in the \pkg{nlme} package \citep{nlme_pkg}, which goes to some
lengths to use an unconstrained variance-covariance parameterization
(the \emph{log-Cholesky} parameterization:
\cite{pinheiro_unconstrained_1996}), we instead use the Cholesky
parameterization but require the elements of $\bm\theta$ corresponding
to the diagonal elements of the Cholesky factor to be non-negative.
With these constraints, the variance-covariance matrix is singular if
and only if any of the diagonal elements is exactly zero.
Singular fits are common in practical data-analysis situations,
especially with small- to medium-sized data sets and complex
variance-covariance models, so being able to fit a singular model is
an advantage: when the best fitting model lies on the boundary of a
constrained space, approaches that try to remove the constraints by
fitting parameters on a transformed scale will often give rise to
convergence warnings as the algorithm tries to find a maximum on an
asymptotically flat surface \citep{bolker_strategies_2013}.

In principle the likelihood surfaces we are trying to optimize over
are smooth, but in practice using gradient information in optimization
may or may not be worth the effort. In special cases, we have a
closed-form solution for the gradients
(Equations~\ref{eq:grad}--\ref{eq:gradr2}), but in general we will have to
approximate them by finite differences, which is expensive and has
limited accuracy.  (We have considered using automatic
differentiation approaches to compute the gradients more efficiently,
but this strategy requires a great deal of additional machinery, and
would have drawbacks in terms of memory requirements for large
problems.)  This is the primary reason for our switching to
derivative-free optimizers such as BOBYQA and Nelder-Mead in the
current version of \pkg{lme4}, although as discussed above
derivative-based optimizers based on finite differencing are available
as an alternative.

There is most likely further room for improvement in the nonlinear
optimization module; for example, some speed-up could be gained by
using parallel implementations of derivative-free optimizers that
evaluated several trial points at once \citep{klein_nelder_2013}.  In
practice users most often have optimization difficulties with poorly
scaled or centred data --- we are working to implement appropriate
diagnostic tests and warnings to detect these situations.

\section{Output module}
\label{sec:mkMerMod}

Here we describe some of the methods in \pkg{lme4} for exploring
fitted linear mixed models (Table~\ref{tab:methods}), which are
represented as objects of the \code{S4} class \code{lmerMod}.  We
begin by describing the theory underlying these methods
(Section~\ref{sec:outputTheory}) and then continue the
\code{sleepstudy} example introduced in Section~\ref{sec:sleepstudy}
to illustrate these ideas in practice.

\subsection{Theory underlying the output module}
\label{sec:outputTheory}

\subsubsection{Covariance matrix of the fixed effect coefficients}
\label{sec:vcov}

In the \code{lm} function, the variance-covariance matrix of the
coefficients is the inverse of the cross product of the model matrix, times
the residual variance \citep{Chambers:1993}.  The inverse cross-product
matrix is computed by first inverting the upper triangular matrix
resulting from the QR decomposition of the model matrix, and then
taking its cross-product.
\begin{equation}
  \label{eq:varUBeta}
  \Var_{\bm\theta, \sigma}\left(
    \begin{bmatrix}
      \bm\mu_{\mc U|\mc Y=\yobs} \\ \hat\betavec
    \end{bmatrix}
  \right) =
  \sigma^2
  \begin{bmatrix}
    \bm L\trans_{\bm\theta} & \bm R_{ZX}\\
    \bm 0            & \bm R_X
  \end{bmatrix}^{-1}
  \begin{bmatrix}
    \bm L_{\bm\theta}& \bm 0\\
    \bm R_{ZX}\trans & \bm R_X\trans
  \end{bmatrix}^{-1}
\end{equation} 
Because of normality, the marginal covariance matrix of $\hat\betavec$ is
just the lower-right $p$-by-$p$ block of
$\Var_{\bm\theta, \sigma}\left(
    \begin{bmatrix}
      \bm\mu_{\mc U|\mc Y=\yobs} \\ \hat\betavec
    \end{bmatrix}
  \right)$.
This lower-right block is
\begin{equation}
  \label{eq:varBeta}
  \Var_{\bm\theta, \sigma}(\hat\betavec) = \sigma^2 \bm R_X^{-1} (\bm R_X\trans)^{-1}
\end{equation}
which follows from the Schur complement identity (Theorem 1.2 of \citet{zhang2006schur}).  This
matrix can be extracted from a fitted \code{merMod} object as,
\begin{knitrout}
\definecolor{shadecolor}{rgb}{0.969, 0.969, 0.969}\color{fgcolor}\begin{kframe}
\begin{alltt}
\hlstd{> }\hlstd{RX} \hlkwb{<-} \hlkwd{getME}\hlstd{(fm1,} \hlstr{"RX"}\hlstd{)}
\hlstd{> }\hlstd{sigma2} \hlkwb{<-} \hlkwd{sigma}\hlstd{(fm1)}\hlopt{^}\hlnum{2}
\hlstd{> }\hlstd{sigma2}\hlopt{*}\hlkwd{chol2inv}\hlstd{(RX)}
\end{alltt}
\begin{verbatim}
       [,1]   [,2]
[1,] 46.575 -1.451
[2,] -1.451  2.389
\end{verbatim}
\end{kframe}
\end{knitrout}
which could be computed with \pkg{lme4} as \code{vcov(fm1)}.

The square-root diagonal of this covariance matrix contains the
estimates of the standard errors of fixed effects coefficients.  These
standard errors are used to construct Wald confidence intervals with
\code{confint(., method = "Wald")}.  Such confidence intervals are
approximate, and depend on the assumption that the likelihood profile
of the fixed effects is linear on the $\zeta$ scale
(Section~\ref{sec:profile}).

\subsubsection{Profiling}
\label{sec:profile}

\newcommand{\devprof}{-2 {\mc L}_p}

The theory of likelihood profiles is straightforward: the deviance (or
likelihood) profile, $\devprof()$, for a focal model parameter $P$ is
the minimum value of the deviance conditioned on a particular value of
$P$.  For each parameter of interest, our goal is to evaluate the
deviance profile for many points --- optimizing over all of the
non-focal parameters each time --- over a wide enough range and with
high enough resolution to evaluate the shape of the profile (in
particular, whether it is quadratic, which would allow use of Wald
confidence intervals and tests) and to find the values of $P$ such
that $\devprof(P)=-2 {\mc L}(\widehat{P}) + \chi^2(1-\alpha)$, which
represent the profile confidence intervals.  While profile confidence
intervals rely on the asymptotic distribution of the minimum deviance,
this is a much weaker assumption than the log-quadratic likelihood
surface required by Wald tests.

An additional challenge in profiling arises when we want to compute
profiles for quantities of interest that are not parameters of our PLS
function. We have two problems in using the deviance function defined
above to profile linear mixed models. First, a parameterization of the
random effects variance-covariance matrix in terms of standard
deviations and correlations, or variances and covariances, is much
more familiar to users, and much more relevant as output of a
statistical model, than the parameters, $\bm\theta$, of the relative
covariance factor --- users are interested in inferences on variances
or standard deviations, not on $\bm\theta$. Second, the fixed-effect
coefficients and the residual standard deviation, both of which are
also of interest to users, are profiled out (in the sense used above)
of the deviance function (Section~\ref{sec:profdev}), so we have to
find a strategy for estimating the deviance for values of $\bm\beta$
and $\sigma^2$ other than the profiled values.

To handle the first problem we create a new version of the deviance
function that first takes a vector of standard deviations (and
correlations), and a value of the residual standard deviation, and
maps them to a $\bm\theta$ vector, and then computes the PLS as
before; it uses the specified residual standard deviation rather than
the PLS estimate of the standard deviation
(Equation~\ref{eq:residualVarianceEstimate}) in the deviance calculation.
We compute a profile likelihood for the fixed-effect parameters, which
are profiled out of the deviance calculation, by adding an offset to
the linear predictor for the focal element of $\bm\beta$.  The
resulting function is not useful for general nonlinear optimization
--- one can easily wander into parameter regimes corresponding to
infeasible (non-positive semidefinite) variance-covariance matrices ---
but it serves for likelihood profiling, where one focal parameter is
varied at a time and the optimization over the other parameters is
likely to start close to an optimum.

In practice, the \code{profile} method systematically varies the
parameters in a model, assessing the best possible fit that can be
obtained with one parameter fixed at a specific value and comparing
this fit to the globally optimal fit, which is the original model fit
that allowed all the parameters to vary.  The models are compared
according to the change in the deviance, which is the likelihood ratio
test statistic.  We apply a signed square root transformation to this
statistic and plot the resulting function, which we call the
\emph{profile zeta function} or $\zeta$, versus the parameter
value. The signed aspect of this transformation means that $\zeta$ is
positive where the deviation from the parameter estimate is positive
and negative otherwise, leading to a monotonically increasing function
which is zero at the global optimum. A $\zeta$ value can be compared
to the quantiles of the standard normal distribution.  For example, a
95\% profile deviance confidence interval on the parameter consists of
the values for which $-1.96 < \zeta < 1.96$.  Because the process of
profiling a fitted model can be computationally intensive --- it
involves refitting the model many times --- one should exercise
caution with complex models fit to large data sets.

The standard approach to this computational challenge is to compute
$\zeta$ at a limited number of parameter values, and to fill in the
gaps by fitting an interpolation spline.  Often one is able to invert
the spline to obtain a function from $\zeta$ to the focal parameter,
which is necessary in order to construct profile confidence intervals.
However, even if the points themselves are monotonic, it is possible
to obtain non-monotonic interpolation curves.  In such a case,
\pkg{lme4} falls back to linear interpolation, with a warning.

\newcommand{\alphamax}{\alpha_{\textrm{\small max}}}
\newcommand{\pstep}{\epsilon}

The last part of the technical specification for computing profiles is
deciding on a strategy for choosing points to sample.  In one way or
another one wants to start at the estimated value for each parameter
and work outward either until a constraint is reached (i.e., a value of
0 for a standard deviation parameter or a value of $\pm 1$ for a
correlation parameter), or until a sufficiently large deviation from
the minimum deviance is attained.  \pkg{lme4}'s profiler chooses a
cutoff $\phi$ based on the $1-\alphamax$ critical value of the
$\chi^2$ distribution with a number of degrees of freedom equal to the
\emph{total} number of parameters in the model, where $\alphamax$ is
set to 0.05 by default. The profile computation initially adjusts the
focal parameter $p_i$ by an amount $\pstep=1.01 \hat p_i$ from its ML
or REML estimate $\hat p_i$ (or by $\pstep=0.001$ if $\hat p_i$ is
zero, as in the case of a singular variance-covariance model).  The
local slope of the likelihood profile $(\zeta(\hat
p_i+\pstep)-\zeta(\hat p_i))/\pstep$ is used to pick the next point to
evaluate, extrapolating the local slope to find a new $\pstep$ that
would be expected to correspond to the desired step size $\Delta
\zeta$ (equal to $\phi/8$ by default, so that 16 points would be used
to cover the profile in both directions if the log-likelihood surface
were exactly quadratic).  Some fail-safe testing is done to ensure
that the step chosen is always positive, and less than a maximum; if a
deviance is ever detected that is lower than that of the ML deviance,
which can occur if the initial fit was wrong due to numerical
problems, the profiler issues an error and stops.

\subsubsection{Parametric bootstrapping}
\label{sec:pb}

To avoid the asymptotic assumptions of the likelihood ratio test, at
the cost of greatly increased computation time, one can estimate
confidence intervals by parametric bootstrapping --- that is, by
simulating data from the fitted model, refitting the model, and
extracting the new estimated parameters (or any other quantities of
interest).  This task is quite straightforward, since there is already
a \code{simulate} method, and a \code{refit} function which
re-estimates the (RE)ML parameters for new data, starting from the
previous (RE)ML estimates and re-using the previously computed model
structures (including the fill-reducing permutation) for efficiency.
The \code{bootMer} function is thus a fairly thin wrapper around a
\code{simulate}/\code{refit} loop, with a small amount of additional
logic for parallel computation and error-catching.  (Some of the ideas
of \code{bootMer} are adapted from \code{merBoot}, a more ambitious
framework for bootstrapping \pkg{lme4} model fits which unfortunately
seems to be unavailable at present \citep{merBoot}.)

\subsubsection{Conditional variances of random effects}
\label{sec:condVar}

It is useful to clarify that the conditional covariance concept in
\pkg{lme4} is based on a simplification of the linear mixed model.  In
particular, we simplify the model by assuming that the quantities,
$\betavec$, $\bLt$, and $\sigma$, are known (i.e., set at their
estimated values). In fact, the only way to define the conditional
covariance is at fixed parameter values.  Our approximation here is
using the estimated parameter values for the unknown ``true''
parameter values.  In this simplified case, $\mc U$ is the only
quantity in the statistical model that is both random and unknown.

Given this simplified linear mixed model, a standard result in
Bayesian linear regression modelling \citep{gelman2013bayesian}
implies that the conditional distribution of the spherical random
effects given the observed response vector is Gaussian,
\begin{equation}
  \label{eq:condGaussian}
(\mc U | \mc Y = \yobs) \sim \mathcal{N}(\mu_{\mc U | \mc Y = \yobs},
\widehat{\sigma}^2 \bm V)
\end{equation}
where,
\begin{equation}
  \label{eq:condVar}
  \bm V = \left(\bm\Lambda_{\widehat{\bm\theta}}\trans\bm Z\trans \bm W
    \bm Z\bm\Lambda_{\widehat{\bm\theta}} + \bm I_q \right)^{-1} = 
  \left(\bm L_{\widehat{\bm\theta}}^{-1}\right)\trans
  \left(\bm L_{\widehat{\bm\theta}}^{-1}\right)
\end{equation}
is the unscaled conditional variance and,
\begin{equation}
  \label{eq:condMean}
  \mu_{\mc U | \mc Y = \yobs} = \bm V \bm\Lambda_{\widehat{\bm\theta}}\trans\bm Z\trans\bm W
    \left(\yobs-\bm o - X\widehat{\betavec} \right)
\end{equation}  
is the conditional mean/mode. Note that in practice the inverse in
Equation~\ref{eq:condVar} does not get computed directly, but rather an
efficient method is used that exploits the sparse structures.

The random effects coefficient vector, $\mathbf{b}$, is often of
greater interest. The conditional covariance matrix of $\mc B$
may be expressed as,
\begin{equation}
  \label{eq:condVarB}
\widehat{\sigma}^2 \bm\Lambda_{\widehat{\bm\theta}} \bm V \bm\Lambda_{\widehat{\bm\theta}}\trans
\end{equation}

\subsubsection{The hat matrix}
\label{sec:hat}

The hat matrix, $\bm H$, is a useful object in linear model
diagnostics \citep{cook1982residuals}.  In general, $\bm H$ relates
the observed and fitted response vectors, but we specifically define
it as,
\begin{equation}
  \left(\bm\mu_{\mc Y|\mc U=\bm u} - \bm o\right) = 
  \bm H \left(\yobs - \bm o\right)
\end{equation}
To find $\bm H$ we note that,
\begin{equation}
  \left(\bm\mu_{\mc Y|\mc U=\bm u} - \bm o\right) = 
  \begin{bmatrix}
    \bm Z\bm\Lambda & \bm X \\
  \end{bmatrix}
  \begin{bmatrix}
    \mu_{\mc U | \mc Y = \yobs} \\ \widehat{\betavec}_{\bm\theta}
  \end{bmatrix} 
\end{equation}
Next we get an expression for $\begin{bmatrix}
    \mu_{\mc U | \mc Y = \yobs} \\ \widehat{\betavec}_{\bm\theta}
  \end{bmatrix}$ by solving the normal equations
  (Equation~\ref{eq:normalEquations}),
\begin{equation}
 \begin{bmatrix}
    \mu_{\mc U | \mc Y = \yobs} \\ \widehat{\betavec}_{\bm\theta}
  \end{bmatrix} = 
  \begin{bmatrix}
    \bm L\trans_{\bm\theta} & \bm R_{ZX} \\
    \bm 0 & \bm R_{X}
  \end{bmatrix}^{-1}
    \begin{bmatrix}
    \bm L_{\bm\theta} & \bm 0 \\
    \bm R\trans_{ZX} & \bm R\trans_{X} \\
  \end{bmatrix}^{-1}
    \begin{bmatrix}
    \bLt\trans \bm Z\trans \\
    \bm X\trans \\
  \end{bmatrix} \bm W (\yobs - \bm o)
\end{equation} 
By the Schur complement identity \citep{zhang2006schur},
\begin{equation}
  \begin{bmatrix}
    \bm L\trans_{\bm\theta} & \bm R_{ZX} \\
    \bm 0 & \bm R_{X}
  \end{bmatrix}^{-1} = 
  \begin{bmatrix}
    \left(\bm L\trans_{\bm\theta}\right)^{-1} & 
    - \left(\bm L\trans_{\bm\theta}\right)^{-1} \bm R_{ZX} \bm R_{X}^{-1}\\
    \bm 0 & \bm R_{X}^{-1}
  \end{bmatrix}  
\end{equation}
Therefore, we may write the hat matrix as,
\begin{equation}
  \label{eq:hatComputational}
  \bm H = (\bm C_L\trans \bm C_L + \bm C_R\trans \bm C_R)
\end{equation}
where,
\begin{equation}
  \label{eq:CL}
  \bm L_{\bm\theta} \bm C_L = \bLt\trans \bm Z\trans\bm W^{1/2}
\end{equation}
and,
\begin{equation}
  \label{eq:CR}
  \bm R\trans_{X} \bm C_R = \bm X\trans\bm W^{1/2} - \bm R\trans_{ZX}\bm C_L
\end{equation}

The trace of the hat matrix is often used as a measure of the
effective degrees of freedom (e.g., \citet{vaida2005conditional}).
Using a relationship between the trace and vec operators, the trace of
$\bm H$ may be expressed as,
\begin{equation}
  \label{eq:traceHat}
  \tr(\bm H) = \left\| \VEC(\bm C_L) \right\|^2
  + \left\| \VEC(\bm C_R) \right\|^2
\end{equation}

\subsection{Using the output module}
\label{sec:outputPractice}

The user interface for the output module largely consists of a set of
methods (Table~\ref{tab:methods}) for objects of class \code{merMod},
which is the class of objects returned by
\code{lmer} (In addition to these methods, the \code{getME}
  function can be used to extract various objects from a fitted mixed
  model in \pkg{lme4}.) Here we illustrate the use of several of
these methods.

\begin{table}
  \begin{tabular}{ll}
    \hline
    Generic (Section) & Brief description of return value \\
    \hline
    \code{anova} (\ref{sec:modComp}) &
    Decomposition of fixed-effects contributions \\
    & or model comparison. \\
    \code{as.function} &
    Function returning profiled deviance or REML criterion. \\
    \code{coef} &
    Sum of the random and fixed effects for each level. \\
    \code{confint} (\ref{sec:CI}) &
    Confidence intervals on linear mixed-model parameters. \\
    \code{deviance} (\ref{sec:summary}) &
    Minus twice maximum log-likelihood. \\
    & (Use \code{REMLcrit} for the REML criterion.) \\
    \code{df.residual} &
    Residual degrees of freedom. \\
    \code{drop1} &
    Drop allowable single terms from the model. \\
    \code{extractAIC} &
    Generalized Akaike information criterion \\
    \code{fitted} &
    Fitted values given conditional modes (Equation~\ref{eq:meanYgivenU}). \\
    \code{fixef} (\ref{sec:summary}) &
    Estimates of the fixed effects coefficients, $\widehat{\bm\beta}$ \\
    \code{formula} (\ref{sec:uncor}) &
    Mixed-model formula of fitted model. \\
    \code{logLik} & 
    Maximum log-likelihood. \\
    \code{model.frame} & 
    Data required to fit the model. \\
    \code{model.matrix} &
    Fixed effects model matrix, $\bm X$.\\
    \code{ngrps (\ref{sec:summary})} &
    Number of levels in each grouping factor. \\
    \code{nobs (\ref{sec:summary})} &
    Number of observations.\\
    \code{plot} &
    Diagnostic plots for mixed-model fits. \\
    \code{predict} (\ref{sec:predict}) &
    Various types of predicted values. \\
    \code{print} &
    Basic printout of mixed-model objects. \\
    \code{profile} (\ref{sec:profile}) &
    Profiled likelihood over various model parameters. \\
    \code{ranef} (\ref{sec:summary}) &
    Conditional modes of the random effects. \\
    \code{refit} (\ref{sec:pb}) &
    A model (re)fitted to a new set of observations of the response variable. \\
    \code{refitML} (\ref{sec:modComp}) &
    A model (re)fitted by maximum likelihood.  \\
    \code{residuals}  (\ref{sec:summary}) &
    Various types of residual values. \\
    \code{sigma} (\ref{sec:summary}) &
    Residual standard deviation. \\
    \code{simulate} (\ref{sec:predict}) &
    Simulated data from a fitted mixed model. \\
    \code{summary} (\ref{sec:summary}) &
    Summary of a mixed model. \\
    \code{terms} &
    Terms representation of a mixed model. \\
    \code{update} (\ref{sec:update}) &
    An updated model using a revised formula or other arguments. \\
    \code{VarCorr} (\ref{sec:summary})  &
    Estimated random-effects variances, standard deviations, and correlations. \\
    \code{vcov} (\ref{sec:summary}) &
    Covariance matrix of the fixed effect estimates. \\
    \code{weights} &
    Prior weights used in model fitting. \\
    \hline
  \end{tabular}
  \caption{List of currently available methods for objects of the class \code{merMod}.}
  \label{tab:methods}
\end{table}

\subsubsection{Updating fitted mixed models}
\label{sec:update}

To illustrate the \code{update} method for \code{merMod} objects we
construct a random intercept only model. This task could be done in
several ways, but we choose to first remove the random effects term
\code{(Days|Subject)} and add a new term with a random intercept,
\begin{knitrout}
\definecolor{shadecolor}{rgb}{0.969, 0.969, 0.969}\color{fgcolor}\begin{kframe}
\begin{alltt}
\hlstd{> }\hlstd{fm3} \hlkwb{<-} \hlkwd{update}\hlstd{(fm1, .}\hlopt{~}\hlstd{.} \hlopt{-} \hlstd{(Days}\hlopt{|}\hlstd{Subject)} \hlopt{+} \hlstd{(}\hlnum{1}\hlopt{|}\hlstd{Subject))}
\hlstd{> }\hlkwd{formula}\hlstd{(fm3)} \hlcom{# how the updated formula is parsed}
\end{alltt}
\begin{verbatim}
Reaction ~ Days + (1 | Subject)
\end{verbatim}
\end{kframe}
\end{knitrout}

\subsubsection{Model summary and associated accessors}
\label{sec:summary}

The \code{summary} method for \code{merMod} objects provides
information about the model fit.  Here we consider the output of
\code{summary(fm1)} in four steps.  The first few lines of output
indicate that the model was fitted by restricted maximum likelihood
(REML) as well as the value of the REML criterion
(Equation~\ref{eq:REMLdeviance}) at convergence (which may also be
extracted using \code{deviance(fm1)}).  The beginning of the summary
also reproduces the model formula and the scaled Pearson residuals,
\begin{knitrout}
\definecolor{shadecolor}{rgb}{0.969, 0.969, 0.969}\color{fgcolor}\begin{kframe}
\begin{verbatim}
Linear mixed model fit by REML ['lmerMod']
Formula: Reaction ~ Days + (Days | Subject)
   Data: sleepstudy

REML criterion at convergence: 1744

Scaled residuals: 
   Min     1Q Median     3Q    Max 
-3.954 -0.463  0.023  0.463  5.179 
\end{verbatim}
\end{kframe}
\end{knitrout}
This information may also be obtained using standard accessor
functions,
\begin{knitrout}
\definecolor{shadecolor}{rgb}{0.969, 0.969, 0.969}\color{fgcolor}\begin{kframe}
\begin{alltt}
\hlstd{> }\hlkwd{formula}\hlstd{(fm1)}
\hlstd{> }\hlkwd{REMLcrit}\hlstd{(fm1)}
\hlstd{> }\hlkwd{quantile}\hlstd{(}\hlkwd{residuals}\hlstd{(fm1,} \hlstr{"pearson"}\hlstd{,} \hlkwc{scaled} \hlstd{=} \hlnum{TRUE}\hlstd{))}
\end{alltt}
\end{kframe}
\end{knitrout}
The second piece of \code{summary} output provides information
regarding the random effects and residual variation,
\begin{knitrout}
\definecolor{shadecolor}{rgb}{0.969, 0.969, 0.969}\color{fgcolor}\begin{kframe}
\begin{verbatim}
Random effects:
 Groups   Name        Variance Std.Dev. Corr
 Subject  (Intercept) 612.1    24.74        
          Days         35.1     5.92    0.07
 Residual             654.9    25.59        
Number of obs: 180, groups:  Subject, 18
\end{verbatim}
\end{kframe}
\end{knitrout}
which can also be accessed using,
\begin{knitrout}
\definecolor{shadecolor}{rgb}{0.969, 0.969, 0.969}\color{fgcolor}\begin{kframe}
\begin{alltt}
\hlstd{> }\hlkwd{print}\hlstd{(vc} \hlkwb{<-} \hlkwd{VarCorr}\hlstd{(fm1),} \hlkwc{comp} \hlstd{=} \hlkwd{c}\hlstd{(}\hlstr{"Variance"}\hlstd{,}\hlstr{"Std.Dev."}\hlstd{))}
\hlstd{> }\hlkwd{nobs}\hlstd{(fm1)}  \hlcom{# number of rows in sleepstudy data}
\hlstd{> }\hlkwd{ngrps}\hlstd{(fm1)} \hlcom{# number of levels of the Subject grouping factor}
\hlstd{> }\hlkwd{sigma}\hlstd{(fm1)} \hlcom{# residual standard deviation}
\end{alltt}
\end{kframe}
\end{knitrout}
The print method for \code{VarCorr} hides the internal structure of
\code{VarCorr.merMod} objects. The internal structure of an object of
this class is a list of matrices, one for each random effects term.
The standard deviations and correlation matrices for each term are
stored as attributes, \code{stddev} and \code{correlation},
respectively, of the variance-covariance matrix, and the residual
standard deviation is stored as attribute \code{sc}.  For programming
use, these objects can be summarized differently using their
\code{as.data.frame} method,
\begin{knitrout}
\definecolor{shadecolor}{rgb}{0.969, 0.969, 0.969}\color{fgcolor}\begin{kframe}
\begin{alltt}
\hlstd{> }\hlkwd{as.data.frame}\hlstd{(}\hlkwd{VarCorr}\hlstd{(fm1))}
\end{alltt}
\begin{verbatim}
       grp        var1 var2    vcov    sdcor
1  Subject (Intercept) <NA> 612.090 24.74045
2  Subject        Days <NA>  35.072  5.92213
3  Subject (Intercept) Days   9.604  0.06555
4 Residual        <NA> <NA> 654.941 25.59182
\end{verbatim}
\end{kframe}
\end{knitrout}
which contains one row for each variance or covariance parameter. The
\code{grp} column gives the grouping factor associated with this
parameter. The \code{var1} and \code{var2} columns give the names of
the variables associated with the parameter (\code{var2} is \code{<NA>}
unless it is a covariance parameter). The \code{vcov} column gives the
variances and covariances, and the \code{sdcor} column gives these
numbers on the standard deviation and correlation scales.

The next chunk of output gives the fixed effect estimates,
\begin{knitrout}
\definecolor{shadecolor}{rgb}{0.969, 0.969, 0.969}\color{fgcolor}\begin{kframe}
\begin{verbatim}
Fixed effects:
            Estimate Std. Error t value
(Intercept)   251.41       6.82    36.8
Days           10.47       1.55     6.8
\end{verbatim}
\end{kframe}
\end{knitrout}
Note that there are no $p$~values (see Section~\ref{sec:pvalues}).
The fixed effect point estimates may be obtained separately via
\code{fixef(fm1)}.
Conditional modes of the random effects coefficients can be obtained
with \code{ranef} (see section~\ref{sec:condVar} for information on
the theory).  Finally, we have the correlations between the fixed
effect estimates
\begin{knitrout}
\definecolor{shadecolor}{rgb}{0.969, 0.969, 0.969}\color{fgcolor}\begin{kframe}
\begin{verbatim}
Correlation of Fixed Effects:
     (Intr)
Days -0.138
\end{verbatim}
\end{kframe}
\end{knitrout}
The full variance-covariance matrix of the fixed effects estimates can
be obtained in the usual \proglang{R} way with the \code{vcov} method
(Section~\ref{sec:vcov}).  Alternatively, \code{coef(summary(fm1))} will
return the full fixed-effects parameter table as shown in the summary.

\subsubsection{Diagnostic plots}
\label{sec:diagnostics}

\pkg{lme4} provides tools for generating most of the standard
graphical diagnostic plots (with the exception of those incorporating
influence measures, e.g., Cook's distance and
leverage plots), in a way modeled on the diagnostic graphics of
\pkg{nlme} \citep{R:Pinheiro+Bates:2000}.  In particular, the following
commands will show the standard fitted vs. residual, scale-location,
and quantile-quantile plots familiar from the
\code{plot} method in base \proglang{R}
for linear models (objects of class \code{lm}):
\begin{knitrout}
\definecolor{shadecolor}{rgb}{0.969, 0.969, 0.969}\color{fgcolor}\begin{kframe}
\begin{alltt}
\hlstd{> }\hlkwd{plot}\hlstd{(fm1,}\hlkwc{type}\hlstd{=}\hlkwd{c}\hlstd{(}\hlstr{"p"}\hlstd{,}\hlstr{"smooth"}\hlstd{))}          \hlcom{## fitted vs residual}
\hlstd{> }\hlkwd{plot}\hlstd{(fm1,}\hlkwd{sqrt}\hlstd{(}\hlkwd{abs}\hlstd{(}\hlkwd{resid}\hlstd{(.)))}\hlopt{~}\hlkwd{fitted}\hlstd{(.),} \hlcom{## scale-location}
\hlstd{+ }     \hlkwc{type}\hlstd{=}\hlkwd{c}\hlstd{(}\hlstr{"p"}\hlstd{,}\hlstr{"smooth"}\hlstd{))}
\hlstd{> }\hlkwd{qqmath}\hlstd{(fm1,}\hlkwc{id}\hlstd{=}\hlnum{0.05}\hlstd{)}                     \hlcom{## quantile-quantile}
\end{alltt}
\end{kframe}
\end{knitrout}
(In contrast to \code{plot.lm}, these scale-location and Q-Q plots
are based on raw rather than standardized residuals.)

In addition to these standard diagnostic plots, which
examine the validity of various assumptions (linearity,
homoscedasticity, normality) at the level of the residuals,
one can also use the \code{dotplot} and \code{qqmath} methods
for the conditional modes (i.e., \code{ranef.mer}
objects generated by \code{ranef(fit)})
to check for interesting patterns and normality in the conditional
modes.  \pkg{lme4} does not provide influence diagnostics, but
these (and other useful diagnostic procedures) are available
in the dependent packages \pkg{HLMdiag} and \pkg{influence.ME}
\citep{HLMdiag_pkg,influenceME_pkg}.

Finally, \emph{posterior predictive simulation}
\citep{gelman_data_2006} is a generally useful diagnostic tool,
adapted from Bayesian methods, for exploring model fit. The user picks
some summary statistic of interest.  They then compute the summary
statistic for an ensemble of simulations (Section~\ref{sec:predict}),
and see where the observed data falls within the simulated
distribution; if the observed data is extreme, we might conclude that
the model is a poor representation of reality.  For example, using the
sleep study fit and choosing the interquartile range 
of the reaction times as the summary
statistic:
\begin{knitrout}
\definecolor{shadecolor}{rgb}{0.969, 0.969, 0.969}\color{fgcolor}\begin{kframe}
\begin{alltt}
\hlstd{> }\hlstd{iqrvec} \hlkwb{<-} \hlkwd{sapply}\hlstd{(}\hlkwd{simulate}\hlstd{(fm1,}\hlnum{1000}\hlstd{),IQR)}
\hlstd{> }\hlstd{obsval} \hlkwb{<-} \hlkwd{IQR}\hlstd{(sleepstudy}\hlopt{$}\hlstd{Reaction)}
\hlstd{> }\hlstd{post.pred.p} \hlkwb{<-} \hlkwd{mean}\hlstd{(obsval}\hlopt{>=}\hlkwd{c}\hlstd{(obsval,iqrvec))}
\end{alltt}
\end{kframe}
\end{knitrout}
The (one-tailed) posterior predictive $p$~value of 0.78
indicates that the model represents the data adequately, at least for
this summary statistic.  In contrast to the full Bayesian case, the
procedure described here does not allow for the uncertainty in the
estimated parameters. However, it should be a reasonable approximation
when the residual variation is large.

\subsubsection{Sequential decomposition and model comparison}
\label{sec:modComp}

Objects of class \code{merMod} have an \code{anova} method which
returns $F$~statistics corresponding to the sequential decomposition
of the contributions of fixed-effects terms.  In order to illustrate
this sequential ANOVA decomposition, we fit a model with orthogonal
linear and quadratic \code{Days} terms,
\begin{knitrout}
\definecolor{shadecolor}{rgb}{0.969, 0.969, 0.969}\color{fgcolor}\begin{kframe}
\begin{alltt}
\hlstd{> }\hlstd{fm4} \hlkwb{<-} \hlkwd{lmer}\hlstd{(Reaction} \hlopt{~} \hlstd{polyDays[,}\hlnum{1}\hlstd{]} \hlopt{+} \hlstd{polyDays[,}\hlnum{2}\hlstd{]} \hlopt{+}
\hlstd{+ }            \hlstd{(polyDays[,}\hlnum{1}\hlstd{]} \hlopt{+} \hlstd{polyDays[,}\hlnum{2}\hlstd{]} \hlopt{|} \hlstd{Subject),}
\hlstd{+ }            \hlkwd{within}\hlstd{(sleepstudy, polyDays} \hlkwb{<-} \hlkwd{poly}\hlstd{(Days,} \hlnum{2}\hlstd{)))}
\hlstd{> }\hlkwd{anova}\hlstd{(fm4)}
\end{alltt}
\begin{verbatim}
Analysis of Variance Table
              Df Sum Sq Mean Sq F value
polyDays[, 1]  1  23874   23874   46.08
polyDays[, 2]  1    340     340    0.66
\end{verbatim}
\end{kframe}
\end{knitrout}

The relative magnitudes of the two sums of squares indicate that the
quadratic term explains much less variation than the linear term.
Furthermore, the magnitudes of the two $F$~statistics strongly
suggest significance of the linear term, but not the quadratic term.
Notice that this is only an informal assessment of significance as
there are no $p$~values associated with these $F$~statistics, an issue
discussed in more detail in Section~\ref{sec:pvalues}.

To understand how these quantities are computed, let $\bm R_i$ contain
the rows of $\bm R_X$ (Equation~\ref{eq:blockCholeskyDecomp}) associated
with the $i$th fixed-effects term.  Then the sum of squares for term
$i$ is,
\begin{equation}
  \label{eq:SS}
  SS_i = \widehat{\bm\beta}\trans\bm R_i\trans \bm R_i \widehat{\bm\beta}
\end{equation}
If $DF_i$ is the number of columns in $\bm R_i$, then the
$F$~statistic for term $i$ is,
\begin{equation}
  \label{eq:Fstat}
  F_i = \frac{SS_i}{\widehat{\sigma}^2 DF_i}
\end{equation}

For multiple arguments, the \code{anova} method returns model
comparison statistics,
\begin{knitrout}
\definecolor{shadecolor}{rgb}{0.969, 0.969, 0.969}\color{fgcolor}\begin{kframe}
\begin{alltt}
\hlstd{> }\hlkwd{anova}\hlstd{(fm1,fm2,fm3)}
\end{alltt}

{\ttfamily\noindent\itshape\color{messagecolor}{refitting model(s) with ML (instead of REML)}}\begin{verbatim}
Data: sleepstudy
Models:
fm3: Reaction ~ Days + (1 | Subject)
fm2: Reaction ~ Days + ((1 | Subject) + (0 + Days | Subject))
fm1: Reaction ~ Days + (Days | Subject)
    Df  AIC  BIC logLik deviance Chisq Chi Df Pr(>Chisq)
fm3  4 1802 1815   -897     1794                        
fm2  5 1762 1778   -876     1752 42.08      1    8.8e-11
fm1  6 1764 1783   -876     1752  0.06      1        0.8
\end{verbatim}
\end{kframe}
\end{knitrout}

The output shows $\chi^2$ statistics representing the difference in
deviance between successive models, as well as $p$~values based on
likelihood ratio test comparisons.  In this case, the sequential
comparison shows that adding a linear effect of time uncorrelated with
the intercept leads to an enormous and significant drop in deviance
($\Delta\mbox{deviance} \approx 42$, $p \approx
\ensuremath{10^{-10}}$), while the further addition of correlation
between the slope and intercept leads to a trivial and non-significant
change in deviance ($\Delta\mbox{deviance} \approx
0.06$).  For objects of class \code{lmerMod} the
default behavior is to refit the models with ML if fitted with
\code{REML = TRUE}, which is necessary in order to get sensible
answers when comparing models that differ in their fixed effects; this
can be controlled via the \code{refit} argument.

\subsubsection{Computing $p$~values}
\label{sec:pvalues}

One of the more controversial design decisions of \pkg{lme4} has been
to omit the output of $p$~values associated with sequential ANOVA
decompositions of fixed effects.  The absence of analytical results for null
distributions of parameter estimates in complex situations
(e.g., unbalanced or partially crossed designs) is a long-standing
problem in mixed-model inference.  While the null distributions (and
the sampling distributions of non-null estimates) are asymptotically
normal, these distributions are not $t$~distributed for finite size
samples --- nor are the corresponding null distributions of
differences in scaled deviances $F$~distributed.  Thus approximate methods
for computing the approximate degrees of freedom for
$t$~distributions, or the denominator degrees of freedom for
$F$~statistics \citep{Satterthwaite_1946,kenward_small_1997}, are at
best \emph{ad hoc} solutions.

However, computing finite-size-corrected $p$~values is
sometimes necessary.  Therefore, although the package does not provide
them (except via parametric bootstrapping,
Section~\ref{sec:pb}), we have provided a help page to guide users in
finding appropriate methods:
\begin{knitrout}
\definecolor{shadecolor}{rgb}{0.969, 0.969, 0.969}\color{fgcolor}\begin{kframe}
\begin{alltt}
\hlstd{> }\hlkwd{help}\hlstd{(}\hlstr{"pvalues"}\hlstd{)}
\end{alltt}
\end{kframe}
\end{knitrout}
This help page provides pointers to other packages that provide
machinery for calculating $p$~values associated with \code{merMod}
objects. It also suggests framing the inference problem as a
likelihood ratio test (achieved by supplying multiple \code{merMod}
objects to the \code{anova} method (Section~\ref{sec:modComp}), as
well as alternatives to $p$~values such as confidence intervals
(Section~\ref{sec:CI}).

Previous versions of \pkg{lme4} provided the \code{mcmcsamp} function,
which generated a Markov chain Monte Carlo sample from the posterior
distribution of the parameters (assuming flat priors). Due to
difficulty in constructing a version of \code{mcmcsamp} that was
reliable even in cases where the estimated random effect variances
were near zero, \code{mcmcsamp} has been withdrawn.

\subsubsection{Computing confidence intervals}
\label{sec:CI}

As described above (Sections~\ref{sec:vcov}, \ref{sec:profile},
\ref{sec:pb}), \pkg{lme4} provides confidence intervals (using
\code{confint}) via Wald approximations (for fixed-effect parameters
only), likelihood profiling, or parametric bootstrapping (the
\code{type} argument selects the method).

As is typical for computationally intensive profile confidence
intervals in R, the intervals can be computed either directly from
fitted \code{merMod} objects (in which case profiling is done as an
interim step), or from a previously computed likelihood profile (of
class \code{thpr}, for ``theta profile'').  Parametric bootstrapping
confidence intervals use a thin wrapper around the \code{bootMer}
function that passes the results to \code{boot.ci} from the \pkg{boot}
package \citep{boot_pkg,DavisonHinkley1997} for confidence interval calculation.

In the running sleep study examples, the 95\% confidence intervals
estimated by all three methods are quite similar.  The largest
change is a 26\% difference in confidence
interval widths between profile and parametric bootstrap methods for
the correlation between the intercept and slope random effects
(\{\ensuremath{-0.54},0.98\} vs.
\{\ensuremath{-0.48},0.68\}).

\subsubsection{General profile zeta and related plots}
\label{sec:profzeta}

\pkg{lme4} provides several functions (built on \pkg{lattice}
graphics) for plotting the profile zeta functions
(Section~\ref{sec:profile}) and other related quantities.

\begin{itemize}
\item The \emph{profile zeta} plot
  (Figure~\ref{fig:profile_zeta_plot}; \code{xyplot}) is simply a plot
  of the profile zeta function for each model parameter; linearity of
  this plot for a given parameter implies that the likelihood profile
  is quadratic (and thus that Wald approximations would be reasonably
  accurate).
\item The \emph{profile density plot}
  (Figure~\ref{fig:profile_density_plot}; \code{densityplot}) displays
  an approximation of the probability density function of the sampling
  distribution for each parameter.  These densities are derived by
  setting the cumulative distribution function (c.d.f) to be
  $\Phi(\zeta)$ where $\Phi$ is the c.d.f.{} of the standard normal
  distribution.  This is not quite the same as evaluating the
  distribution of the estimator of the parameter, which for mixed
  models can be very difficult, but it gives a reasonable
  approximation.  If the profile zeta plot is linear, then the profile
  density plot will be Gaussian.
\item The \emph{profile pairs plot}
  (Figure~\ref{fig:profile_pairs_plot}; \code{splom}) gives an
  approximation of the two-dimensional profiles of pairs of
  parameters, interpolated from the univariate profiles as described
  in \citet[Chapter 6]{bateswatts88:_nonlin}.  The profile pairs plot
  shows two-dimensional 50\%, 80\%, 90\%, 95\% and 99\% marginal
  confidence regions based on the likelihood ratio, as well as the
  \emph{profile traces}, which indicate the conditional estimates of
  each parameter for fixed values of the other parameter.  While
  panels above the diagonal show profiles with respect to the original
  parameters (with random effects parameters on the standard
  deviation\slash{}correlation scale, as for all profile plots), the
  panels below the diagonal show plots on the $(\zeta_i,\zeta_j)$
  scale.  The below-diagonal panels allow us to see distortions from
  an elliptical shape due to nonlinearity of the traces, separately
  from the one-dimensional distortions caused by a poor choice of
  scale for the parameter.  The $\zeta$ scales provide, in some sense,
  the best possible set of single-parameter transformations for
  assessing the contours.  On the $\zeta$ scales the extent of a
  contour on the horizontal axis is exactly the same as the extent on
  the vertical axis and both are centered about zero.
\end{itemize}

\begin{knitrout}
\definecolor{shadecolor}{rgb}{0.969, 0.969, 0.969}\color{fgcolor}\begin{figure}[tb]

{\centering \includegraphics[width=\maxwidth]{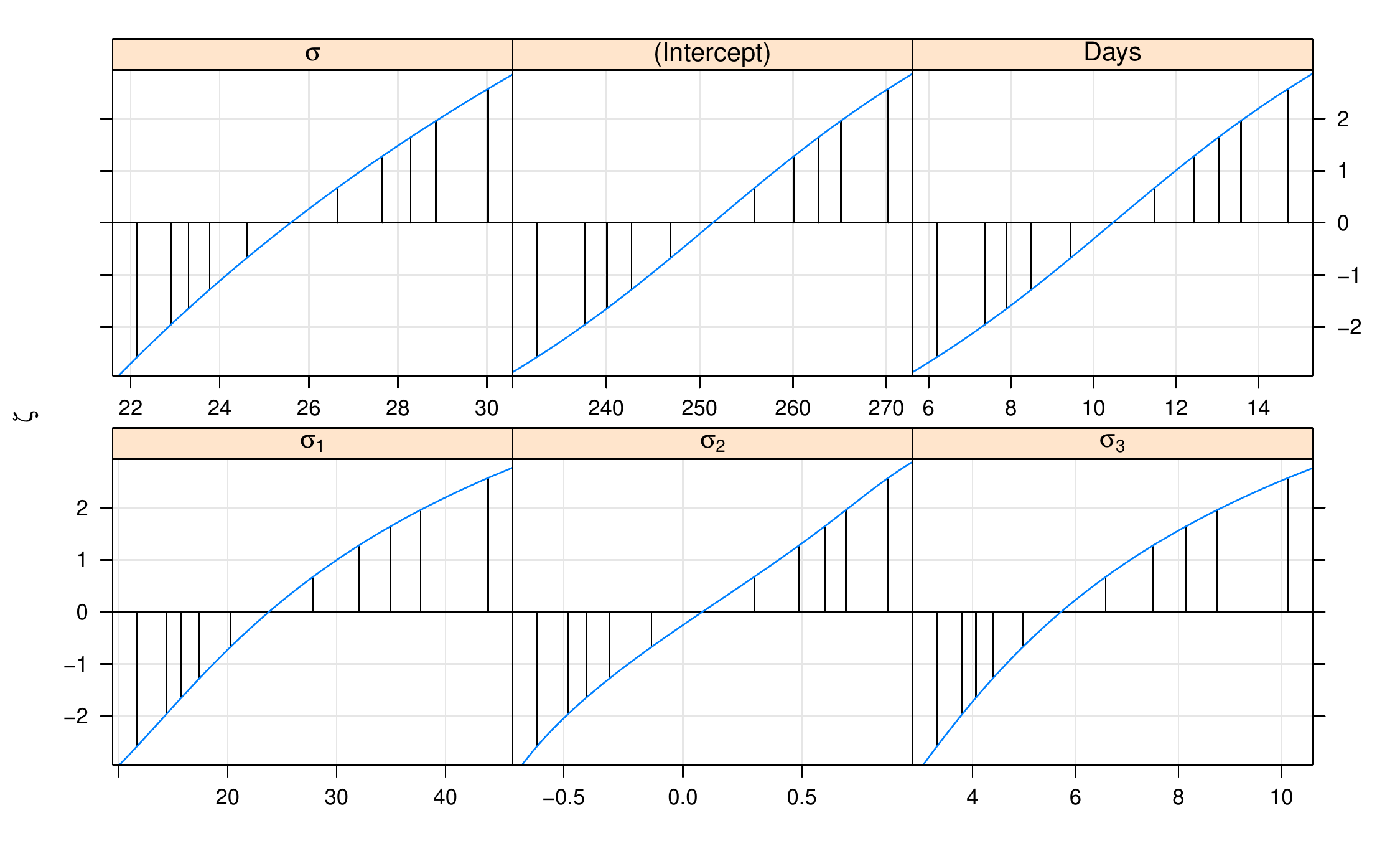} 

}

\caption[Profile zeta plot]{Profile zeta plot: \code{xyplot(prof.obj)}\label{fig:profile_zeta_plot}}
\end{figure}

\end{knitrout}
\begin{knitrout}
\definecolor{shadecolor}{rgb}{0.969, 0.969, 0.969}\color{fgcolor}\begin{figure}[tb]

{\centering \includegraphics[width=\maxwidth]{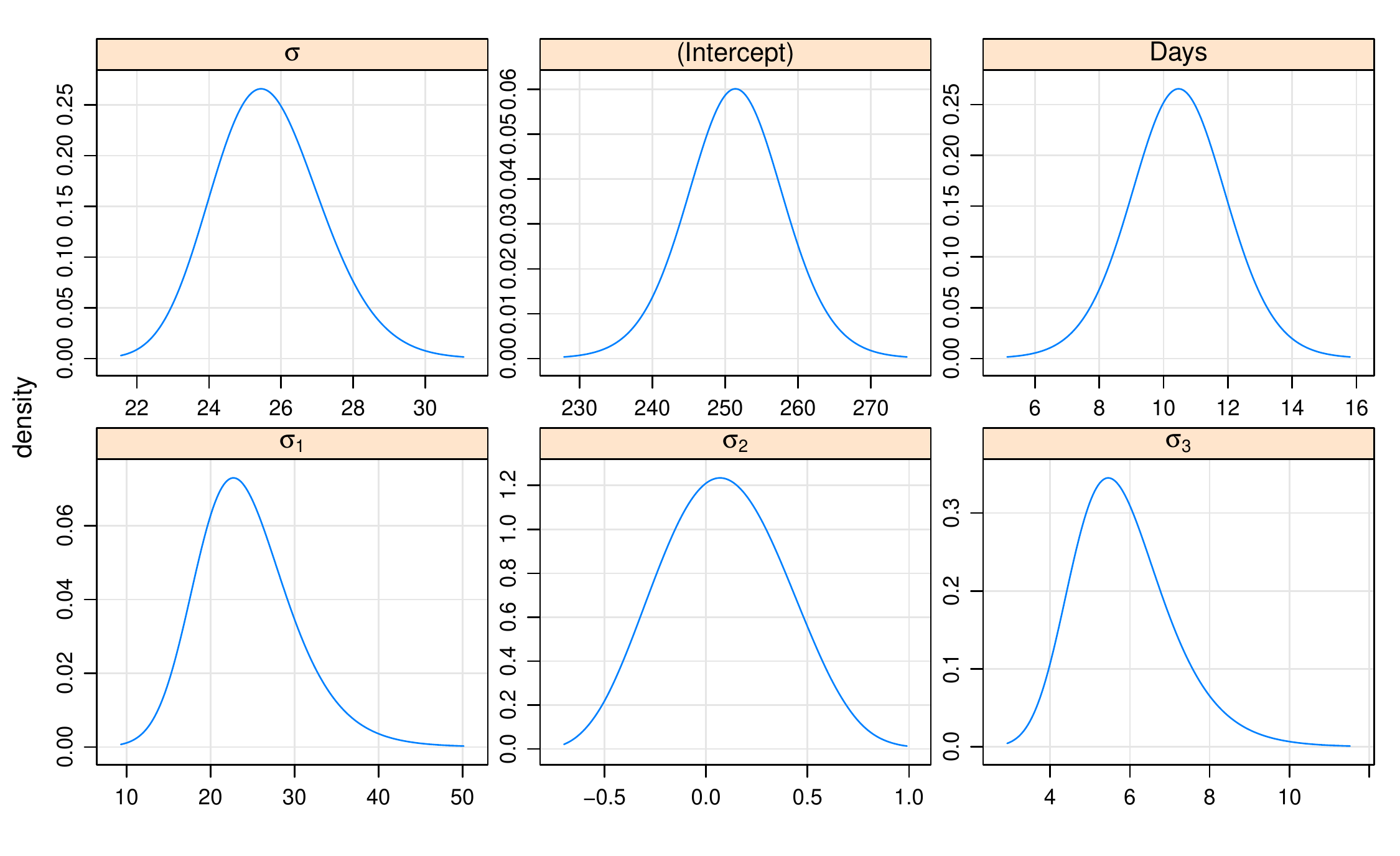} 

}

\caption[Profile density plot]{Profile density plot: \code{densityplot(prof.obj)}\label{fig:profile_density_plot}}
\end{figure}

\end{knitrout}
\begin{knitrout}
\definecolor{shadecolor}{rgb}{0.969, 0.969, 0.969}\color{fgcolor}\begin{figure}[htb]

{\centering \includegraphics[width=\maxwidth,height=5.5in]{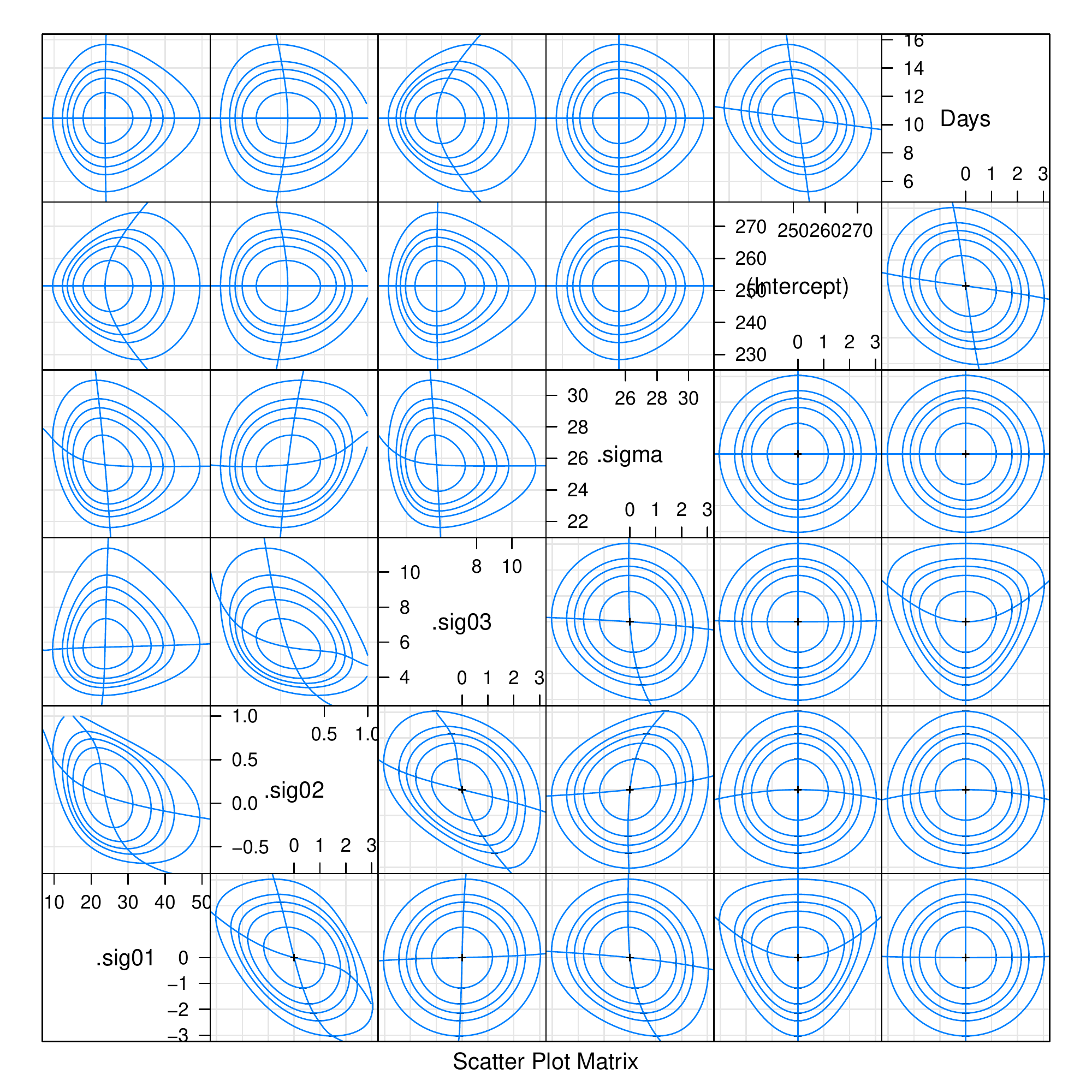} 

}

\caption[Profile pairs plot]{Profile pairs plot: \code{splom(prof.obj)}\label{fig:profile_pairs_plot}}
\end{figure}

\end{knitrout}

For users who want to build their own graphical displays of the
profile, there is an \code{as.data.frame} method that converts profile
(\code{thpr}) objects to a more convenient format.

\subsubsection{Computing fitted and predicted values; simulating}
\label{sec:predict}

Because mixed models involve random coefficients, one must always
clarify whether predictions should be based on the marginal
distribution of the response variable or on the distribution that is
conditional on the modes of the random effects
(Equation~\ref{eq:distYgivenU}).  The \code{fitted} method retrieves fitted
values that are conditional on all of the modes of the random effects;
the \code{predict} method returns the same values by default, but also
allows for predictions to be made conditional on different sets of
random effects.  For example, if the \code{re.form} argument is set to
\code{NULL} (the default), then the predictions are conditional on all
random effects in the model; if \code{re.form} is \verb+~0+ or
\code{NA}, then the predictions are made at the population level (all
random effect values set to zero).  In models with multiple random
effects, the user can give \code{re.form} as a formula that specifies
which random effects are conditioned on.

\pkg{lme4} also provides a \code{simulate} method, which allows
similar flexibility in conditioning on random effects; in addition it
allows the user to choose (via the \code{use.u} argument) between
conditioning on the fitted conditional modes or choosing a new set of
simulated conditional modes (zero-mean Normal deviates chosen
according to the estimated random effects variance-covariance
matrices).  Finally, the \code{simulate} method has a method for
\code{formula} objects, which allows for \emph{de novo} simulation in
the absence of a fitted model.  In this case, the user must specify
the random effects ($\bm\theta$), fixed effects ($\bm\beta$), and
residual standard deviation ($\sigma$) parameters via the
\code{newparams} argument.  The standard simulation method (based on a
\code{merMod} object) is the basis of parametric bootstrapping
(Section~\ref{sec:pb}) and posterior predictive simulation
(Section~\ref{sec:diagnostics}); \emph{de novo} simulation based on a
formula provides a flexible framework for power analysis.

\section{Conclusion}

Mixed models are an extremely useful but computationally intensive
approach. Computational limitations are especially important because
mixed models are commonly applied to moderately large data sets
($10^4$--$10^6$ observations).  By developing an efficient, general,
and (now) well-documented platform for fitted mixed models in
\proglang{R}, we hope to provide both a practical tool for end users
interested in analyzing data and a reusable, modular framework for
downstream developers interested in extending the class of models that
can be easily and efficiently analyzed in \proglang{R}.


We have learned much about linear mixed models in the process of
developing \pkg{lme4}, both from our own attempts to develop robust
and reliable procedures and from the broad community of \pkg{lme4}
users; we have attempted to describe many of these lessons here.  In
moving forward, our main priorities are (1) to maintain the reference
implementation of \pkg{lme4} on CRAN, developing relatively few new
features; (2) to improve the flexibility, efficiency and scalability
of mixed-model analysis across multiple compatible implementations,
including both the \pkg{MixedModels} package for \proglang{Julia} and
the experimental \code{flexLambda} branch of \pkg{lme4}.

\section*{Acknowledgements}

Rune Haubo Christensen, Henrik Singmann, Fabian Scheipl,
Vincent Dorie, and Bin Dai contributed ideas
and code to the current version of \pkg{lme4}; the
large community of \pkg{lme4} users has exercised the software,
discovered bugs, and made many useful suggestions.
S{\o}ren H{\o}jsgaard participated in useful discussions
and Xi Xia  and Christian Zingg exercised the code
and reported problems.
We would like to thank the Banff International Research Station
for hosting a working group on \pkg{lme4}, and 
an NSERC Discovery grant and NSERC postdoctoral fellowship for funding.

\bibliography{lmer}

\appendix

\section{Appendix: modularization examples}
\label{sec:modularExamples}

\subsection{Homogeneous covariance among random effects terms}
\label{sec:homoLmer}

Consider the following mixed-model formula,
\begin{knitrout}
\definecolor{shadecolor}{rgb}{0.969, 0.969, 0.969}\color{fgcolor}\begin{kframe}
\begin{alltt}
\hlstd{> }\hlstd{form} \hlkwb{<-} \hlstd{respVar} \hlopt{~} \hlnum{1} \hlopt{+} \hlstd{(explVar1}\hlopt{|}\hlstd{groupFac1)} \hlopt{+} \hlstd{(explVar2}\hlopt{|}\hlstd{groupFac2)}
\end{alltt}
\end{kframe}
\end{knitrout}
in which we have a fixed intercept and two random effects terms, each
with a random slope and intercept.  The \code{lmer}  function would
fit this formula to data by estimating two $2 \times 2$ covariance
matrices---one for each random effects term.  It is not possible to
use the current version of
\code{lmer} to fit a model in which both terms share the same
$2 \times 2$ covariance matrix.  We illustrate the use of the modular
functions in \pkg{lme4} to fit such a model.

We simulate data from this homogeneous covariance model using a
balanced design that crosses both grouping factors,
\begin{knitrout}
\definecolor{shadecolor}{rgb}{0.969, 0.969, 0.969}\color{fgcolor}\begin{kframe}
\begin{alltt}
\hlstd{> }\hlkwd{set.seed}\hlstd{(}\hlnum{1}\hlstd{)}
\hlstd{> }                                        \hlcom{# number of groups per}
\hlstd{> }                                        \hlcom{# grouping factor}
\hlstd{> }\hlstd{nGrps} \hlkwb{<-} \hlnum{50}
\hlstd{> }                                        \hlcom{# two explanatory variables}
\hlstd{> }\hlstd{explVar} \hlkwb{<-} \hlkwd{data.frame}\hlstd{(}\hlkwc{explVar1} \hlstd{=} \hlkwd{rnorm}\hlstd{(nGrps}\hlopt{^}\hlnum{2}\hlstd{),}
\hlstd{+ }                      \hlkwc{explVar2} \hlstd{=} \hlkwd{rnorm}\hlstd{(nGrps}\hlopt{^}\hlnum{2}\hlstd{))}
\hlstd{> }                                        \hlcom{# two group factors}
\hlstd{> }\hlstd{groupFac} \hlkwb{<-} \hlkwd{expand.grid}\hlstd{(}\hlkwc{groupFac1} \hlstd{=} \hlkwd{as.factor}\hlstd{(}\hlnum{1}\hlopt{:}\hlstd{nGrps),}
\hlstd{+ }                        \hlkwc{groupFac2} \hlstd{=} \hlkwd{as.factor}\hlstd{(}\hlnum{1}\hlopt{:}\hlstd{nGrps))}
\hlstd{> }                                        \hlcom{# random intercepts}
\hlstd{> }\hlstd{randomIntercept} \hlkwb{<-} \hlkwd{expand.grid}\hlstd{(}\hlkwc{randomIntercept1} \hlstd{=} \hlkwd{rnorm}\hlstd{(nGrps),}
\hlstd{+ }                               \hlkwc{randomIntercept2} \hlstd{=} \hlkwd{rnorm}\hlstd{(nGrps))}
\hlstd{> }                                        \hlcom{# random slopes, negatively}
\hlstd{> }                                        \hlcom{# correlated with the}
\hlstd{> }                                        \hlcom{# intercepts}
\hlstd{> }\hlstd{rnmdSlope} \hlkwb{<-} \hlkwd{expand.grid}\hlstd{(}\hlkwc{randomSlope1} \hlstd{=} \hlkwd{rnorm}\hlstd{(nGrps),}
\hlstd{+ }                         \hlkwc{randomSlope2} \hlstd{=} \hlkwd{rnorm}\hlstd{(nGrps))} \hlopt{-} \hlstd{randomIntercept}
\hlstd{> }                                        \hlcom{# linear predictor}
\hlstd{> }\hlstd{linearPredictor} \hlkwb{<-} \hlkwd{apply}\hlstd{(randomIntercept} \hlopt{+} \hlstd{rnmdSlope}\hlopt{*}\hlstd{explVar,} \hlnum{1}\hlstd{, sum)}
\hlstd{> }                                        \hlcom{# residual error}
\hlstd{> }\hlstd{residError} \hlkwb{<-} \hlkwd{rnorm}\hlstd{(nGrps}\hlopt{^}\hlnum{2}\hlstd{)}
\hlstd{> }                                        \hlcom{# resp variable}
\hlstd{> }\hlstd{respVar} \hlkwb{<-} \hlstd{residError} \hlopt{+} \hlstd{linearPredictor}
\hlstd{> }                                        \hlcom{# data set}
\hlstd{> }\hlstd{dat} \hlkwb{<-} \hlkwd{data.frame}\hlstd{(respVar, explVar, groupFac)}
\hlstd{> }\hlkwd{head}\hlstd{(dat)}
\end{alltt}
\begin{verbatim}
  respVar explVar1 explVar2 groupFac1 groupFac2
1 -1.9019  -0.6265  -1.8055         1         1
2  2.2711   0.1836  -0.6780         2         1
3 -3.1781  -0.8356  -0.4734         3         1
4 -3.0882   1.5953   1.0274         4         1
5 -0.2488   0.3295  -0.5974         5         1
6 -3.6511  -0.8205   1.1598         6         1
\end{verbatim}
\end{kframe}
\end{knitrout}

To fit \code{form} to \code{dat} using a homogeneous covariance model,
we construct the \code{homoLmer} function, which we describe in
sections.  To simplify the problem, we only allow the specification of
three arguments: \code{formula}, \code{data}, and \code{use.mkMerMod}.
This last argument specifies how to process the fitted model, which we
discuss below.  All other relevant parameters are set to \code{lmer}
defaults, unless otherwise noted below.  Here is the argument list for
\code{homoLmer},
\begin{knitrout}
\definecolor{shadecolor}{rgb}{0.969, 0.969, 0.969}\color{fgcolor}\begin{kframe}
\begin{alltt}
\hlstd{> }\hlstd{homoLmer} \hlkwb{<-} \hlkwa{function}\hlstd{(}\hlkwc{formula}\hlstd{,} \hlkwc{data}\hlstd{,} \hlkwc{use.mkMerMod} \hlstd{=} \hlnum{FALSE}\hlstd{) \{\}}
\end{alltt}
\end{kframe}
\end{knitrout}
After saving the \code{call}, the next step is to parse the formula
and data using the \code{lFormula} modular function (setting
\code{REML} to \code{FALSE} for simplicity),
\begin{knitrout}
\definecolor{shadecolor}{rgb}{0.969, 0.969, 0.969}\color{fgcolor}\begin{kframe}
\begin{alltt}
\hlstd{> }    \hlstd{mc} \hlkwb{<-} \hlkwd{match.call}\hlstd{()}
\hlstd{> }    \hlstd{lfHetero} \hlkwb{<-} \hlstd{lfHomo} \hlkwb{<-} \hlkwd{lFormula}\hlstd{(}\hlkwc{formula} \hlstd{= formula,} \hlkwc{data} \hlstd{= data,} \hlkwc{REML} \hlstd{=} \hlnum{FALSE}\hlstd{)}
\end{alltt}
\end{kframe}
\end{knitrout}
We then check to make sure that this \code{formula} is a candidate for
a homogeneous covariance model.  In particular, a homogeneous
covariance model only makes sense if each random effects term has the
same number of random effects coefficients,
\begin{knitrout}
\definecolor{shadecolor}{rgb}{0.969, 0.969, 0.969}\color{fgcolor}\begin{kframe}
\begin{alltt}
\hlstd{> }\hlkwa{if}\hlstd{(}\hlkwd{length}\hlstd{(pRE} \hlkwb{<-} \hlkwd{unique}\hlstd{(}\hlkwd{sapply}\hlstd{(cnms} \hlkwb{<-} \hlstd{lfHomo}\hlopt{$}\hlstd{reTrms}\hlopt{$}\hlstd{cnms, length)))} \hlopt{>} \hlnum{1L}\hlstd{) \{}
\hlstd{+ }        \hlkwd{stop}\hlstd{(}\hlstr{"each random effects term must have the same number\textbackslash{}n"}\hlstd{,}
\hlstd{+ }             \hlstr{"of model matrix columns for a homogeneous structure"}\hlstd{)}
\hlstd{+ }    \hlstd{\}}
\end{alltt}
\end{kframe}
\end{knitrout}
We save the number of fixed effects coefficients, \code{p}, number of
covariance parameters per term, \code{nth}, and number random effects
terms, \code{n_trms},
\begin{knitrout}
\definecolor{shadecolor}{rgb}{0.969, 0.969, 0.969}\color{fgcolor}\begin{kframe}
\begin{alltt}
\hlstd{> }    \hlstd{p} \hlkwb{<-} \hlkwd{ncol}\hlstd{(lfHomo}\hlopt{$}\hlstd{X)}
\hlstd{> }    \hlstd{nth} \hlkwb{<-} \hlkwd{choose}\hlstd{(pRE} \hlopt{+} \hlnum{1}\hlstd{,} \hlnum{2}\hlstd{)}
\hlstd{> }    \hlstd{n_trms} \hlkwb{<-} \hlkwd{length}\hlstd{(cnms)}
\end{alltt}
\end{kframe}
\end{knitrout}
We now modify the parsed random effects terms to have homogeneous
covariance structure, which requires three lines.  First, use only the
first \code{nth} elements of the covariance parameters, \code{theta},
as these will be repeated to generate common covariance structure
across random effects terms.  Second, match the lower bounds of the
covariance parameters to the new \code{theta}.  Third, adjust the
mapping from the new \code{theta} to the nonzero elements of the
transposed relative covariance factor, \code{Lambdat}.
\begin{knitrout}
\definecolor{shadecolor}{rgb}{0.969, 0.969, 0.969}\color{fgcolor}\begin{kframe}
\begin{alltt}
\hlstd{> }    \hlstd{lfHomo}\hlopt{$}\hlstd{reTrms} \hlkwb{<-} \hlkwd{within}\hlstd{(lfHomo}\hlopt{$}\hlstd{reTrms, \{}
\hlstd{+ }        \hlstd{theta} \hlkwb{<-} \hlstd{theta[}\hlnum{1}\hlopt{:}\hlstd{nth]}
\hlstd{+ }        \hlstd{lower} \hlkwb{<-} \hlstd{lower[}\hlnum{1}\hlopt{:}\hlstd{nth]}
\hlstd{+ }        \hlstd{Lind} \hlkwb{<-} \hlkwd{rep}\hlstd{(}\hlnum{1}\hlopt{:}\hlstd{nth,} \hlkwc{length} \hlstd{=} \hlkwd{length}\hlstd{(lfHomo}\hlopt{$}\hlstd{reTrms}\hlopt{$}\hlstd{Lambdat}\hlopt{@}\hlkwc{x}\hlstd{))}
\hlstd{+ }    \hlstd{\})}
\end{alltt}
\end{kframe}
\end{knitrout}
Then we construct and minimize a deviance function for the homogeneous
model,
\begin{knitrout}
\definecolor{shadecolor}{rgb}{0.969, 0.969, 0.969}\color{fgcolor}\begin{kframe}
\begin{alltt}
\hlstd{> }    \hlstd{devf} \hlkwb{<-} \hlkwd{do.call}\hlstd{(mkLmerDevfun, lfHomo)}
\hlstd{> }    \hlstd{opt} \hlkwb{<-} \hlkwd{optimizeLmer}\hlstd{(devf)}
\end{alltt}
\end{kframe}
\end{knitrout}
The object \code{opt} contains the optimized values of the homogeneous
covariance parameters.  However, in order to interpret and make
inferences about this model, we need to do more work.  We highlight
two general approaches to doing this, which in this example are
selected with the \code{use.mkMerMod} logical argument.  If
\code{use.mkMerMod} is set to \code{FALSE}, then the parsed data and
formula object, \code{lfHomo}, deviance function, \code{devf}, and
optimization results, \code{opt}, are returned,
\begin{knitrout}
\definecolor{shadecolor}{rgb}{0.969, 0.969, 0.969}\color{fgcolor}\begin{kframe}
\begin{alltt}
\hlstd{> }    \hlkwa{if}\hlstd{(}\hlopt{!}\hlstd{use.mkMerMod) \{}
\hlstd{+ }        \hlkwd{return}\hlstd{(}\hlkwd{list}\hlstd{(}\hlkwc{lf} \hlstd{= lfHomo,} \hlkwc{devf} \hlstd{= devf,} \hlkwc{opt} \hlstd{= opt))}
\hlstd{+ }    \hlstd{\}}
\end{alltt}
\end{kframe}
\end{knitrout}
This option means that the convenience functions of the \code{lme4}
output module (Section~\ref{sec:mkMerMod}) are not available because the resulting
object is not a \code{merMod} object.  However, constructing a
\code{merMod} object that will be treated appropriately by the
convenience functions of the output module is rather difficult, and
sometimes impossible, and should therefore always be done with
caution.  In particular, \code{mkMerMod} should be used to work with
specific functions from the output module; users should never
assume that because some aspects of the output module give appropriate
results with a \code{merMod} object constructed from a modular fit,
all aspects will work.  It will sometimes be
necessary, as we next illustrate, to extend the \code{merMod} class
and rewrite some methods from the output module that are specialized
for this new class.

Next we update the unmodified model by capturing the modified model
estimates of the covariance parameters,
\begin{knitrout}
\definecolor{shadecolor}{rgb}{0.969, 0.969, 0.969}\color{fgcolor}\begin{kframe}
\begin{alltt}
\hlstd{> }    \hlstd{th} \hlkwb{<-} \hlkwd{rep}\hlstd{(opt}\hlopt{$}\hlstd{par, n_trms)}
\end{alltt}
\end{kframe}
\end{knitrout}
constructing a deviance function for the unmodified model,
\begin{knitrout}
\definecolor{shadecolor}{rgb}{0.969, 0.969, 0.969}\color{fgcolor}\begin{kframe}
\begin{alltt}
\hlstd{> }    \hlstd{devf0} \hlkwb{<-} \hlkwd{do.call}\hlstd{(mkLmerDevfun, lfHetero)}
\end{alltt}
\end{kframe}
\end{knitrout}
and installing the parameters in the environment of the unmodified
deviance function,
\begin{knitrout}
\definecolor{shadecolor}{rgb}{0.969, 0.969, 0.969}\color{fgcolor}\begin{kframe}
\begin{alltt}
\hlstd{> }    \hlkwd{devf0}\hlstd{(opt}\hlopt{$}\hlstd{par} \hlkwb{<-} \hlstd{th)}
\end{alltt}
\end{kframe}
\end{knitrout}

Finally, we need to extend the \code{merMod} class,
\begin{knitrout}
\definecolor{shadecolor}{rgb}{0.969, 0.969, 0.969}\color{fgcolor}\begin{kframe}
\begin{alltt}
\hlstd{> }\hlkwd{setClass}\hlstd{(}\hlstr{"homoLmerMod"}\hlstd{,} \hlkwd{representation}\hlstd{(}\hlkwc{thetaUnique} \hlstd{=} \hlstr{"numeric"}\hlstd{),}
\hlstd{+ }         \hlkwc{contains} \hlstd{=} \hlstr{"lmerMod"}\hlstd{)}
\end{alltt}
\end{kframe}
\end{knitrout}
and redefine some methods for this extension,
\begin{knitrout}
\definecolor{shadecolor}{rgb}{0.969, 0.969, 0.969}\color{fgcolor}\begin{kframe}
\begin{alltt}
\hlstd{> }\hlstd{logLik.homoLmerMod} \hlkwb{<-} \hlkwa{function}\hlstd{(}\hlkwc{object}\hlstd{,} \hlkwc{...}\hlstd{) \{}
\hlstd{+ }    \hlstd{ll} \hlkwb{<-} \hlstd{lme4:::}\hlkwd{logLik.merMod}\hlstd{(object, ...)}
\hlstd{+ }    \hlkwd{attr}\hlstd{(ll,} \hlstr{"df"}\hlstd{)} \hlkwb{<-} \hlkwd{length}\hlstd{(object}\hlopt{@}\hlkwc{beta}\hlstd{)} \hlopt{+}
\hlstd{+ }        \hlkwd{length}\hlstd{(object}\hlopt{@}\hlkwc{thetaUnique}\hlstd{)} \hlopt{+}
\hlstd{+ }            \hlstd{object}\hlopt{@}\hlkwc{devcomp}\hlstd{[[}\hlstr{"dims"}\hlstd{]][[}\hlstr{"useSc"}\hlstd{]]}
\hlstd{+ }    \hlkwd{return}\hlstd{(ll)}
\hlstd{+ }\hlstd{\}}
\hlstd{> }\hlstd{refitML.homoLmerMod} \hlkwb{<-} \hlkwa{function}\hlstd{(}\hlkwc{object}\hlstd{,} \hlkwc{newresp}\hlstd{,} \hlkwc{...}\hlstd{) \{}
\hlstd{+ }    \hlkwa{if}\hlstd{(}\hlopt{!}\hlkwd{isREML}\hlstd{(object)} \hlopt{&&} \hlkwd{missing}\hlstd{(newresp))} \hlkwd{return}\hlstd{(object)}
\hlstd{+ }    \hlkwd{stop}\hlstd{(}\hlstr{"can't refit homoLmerMod objects yet"}\hlstd{)}
\hlstd{+ }\hlstd{\}}
\end{alltt}
\end{kframe}
\end{knitrout}
\begin{knitrout}
\definecolor{shadecolor}{rgb}{0.969, 0.969, 0.969}\color{fgcolor}\begin{kframe}
\begin{alltt}
\hlstd{> }\hlstd{mod} \hlkwb{<-} \hlkwd{homoLmer}\hlstd{(form, dat,} \hlkwc{use.mkMerMod} \hlstd{=} \hlnum{TRUE}\hlstd{)}
\hlstd{> }\hlkwd{summary}\hlstd{(mod)}
\end{alltt}
\begin{verbatim}
Linear mixed model fit by maximum likelihood  ['homoLmerMod']
Formula: form
   Data: dat

     AIC      BIC   logLik deviance df.resid 
    7920     7950    -3955     7910     2492 

Scaled residuals: 
   Min     1Q Median     3Q    Max 
-3.094 -0.640 -0.008  0.648  3.426 

Random effects:
 Groups    Name        Variance Std.Dev. Corr 
 groupFac1 (Intercept) 0.955    0.977         
           explVar1    1.878    1.370    -0.76
 groupFac2 (Intercept) 0.955    0.977         
           explVar2    1.878    1.370    -0.76
 Residual              1.026    1.013         
Number of obs: 2500, groups:  groupFac1, 50 groupFac2, 50

Fixed effects:
            Estimate Std. Error t value
(Intercept)  -0.0477     0.1291   -0.37
\end{verbatim}
\end{kframe}
\end{knitrout}

\subsection{Homogeneous variance over all random effects}
\label{sec:homoVar}

We modify the \code{sleepstudy} example,
\begin{knitrout}
\definecolor{shadecolor}{rgb}{0.969, 0.969, 0.969}\color{fgcolor}\begin{kframe}
\begin{alltt}
\hlstd{> }\hlstd{parsedFormula} \hlkwb{<-} \hlkwd{lFormula}\hlstd{(}\hlkwc{formula} \hlstd{= Reaction} \hlopt{~} \hlstd{Days} \hlopt{+} \hlstd{(Days}\hlopt{|}\hlstd{Subject),}
\hlstd{+ }                          \hlkwc{data} \hlstd{= sleepstudy)}
\end{alltt}
\end{kframe}
\end{knitrout}
such that the random slope and intercept are uncorrelated and have
identical variance
(because the slope and intercept have different units,
this example is not entirely sensible --- it is offered for
technical illustration only).
  The \code{reTrms} component of
\code{parsedFormula} contains the relevant information, and can be
modified using the \code{within} function,
\begin{knitrout}
\definecolor{shadecolor}{rgb}{0.969, 0.969, 0.969}\color{fgcolor}\begin{kframe}
\begin{alltt}
\hlstd{> }\hlstd{parsedFormula}\hlopt{$}\hlstd{reTrms} \hlkwb{<-} \hlkwd{within}\hlstd{(parsedFormula}\hlopt{$}\hlstd{reTrms, \{}
\hlstd{+ }                                        \hlcom{# capture the dimension of}
\hlstd{+ }                                        \hlcom{# Lambda}
\hlstd{+ }    \hlstd{q} \hlkwb{<-} \hlkwd{nrow}\hlstd{(Lambdat)}
\hlstd{+ }                                        \hlcom{# alter the mapping from theta}
\hlstd{+ }                                        \hlcom{# to the non-zero elements of}
\hlstd{+ }                                        \hlcom{# Lambda}
\hlstd{+ }    \hlstd{Lind} \hlkwb{<-} \hlkwd{rep}\hlstd{(}\hlnum{1}\hlstd{, q)}
\hlstd{+ }                                        \hlcom{# construct a diagonal Lambda}
\hlstd{+ }    \hlstd{Lambdat} \hlkwb{<-} \hlkwd{sparseMatrix}\hlstd{(}\hlnum{1}\hlopt{:}\hlstd{q,} \hlnum{1}\hlopt{:}\hlstd{q,} \hlkwc{x} \hlstd{= Lind)}
\hlstd{+ }                                        \hlcom{# initialize 1-dimensional}
\hlstd{+ }                                        \hlcom{# theta}
\hlstd{+ }    \hlstd{theta} \hlkwb{<-} \hlnum{1}
\hlstd{+ }\hlstd{\})}
\end{alltt}
\end{kframe}
\end{knitrout}
We fit this modified model using the modularised functions,
\begin{knitrout}
\definecolor{shadecolor}{rgb}{0.969, 0.969, 0.969}\color{fgcolor}\begin{kframe}
\begin{alltt}
\hlstd{> }\hlstd{devianceFunction} \hlkwb{<-} \hlkwd{do.call}\hlstd{(mkLmerDevfun, parsedFormula)}
\hlstd{> }\hlstd{optimizerOutput} \hlkwb{<-} \hlkwd{optimizeLmer}\hlstd{(devianceFunction)}
\end{alltt}
\end{kframe}
\end{knitrout}
Because \code{lme4} is not designed for post-processing of such
models, \code{mkMerMod} cannot be expected to give sensible results.
To illustrate this point, we use it anyway and point out where it
makes mistakes,
\begin{knitrout}
\definecolor{shadecolor}{rgb}{0.969, 0.969, 0.969}\color{fgcolor}\begin{kframe}
\begin{alltt}
\hlstd{> }\hlkwd{mkMerMod}\hlstd{(}   \hlkwc{rho} \hlstd{=} \hlkwd{environment}\hlstd{(devianceFunction),}
\hlstd{+ }            \hlkwc{opt} \hlstd{= optimizerOutput,}
\hlstd{+ }         \hlkwc{reTrms} \hlstd{= parsedFormula}\hlopt{$}\hlstd{reTrms,}
\hlstd{+ }             \hlkwc{fr} \hlstd{= parsedFormula}\hlopt{$}\hlstd{fr)}
\end{alltt}
\begin{verbatim}
Linear mixed model fit by REML ['lmerMod']
REML criterion at convergence: 1759
\end{verbatim}

{\ttfamily\noindent\color{warningcolor}{Warning: data length is not a multiple of split variable}}\begin{verbatim}
Random effects:
 Groups   Name        Std.Dev. Corr
 Subject  (Intercept)  8.9         
          Days        12.6     0.71
 Residual             27.4         
Number of obs: 180, groups:  Subject, 18
Fixed Effects:
(Intercept)         Days  
      251.4         10.5  
\end{verbatim}
\end{kframe}
\end{knitrout}
Note that \code{mkMerMod} is attempting to estimate a correlation
between the slope and intercept, which makes no sense within the
current model.  When using modular functions, the user is responsible
for producing interpretable output.  For example, to find the residual
and random-effects standard deviation, we may use,
\begin{knitrout}
\definecolor{shadecolor}{rgb}{0.969, 0.969, 0.969}\color{fgcolor}\begin{kframe}
\begin{alltt}
\hlstd{> }\hlkwd{with}\hlstd{(}\hlkwd{environment}\hlstd{(devianceFunction), \{}
\hlstd{+ }                                        \hlcom{# dimensions of the problem}
\hlstd{+ }    \hlstd{n} \hlkwb{<-} \hlkwd{length}\hlstd{(resp}\hlopt{$}\hlstd{y)}
\hlstd{+ }    \hlstd{p} \hlkwb{<-} \hlkwd{length}\hlstd{(pp}\hlopt{$}\hlstd{beta0)}
\hlstd{+ }                                        \hlcom{# penalized weighted residual}
\hlstd{+ }                                        \hlcom{# sum-of-squares}
\hlstd{+ }    \hlstd{pwrss} \hlkwb{<-} \hlstd{resp}\hlopt{$}\hlkwd{wrss}\hlstd{()} \hlopt{+} \hlstd{pp}\hlopt{$}\hlkwd{sqrL}\hlstd{(}\hlnum{1}\hlstd{)}
\hlstd{+ }                                        \hlcom{# residual standard deviation}
\hlstd{+ }    \hlstd{sig} \hlkwb{<-} \hlkwd{sqrt}\hlstd{(pwrss}\hlopt{/}\hlstd{(n}\hlopt{-}\hlstd{p))}
\hlstd{+ }    \hlkwd{c}\hlstd{(}\hlkwc{Residual} \hlstd{= sig,}
\hlstd{+ }                                        \hlcom{# random effects variance}
\hlstd{+ }                                        \hlcom{# requires no matrix algebra}
\hlstd{+ }                                        \hlcom{# in this case because the}
\hlstd{+ }                                        \hlcom{# covariance matrix of the}
\hlstd{+ }                                        \hlcom{# random effects}
\hlstd{+ }                                        \hlcom{# is proportional to the}
\hlstd{+ }                                        \hlcom{# identity matrix}
\hlstd{+ }       \hlkwc{Subject} \hlstd{= sig}\hlopt{*}\hlstd{pp}\hlopt{$}\hlstd{theta)}
\hlstd{+ }\hlstd{\})}
\end{alltt}
\begin{verbatim}
Residual  Subject 
  27.374    8.899 
\end{verbatim}
\end{kframe}
\end{knitrout}

\subsection{Additive models}
\label{sec:additive}

It is possible to interpret additive models as a particular class of
mixed models \citep{gamm4}.  The main benefit of this approach is
that it bypasses the need to select a smoothness parameter using
cross-validation or generalized cross-validation.  However, it is
inconvenient to specify additive models using the \pkg{lme4} formula
interface.  The \pkg{gamm4} package wraps the modularized functions of
\pkg{lme4} within a more convenient interface \citep{gamm4}.  In
particular, the strategy involves the following steps,
\begin{enumerate}
\item Convert a \pkg{gamm4} formula into an \pkg{lme4} formula that
  approximates the intended model.
\item Parse this formula using \code{lFormula}.
\item Modify the resulting transposed random effects model matrix, $\bm
  Z\trans$, so that the intended additive model results.
\item Fit the resulting model using the remaining modularized
  functions (Table~\ref{tab:modular}).
\end{enumerate}
Here we illustrate this general strategy using a simple simulated data
set,
\begin{knitrout}
\definecolor{shadecolor}{rgb}{0.969, 0.969, 0.969}\color{fgcolor}\begin{kframe}
\begin{alltt}
\hlstd{> }\hlkwd{library}\hlstd{(}\hlstr{"gamm4"}\hlstd{)}
\hlstd{> }\hlkwd{library}\hlstd{(}\hlstr{"splines"}\hlstd{)}
\hlstd{> }
\hlstd{> }\hlkwd{set.seed}\hlstd{(}\hlnum{1}\hlstd{)}
\hlstd{> }                                        \hlcom{# sample size}
\hlstd{> }\hlstd{n} \hlkwb{<-} \hlnum{100}
\hlstd{> }                                        \hlcom{# number of columns in the}
\hlstd{> }                                        \hlcom{# simulation model matrix}
\hlstd{> }\hlstd{pSimulation} \hlkwb{<-} \hlnum{3}
\hlstd{> }                                        \hlcom{# number of columns in the}
\hlstd{> }                                        \hlcom{# statistical model matrix}
\hlstd{> }\hlstd{pStatistical} \hlkwb{<-} \hlnum{8}
\hlstd{> }                                        \hlcom{# explanatory variable}
\hlstd{> }\hlstd{x} \hlkwb{<-} \hlkwd{rnorm}\hlstd{(n)}
\hlstd{> }                                        \hlcom{# simulation model matrix}
\hlstd{> }\hlstd{Bsimulation} \hlkwb{<-} \hlkwd{ns}\hlstd{(x, pSimulation)}
\hlstd{> }                                        \hlcom{# statistical model matrix}
\hlstd{> }\hlstd{Bstatistical} \hlkwb{<-} \hlkwd{ns}\hlstd{(x, pStatistical)}
\hlstd{> }                                        \hlcom{# simulation model coefficients}
\hlstd{> }\hlstd{beta} \hlkwb{<-} \hlkwd{rnorm}\hlstd{(pSimulation)}
\hlstd{> }                                        \hlcom{# response variable}
\hlstd{> }\hlstd{y} \hlkwb{<-} \hlkwd{as.numeric}\hlstd{(Bsimulation}\hlopt{%*%}\hlstd{beta} \hlopt{+} \hlkwd{rnorm}\hlstd{(n,} \hlkwc{sd} \hlstd{=} \hlnum{0.3}\hlstd{))}
\end{alltt}
\end{kframe}
\end{knitrout}
We plot the resulting data along with the predictions from the
generating model,
\begin{knitrout}
\definecolor{shadecolor}{rgb}{0.969, 0.969, 0.969}\color{fgcolor}\begin{kframe}
\begin{alltt}
\hlstd{> }\hlkwd{par}\hlstd{(}\hlkwc{mar} \hlstd{=} \hlkwd{c}\hlstd{(}\hlnum{4}\hlstd{,} \hlnum{4}\hlstd{,} \hlnum{1}\hlstd{,} \hlnum{1}\hlstd{),} \hlkwc{las}\hlstd{=}\hlnum{1}\hlstd{,} \hlkwc{bty}\hlstd{=}\hlstr{"l"}\hlstd{)}
\hlstd{> }\hlkwd{plot}\hlstd{(x, y,} \hlkwc{las} \hlstd{=} \hlnum{1}\hlstd{)}
\hlstd{> }\hlkwd{lines}\hlstd{(x[}\hlkwd{order}\hlstd{(x)], (Bsimulation}\hlopt{%*%}\hlstd{beta)[}\hlkwd{order}\hlstd{(x)])}
\end{alltt}
\end{kframe}

{\centering \includegraphics[width=\maxwidth]{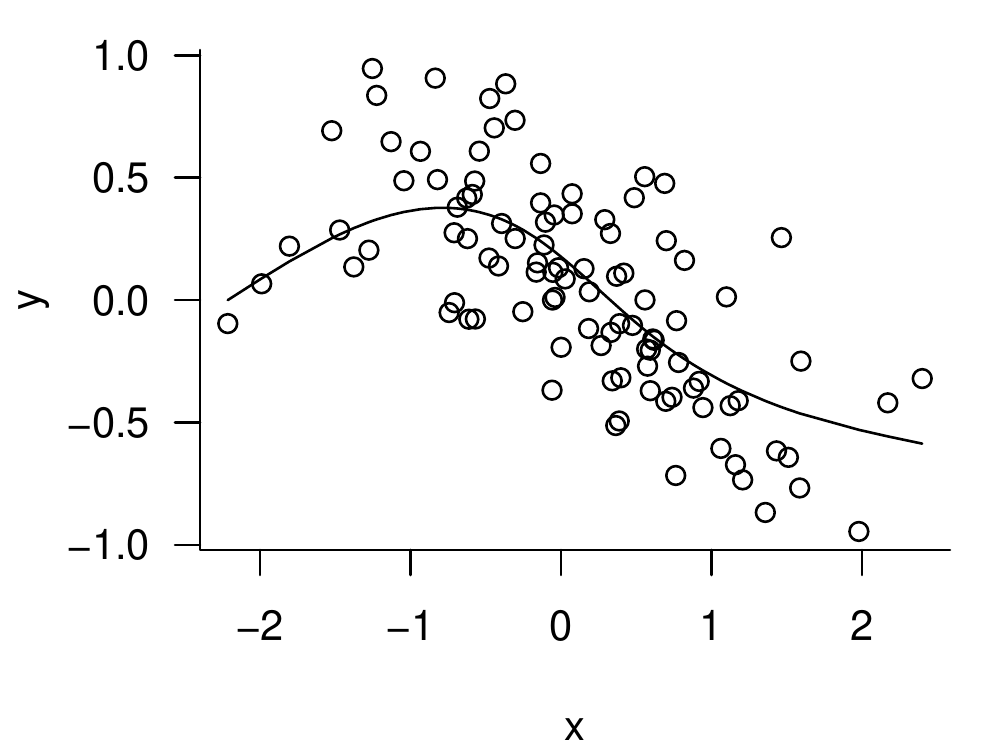} 

}

\end{knitrout}
We now set up an approximate \pkg{lme4} model formula, and parse it
using \code{lFormula},
\begin{knitrout}
\definecolor{shadecolor}{rgb}{0.969, 0.969, 0.969}\color{fgcolor}\begin{kframe}
\begin{alltt}
\hlstd{> }\hlstd{pseudoGroups} \hlkwb{<-} \hlkwd{as.factor}\hlstd{(}\hlkwd{rep}\hlstd{(}\hlnum{1}\hlopt{:}\hlstd{pStatistical,} \hlkwc{length} \hlstd{= n))}
\hlstd{> }\hlstd{parsedFormula} \hlkwb{<-} \hlkwd{lFormula}\hlstd{(y} \hlopt{~} \hlstd{x} \hlopt{+} \hlstd{(}\hlnum{1}\hlopt{|}\hlstd{pseudoGroups))}
\end{alltt}
\end{kframe}
\end{knitrout}
We now insert a spline basis into the \code{parsedFormula} object,
\begin{knitrout}
\definecolor{shadecolor}{rgb}{0.969, 0.969, 0.969}\color{fgcolor}\begin{kframe}
\begin{alltt}
\hlstd{> }\hlstd{parsedFormula}\hlopt{$}\hlstd{reTrms} \hlkwb{<-} \hlkwd{within}\hlstd{(parsedFormula}\hlopt{$}\hlstd{reTrms, \{}
\hlstd{+ }    \hlstd{Bt} \hlkwb{<-} \hlkwd{t}\hlstd{(}\hlkwd{as.matrix}\hlstd{(Bstatistical))[]}
\hlstd{+ }    \hlstd{cnms}\hlopt{$}\hlstd{pseudoGroups} \hlkwb{<-} \hlstr{"spline"}
\hlstd{+ }    \hlstd{Zt} \hlkwb{<-} \hlkwd{as}\hlstd{(Bt,} \hlkwd{class}\hlstd{(Zt))}
\hlstd{+ }\hlstd{\})}
\end{alltt}
\end{kframe}
\end{knitrout}
Finally we continue with the remaining modular steps,
\begin{knitrout}
\definecolor{shadecolor}{rgb}{0.969, 0.969, 0.969}\color{fgcolor}\begin{kframe}
\begin{alltt}
\hlstd{> }\hlstd{devianceFunction} \hlkwb{<-} \hlkwd{do.call}\hlstd{(mkLmerDevfun, parsedFormula)}
\hlstd{> }\hlstd{optimizerOutput} \hlkwb{<-} \hlkwd{optimizeLmer}\hlstd{(devianceFunction)}
\hlstd{> }\hlstd{mSpline} \hlkwb{<-} \hlkwd{mkMerMod}\hlstd{(}   \hlkwc{rho} \hlstd{=} \hlkwd{environment}\hlstd{(devianceFunction),}
\hlstd{+ }                       \hlkwc{opt} \hlstd{= optimizerOutput,}
\hlstd{+ }                    \hlkwc{reTrms} \hlstd{= parsedFormula}\hlopt{$}\hlstd{reTrms,}
\hlstd{+ }                        \hlkwc{fr} \hlstd{= parsedFormula}\hlopt{$}\hlstd{fr)}
\hlstd{> }\hlstd{mSpline}
\end{alltt}
\begin{verbatim}
Linear mixed model fit by REML ['lmerMod']
REML criterion at convergence: 60.73
Random effects:
 Groups       Name   Std.Dev.
 pseudoGroups spline 0.293   
 Residual            0.300   
Number of obs: 100, groups:  pseudoGroups, 8
Fixed Effects:
(Intercept)            x  
    -0.0371      -0.1714  
\end{verbatim}
\end{kframe}
\end{knitrout}
Computing the fitted values of this additive model requires some
custom code, and illustrates the general principle that methods for
\code{merMod} objects constructed from modular fits should only be
used if the user knows what she is doing,
\begin{knitrout}
\definecolor{shadecolor}{rgb}{0.969, 0.969, 0.969}\color{fgcolor}\begin{kframe}
\begin{alltt}
\hlstd{> }                                        \hlcom{# evenly spaced x-values}
\hlstd{> }                                        \hlcom{# for constructing smooth}
\hlstd{> }                                        \hlcom{# curve}
\hlstd{> }\hlstd{xNew} \hlkwb{<-} \hlkwd{seq}\hlstd{(}\hlkwd{min}\hlstd{(x),} \hlkwd{max}\hlstd{(x),} \hlkwc{length} \hlstd{=} \hlnum{100}\hlstd{)}
\hlstd{> }                                        \hlcom{# construct spline matrix}
\hlstd{> }                                        \hlcom{# over these x-values}
\hlstd{> }\hlstd{newBstatistical} \hlkwb{<-} \hlkwd{predict}\hlstd{(Bstatistical, xNew)}
\hlstd{> }                                        \hlcom{# calculate the linear}
\hlstd{> }                                        \hlcom{# predictor}
\hlstd{> }\hlstd{yHat} \hlkwb{<-}   \hlkwd{cbind}\hlstd{(}\hlnum{1}\hlstd{,xNew)} \hlopt{%*%} \hlkwd{getME}\hlstd{(mSpline,} \hlstr{"fixef"}\hlstd{)} \hlopt{+}
\hlstd{+ }        \hlstd{newBstatistical} \hlopt{%*%} \hlkwd{getME}\hlstd{(mSpline,} \hlstr{"u"}\hlstd{)}
\hlstd{> }                                        \hlcom{# plot the results}
\hlstd{> }\hlkwd{par}\hlstd{(}\hlkwc{mar} \hlstd{=} \hlkwd{c}\hlstd{(}\hlnum{4}\hlstd{,} \hlnum{4}\hlstd{,} \hlnum{1}\hlstd{,} \hlnum{1}\hlstd{),} \hlkwc{las}\hlstd{=}\hlnum{1}\hlstd{,} \hlkwc{bty}\hlstd{=}\hlstr{"l"}\hlstd{)}
\hlstd{> }\hlkwd{plot}\hlstd{(x, y)}
\hlstd{> }\hlkwd{lines}\hlstd{(xNew, yHat)}
\hlstd{> }                                        \hlcom{# add true relationship}
\hlstd{> }\hlkwd{lines}\hlstd{(x[}\hlkwd{order}\hlstd{(x)], (Bsimulation}\hlopt{%*%}\hlstd{beta)[}\hlkwd{order}\hlstd{(x)],}\hlkwc{lty}\hlstd{=}\hlnum{2}\hlstd{)}
\hlstd{> }\hlkwd{legend}\hlstd{(}\hlstr{"topright"}\hlstd{,}\hlkwc{bty}\hlstd{=}\hlstr{"n"}\hlstd{,}\hlkwd{c}\hlstd{(}\hlstr{"fitted"}\hlstd{,}\hlstr{"generating"}\hlstd{),}\hlkwc{lty}\hlstd{=}\hlnum{1}\hlopt{:}\hlnum{2}\hlstd{,}\hlkwc{col}\hlstd{=}\hlnum{1}\hlstd{)}
\end{alltt}
\end{kframe}

{\centering \includegraphics[width=\maxwidth]{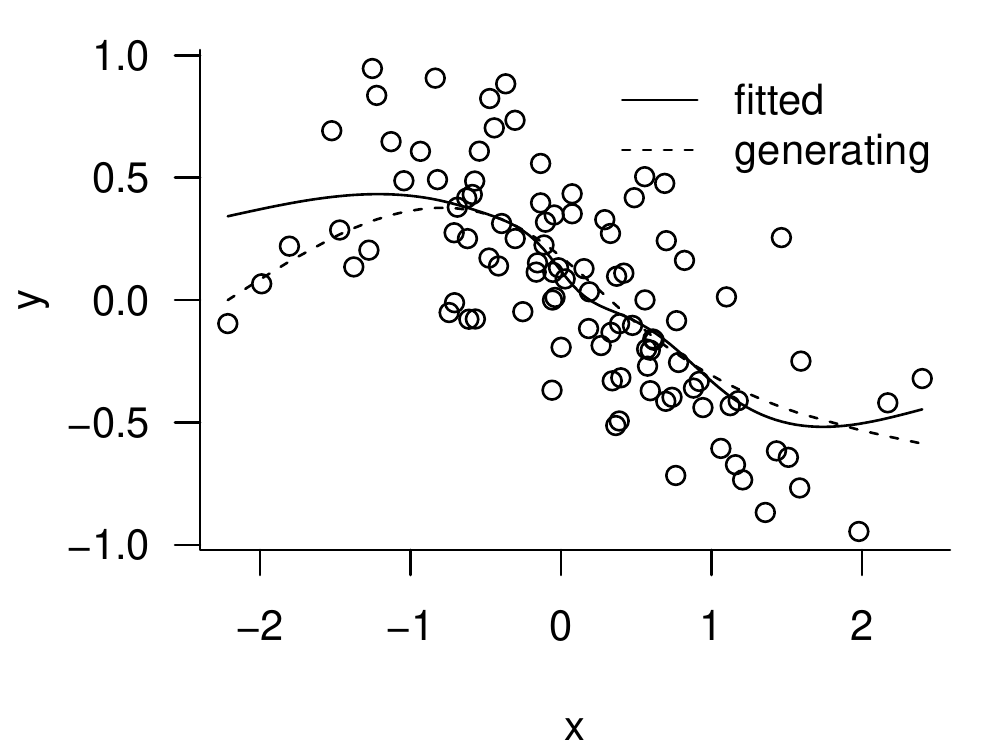} 

}

\end{knitrout}

\end{document}